\documentclass[reqno]{amsart}
\usepackage{amsmath,amsfonts,amsthm,amssymb}

\newtheorem{thm}{Theorem}[section]
\newtheorem{lem}{Lemma}[section]

\newtheorem{prop}[lem]{Proposition}
\theoremstyle{definition}
\newtheorem{defn}{Definition}[section]
\theoremstyle{remark}

\numberwithin{equation}{section}

\newcommand{\eq}[1]{eq.~(\ref{#1})}  
\newcommand{\dn}[1]{(\ref{#1})}  
\newcommand{\Ev}[1]{\E \left( #1 \right)}  
\newcommand{\norm}[1]{\left\Vert#1\right\Vert}

\newcommand{\abs}[1]{\left\vert#1\right\vert}
\newcommand{\set}[1]{\left\{#1\right\}}
\newcommand{\com}[2]{\left[ #1 , #2 \right ]}

\def\const{\mbox{const.}}
\def\eps{\varepsilon}

\def\e{\mathrm e}
\def\im{\mathrm i}
\def\Im{\mathrm{Im}}
\def\half {\frac{1}{2}}
\def\1{{\mathbf 1}}
\def\di{\mathrm d}

\def\Z{\mathbb Z}
\def\N{\mathbb N}
\def\R{\mathbb R}
\def\C{\mathbb C}
\def\E{\mathbb E}  
\def\Fc{{\mathcal F}^c} 

\def\Pr{\operatorname{Prob}} 

\def\dist{\operatorname{dist}}   
\def\tr{\operatorname{Tr}}    
\def\Tr{\operatorname{Tr}}    
\def\supp{\operatorname{supp}}

\def\grad{{\boldsymbol{\nabla}}}
\def\Hi{{\mathcal H}}
\def\L{{\mathcal L}}
\def\I{{\mathcal I}}
\def\J{{\mathcal J}}
\def\A{{\boldsymbol{A}}}
\def\p{{\boldsymbol{D}}}

\begin{document}
\flushbottom
\title[Localization in Random Schr\"odinger Operators]
{Moment Analysis for Localization in Random Schr\"odinger
Operators}
\author[M. Aizenman]{Michael Aizenman}
\address{M. Aizenman \\ Department of Mathematics \\
Princeton University \\ Princeton \\ NJ 08544-1000 \\ USA \\
(also in the Physic Department)}
\email{aizenman@princeton.edu}
\author[A. Elgart]{Alexander Elgart}
\address{A. Elgart \\ Department of Mathematics \\ Stanford University\\ CA 94305-2125 \\ USA}
\email{elgart@math.stanford.edu}
\author[S. Naboko]{Serguei Naboko}
\address{S. Naboko \\ Department of Mathematics \\ St. Petersburg State University \\ 298904 St. Petersburg \\ Russia}
\email{naboko@math.su.se}
\author[J. H. Schenker]{Jeffrey H. Schenker}
\address{J. H. Schenker \\ Theoretische Physik \\ ETH Z\"urich \\ 8093 Z\"urich \\ Switzerland}
\email{jhschenker@itp.phys.ethz.ch}
\author[G. Stolz]{Gunter Stolz}
\address{G. Stolz \\ Department of Mathematics \\ University of Alabama Birmingham \\ Birmingham \\ AL 35294-1170 \\ USA}
\email{stolz@math.uab.edu}

\begin{abstract}
We study localization effects of disorder on the spectral and
dynamical properties of Schr\"odinger operators with random
potentials.   The new results include exponentially decaying
bounds on the transition amplitude and related projection kernels,
including in the mean. These are derived through the analysis of
fractional moments of the resolvent, which are finite due to the
resonance-diffusing effects of the disorder. The main difficulty
which has up to now prevented an extension of this method to the
continuum can be  traced to the lack of a uniform bound on the
Lifshitz-Krein spectral shift associated with the local potential
terms. The difficulty is avoided here through the use of a
weak-$L^1$ estimate concerning the boundary-value distribution of
resolvents of maximally dissipative operators, combined with
standard tools of relative compactness theory.

\end{abstract}

\date{Aug. 19, 2003, revised May 30, 2005}
\maketitle


   \vskip .25truecm         
     \begin{minipage}[t]{\textwidth}
     \tableofcontents
     \end{minipage}
    \vskip .5truecm
\newpage


\section{Introduction}
\label{sec:introduction}

\subsection{Random Schr\"odinger operators}

The addition of disorder can have a profound effect on the
spectral and dynamical properties of a self adjoint differential
operator. We consider here such phenomena for a class of
operators  in $L^2(\R^d)$ of the form
\begin{equation}
   \label{eq:RSO}
   \ H_\omega \ := \  H_0 \ + \ \lambda V_\omega   \, ,
\end{equation}
with the disorder expressed through a random potential $V_\omega
$. In the prototypical example   $H_0 $ is  the Schr\"odinger
operator
\begin{equation}
   H_0  \ = \ - \Delta \ + \ V_0(q) \
\end{equation}
with $\Delta$ the Laplacian  and $V_0(q) $ a bounded periodic
background potential.  The random term $V_\omega$ is given by a
sum of local non-negative ``bumps'', $U_\alpha(q) = U(q-\alpha) $,
centered at the lattice sites $\alpha \in \I=\Z^d$,
\begin{equation} \label{eq-randompot}
  V_{\omega}(q) \ := \  \sum_{\alpha
     \in \I} \eta_{\alpha;\omega}  \,  U_\alpha(q)  \; ,
\end{equation}
with $\set{\eta_\alpha}_{\alpha \in \I} $ a collection of
independent random variables  uniformly distributed in $[0,1]$. It
will be assumed that the space $\R^d$ is covered by the supports
of $\set{U_{\alpha}(\cdot)}$ so that $\inf_q \sum U_\alpha(q)\ge
1$, with the parameter $\lambda \ge 0$ controlling the strength of
the disorder. The subscript $\omega$ indicates a point in a
probability space $(\Omega, \Pr(d\omega))$ and often will be
dropped when it is clear from context we are discussing a random
variable.

More generally, the initial term $H_0$ may incorporate a magnetic
field, i.e., take the form
\begin{equation}
   H_0 \ := \ \p_\A \cdot \p_\A \ + \ V_0(q)
\end{equation}
where $\p_\A = \im \nabla - \A(q)$ with $\A(q)$ the magnetic
vector potential, and the periodicity of $V_0$ and of the bump
potentials  may be replaced by  more relaxed assumptions.   The
required technical conditions, ${\mathcal A}$, are listed in
Section~\ref{sec:model}.

Our objective is to present tools for the study of the phenomenon
known as Anderson localization \cite{Anderson},  which concerns
the potentially drastic effect of the disorder on the dynamical
and spectral properties of the perturbed operator. In general
terms, the effect is that in certain energy ranges the absolutely
continuous spectrum of the unperturbed operator may be modified to
consist of a random dense set of eigenvalues associated with
localized eigenfunctions, and scattering solutions of the
time-dependent Schr\"odinger equation may become dynamically
localized wave packets.

A convenient tool is provided by the Green function $G_E(x,y)$,
which is the kernel of the resolvent operator $(H_{\omega}-E-\im
0)^{-1}$. This kernel is well known to decay exponentially in
$|x-y|$ when $E$ is in the resolvent set \cite{Combes/Thomas}. The
hallmark of localization is rapid (even exponential) decay of
$G_E(x,y)$ at energies in the spectrum, though in this case it
occurs with pre-factors which are not uniform in space and diverge
at a dense countable set of eigenvalues. Rapid decay of the Green
function is related to the non-spreading of wave packets supported
in the corresponding energy regimes and various other
manifestations of localization whose physical implications have
been extensively studied in regards to the conductive properties
of metals \cite{Anderson, Mott/Twose, Thouless,
Abraham/Anderson/Licciardello/Ramakrishnan, Martinelli/Scoppola}
and in particular to the quantum Hall effect \cite{Halperin,
Niu/Thouless/Wu, Avron/Seiler/Simon,
Bellissard/vanElst/Schulz-Baldes, Aizenman/Graf}.

\subsection{Dynamical localization through Green function moment
estimates} \label{sec:1.2}

In presenting our results let us start with a statement which
shows that dynamical localization can be deduced from suitable
bounds on the moments of the Green function.  This relation shows
that moment estimates form a natural and useful tool. For reasons
which will be made apparent later, moments with power $s\ge 1$
diverge in regimes of localization, however, we shall see that
this problem does not affect moments in the fractional range $s\in
(0,1)$, with which we shall work.

We denote here by $H^{(\Lambda)}$ the restrictions of $H$ to open
sets $\Lambda \subset \R^d$.  The default boundary conditions are
Dirichlet, however much of what is said is rather insensitive to
the boundary conditions and can easily be adapted to other
choices, including  Neumann, periodic, or quasi-periodic boundary
conditions.  The latter play a role in our discussion of the
application of density of states bounds
(Section~\ref{sec:smalldensity}).

Throughout we denote the characteristic function of a set
$\Lambda$ by $\1_{\Lambda}$.    It is convenient  to set the
distance unit to $r$\textemdash the size of the ``bumps''
$U_{\alpha}$, as described in  assumption ${\mathcal A}2$ below.
Thus, for $x\in \R^d$ we let $\chi_x = \1_{B_x^r}$, where $B_x^r$
is the ball of radius $r$ centered at $x$.

Decay rates will be expressed below through a distance function
$\dist(x,y) = |x-y|$ for which the choice of the norm on $\R^d$
does not affect our analysis.  It is convenient to interpret it as
$|x| = \sup_j |x_j|$, in which case ``balls'' $B^r_x$ are
hypercubes. We shall also use the domain-adapted distance
\begin{equation}
     \dist_\Lambda (x,y) \ = \ \min
     \set{|x-y|,\dist(x,\Lambda^c)+\dist(y,\Lambda^c)}  \; ,
\end{equation}
for which the boundary of $\Lambda \subset \R^d$ is in effect
regarded  as a single point. As explained in \cite{ASFH} within
the context of discrete operators, the  use of the modified
distance enables the analysis to cover also the cases where
exponential localization in the bulk may possibly coincide with
the occurrence of extended boundary states in certain subdomains.

\begin{thm}\label{cor}
Let $H$ be a random Schr\"odinger operator which satisfies the
regularity assumptions ${\mathcal A}$ (formulated in
Section~\ref{sec:model}). Let $\Omega $ be an open subset of
$\R^d$, and $\Lambda_n$ an increasing sequence of bounded open
subsets of $\Omega $ with $\cup \Lambda_n = \Omega $. Suppose that
for some $0<s < 1$ and an open bounded interval  $\J$ there are
constants $A<\infty$ and $\mu >0$ such that
\begin{equation} \label{eq:corassump}
\int _{\J} \Ev { \|\chi_x \frac{1}{H^{(\Lambda_n)} -E} \chi_y
\|^s} \, \di E  \ \le \  A {e}^{-\mu \dist_{\Lambda_n}(x,y)}
\end{equation}
for all $n\in \N$, $x,y \in \Lambda_n$. Then for every $v<1/(2-s)$
there exists $A_v <\infty$ such that, for all $x,y\in \Omega$,
\begin{equation} \label{eq:corresult}
\Ev {\sup_{g:\,|g|\le 1} \|\chi_x g(H^{(\Omega)})
P_{\J}(H^{(\Omega)})
   \chi_y \|} \ \le A_v \,  {e}^{-v\mu \dist_{\Omega}(x,y)} \; ,
\end{equation}
where the supremum is taken over all Borel measurable functions
$g$ which satisfy $|g|\le 1$ pointwise and $P_{\J}(H^{(\Omega)})$
is the spectral projection for $H^{(\Omega)}$ associated to the
interval $\J$.
\end{thm}

The constant $A_v$ also depends on $s$, $\lambda$, and $E_+=\sup
\J$, as can be seen in the proof, which is in
Section~\ref{sec:proofofthm2}. In particular,  the dependence on
$E_+=\sup \J$ is polynomial with degree slightly larger than
$d/2$. One can also see from the proof that the above result holds
for $s \ge 1$ as well, in which case one can choose $v=1/s$. However,
as explained in Section~\ref{sec:1.3} below, for $s\ge 1$ the
assumption \eqref{eq:corassump} will not be satisfied within the
pure point spectrum.

Of special interest are the following three implications of
\eqref{eq:corresult}.

\noindent     {\em (1) Dynamical localization:\/} With  $g(H) =
\e^{-\im t H}$, eq.~\eqref{eq:corresult} yields for the unitary
evolution operator:
\begin{equation} \label{eq:dynamical}
\Ev {\sup_{t} \|\chi_x e^{-\im t H^{(\Omega)}}
P_{\J}(H^{(\Omega)})
   \chi_y \|} \ \le A_v \, {e}^{-v\mu \dist_{\Omega}(x,y)} \; ,
\end{equation}
which is a strong form of dynamical localization. The result
established here through this criterion is new for continuum
models and has not been obtained with other methods. (The relation
with previous results is  discussed further in
Section~\ref{sec:rel}.)

\noindent     {\em (2) Spectral localization:\/} For $\Omega =
\R^d$ the bound \eqref{eq:corresult}  permits one to further
conclude (using the RAGE theorem as in ~\cite{Graf}) that the
spectrum of $H$ in $\J$ is almost surely pure point with
exponentially decaying eigen-projections, i.e.\ for every $\nu <
\mu/(2-s)$ and $E\in \J$,
\begin{equation} \label{eq:eigprojdecay}
   \norm{ \chi_x \delta_E(H) \chi_y } \ = \ \mathrm{O} (e^{- \nu |x-y|}) \;
   ,
\end{equation}
where $\delta_E(x)=1$ if $E=x$ and $0$ otherwise. An argument
provided in \cite{Combes/Hislop:JFA} shows that almost surely all
eigenvalues of $H$ in $\J$ are finitely degenerate. This allows to
deduce exponential decay of eigenfunctions from
\eqref{eq:eigprojdecay}. The proofs of these results are included
at the end of Section~\ref{sec:infinitevolumes}. Such spectral
localization can also be directly deduced from
\eqref{eq:corassump} using the Simon-Wolff
criterion~\cite{Simon/Wolff} as adapted to continuum operators in
\cite{Combes/Hislop:JFA}.

\noindent {\em (3) Decay of the Fermi-projection kernel:\/}
Another example which plays an important role in physics
applications of the model  involves the Fermi projection
$P_{(-\infty,E_F)}(H^{(\Omega)})$ for $E_F \in \J$. Although not
necessarily of the form $g(H^{(\Omega)}) P_{\J}(H^{(\Omega)})$
since the projection range may be larger than the interval $\J$,
these operators nonetheless satisfy
\begin{equation}\label{eq:fermi}
   \Ev{\sup_{E_F \in \J} \norm{\chi_x P_{(-\infty, E_F)}(H^{(\Omega)})
\chi_y}} \ \le \ \widetilde A
   \e^{-\widetilde \mu \dist_\Omega(x,y)} \; ,
\end{equation}
with constants $\widetilde A < \infty$ and $\widetilde \mu > 0$
whenever \eq{eq:corassump} holds. This may be proved by combining
eqs.\ (\ref{eq:corassump}, \ref{eq:corresult}) and the
Helffer-Sj\"ostrand formula \cite{Helffer/Sjostrand} which is
presented in Appendix A (remark {\rm 11}), or using the argument
of \cite{Aizenman/Graf} \textemdash \ where the issue is discussed
in the context of  lattice operators.

Theorem~\ref{cor} is proven below in
Section~\ref{sec:proofofthm2}.
   Beyond that, the bulk of our
article deals with the derivation of finite volume criteria which
permit to establish the condition \eqref{eq:corassump} for
localization.

\subsection{The reason for fractional moments} \label{sec:1.3}

The criterion  \eqref{eq:corassump} will be of use to us only with
the fractional exponents $s<1$. For $s\ge 1$ the integral over $E$
on the left-hand side of \eqref{eq:corassump} will diverge  even
before the average over the disorder. It is relevant here to note
that in the presence of point spectrum  the Lebesgue measure of
the set of energies at which $\| \chi_x \frac{1}{H^{(\Omega)}-E }
\chi_x \|$ is larger than t exhibits 1/t tails.

For instance, if $\Omega$ is an open subset of $\mathbb{R}^d$ and 
$H^{(\Omega)}$ has only pure-point spectrum in
$\J$, then for any $ \phi \in L^2(\Omega)$:
\begin{equation} \label{eqboole}
\left|  \,  \left \{ E \in \R \, :   \,  |(\phi, \chi_x  \,
\frac{1}{H^{(\Omega)} -E }P_{\J}(H^{(\Omega)})  \, \chi_x  \,
\phi) | \, \ge \, t \right\}  \, \right| \, = \,  \frac{\rm
Const.}{ t}
\end{equation}
with ${\rm Const.} = 2(\phi, \chi_x  \,  P_{\J}(H^{(\Omega)}) \,
\chi_x  \, \phi)$. (To see that, one may use the spectral
representation and the Theorem of Boole~\cite{Boole,A2}.)  Since:
\begin{equation} \label{layercake}
\int_{\R} |Y(E)|^s  \, dE \ = \ \int_0 ^{\infty} |\{ \, E\,  :
|Y(E)| \ge t\}| \   d(t^s) \,
\end{equation}
it follows that the $s$-moments of the Green function seen in
\eqref{eqboole} diverge for all $s\ge 1$, whenever $ (\phi, \chi_x
\, P_{\J}(H^{(\Omega)}) \, \chi_x  \, \phi)\neq 0$, but are
finite for $0<s<1$.

Thus, an important step for our analysis is to show  that  the
left-hand side of \eqref{eq:corassump} is finite.    A highly
instructive statement is the following estimate, which is
formulated in a simplified setting.

\begin{prop} \label{prop:sfinite}
Let $H=-\Delta +V$, with a bounded potential $V$.  Then, for any
$0<s<1$, and  $a,b\in \R$, $x,y\in \R^d$:
\begin{equation} \label{eq:fracs}
   \int_a^b \|  \chi_x   \frac{1}{H -E -\im 0  }
  \chi_y \|^s \, dE \ \le \   C(a,b,d,\|V\|_{\infty},s) \ < \
  \infty \, ,
\end{equation}
where the upper bound holds uniformly in x, y and depends only on
the explicitly listed quantities.
\end{prop}

We will not use Proposition~\ref{prop:sfinite} in the given form,
but rather a related result discussed in the next subsection
below. Still, the following sketch of proof of \eqref{eq:fracs}
may serve to introduce the main ideas behind establishing
finiteness of fractional moments:

As is discussed in Section~\ref{sec:finitemoments}, weak $L^1$
bounds as in \eqref{eqboole} do not have a direct and useful
extension to quantities such as the operator norms considered in
\eqref{eq:fracs}. However the following result, which is valid for
any maximally dissipative operator $A$ and Hilbert-Schmidt
operators $T_1, T_2$, will serve as a key element:
\begin{equation} \label{eq:weakL1}
     \abs{\set{E : \norm{T_1 (A-E -\im 0)^{-1}T_2}_{HS} > t}}
     \ \le \    \frac{C}{t} \norm{T_1}_{HS} \norm{T_2}_{HS}    \, ,
\end{equation}
see Section~\ref{tails} and Appendix~\ref{sec:Weakbound}. Such a
bound implies finiteness of the $s<1$ moments by means of the
``layer-cake'' representation \eqref{layercake} of the integral.
However, first some further work needs to be done since $\chi_x$
and $\chi_y$ are not Hilbert-Schmidt operators. To this end, write
\begin{multline} \label{eq:getHSO}
\chi_x (H -E -\im 0)^{-1} \chi_y \\
= \chi_x (-\Delta+1)^{-1} \chi_y + \chi_x (-\Delta+1)^{-1}
(E-V+1)(H-E-i0)^{-1} \chi_y \, .
\end{multline}
The first term on the right is trivial for the proof of
\eqref{eq:fracs}. The factor $\chi_x (-\Delta+1)^{-1}$ in the
second term is Hilbert-Schmidt if $d\le 3$. As $E-V+1$ is bounded,
the weak bound \eqref{eq:weakL1} becomes applicable after a
similar argument is used on the right of the resolvent. If $d>3$
one can iterate this construction until eventually $(\chi_x
(-\Delta+1)^{-1})^n$ is Hilbert-Schmidt. This also implicitly
justifies the existence of the boundary value in \eqref{eq:fracs},
which is known to exist for Lebesgue-a.e.\ $E$ in the
corresponding expression in \eqref{eq:weakL1}.

\subsection{Finite-volume criteria} \label{sec:1.4}

Our next result deals with finite volume sufficiency criteria for
the localization bounds \dn{eq:corassump}. The basic idea is that
if in some ball $B$ the fractional moments from the center to
$\partial B$ are ``small enough'' then the input criteria of
Theorem~\ref{cor} are satisfied.

This will require an initial step in which the techniques
mentioned above in the context of integrals over $E$ are applied
also to the averages over disorder parameters at fixed energy. By
such means we show that the independent variation of  a local
parameter can resolve a singularity in $\| \chi_x (H + \eta_\alpha
U_\alpha - E)^{-1} \chi_y\|$ which may be present due to the
proximity of the given energy $E$ to an eigenvalue whose
eigenfunction has significant support near $x$ and/or $y$.
Instrumental  for the analysis is the Birman-Schwinger relation:
\begin{equation}\label{eq:BirSch}
U_\alpha^{1/2} \frac{1}{H + \eta_\alpha U_\alpha - z} U_\alpha^{1/2} \  =\
\left[  [U_\alpha^{1/2} \frac{1}{H  - z} U_\alpha^{1/2}]^{-1} +
\eta_\alpha\right]^{-1}
\end{equation}
where $[ \dots ]^{-1}$ is to be interpreted as operator inverse in
$L^2(\supp U_\alpha)$.

The Birman-Schwinger relation makes \eqref{eq:weakL1} applicable
for averages over individual disorder parameters, which take the
role of a ``local energy parameter''.   This strategy motivates
two of our technical assumptions, the covering condition
\eqref{eq:cover} on the single site potentials $U_\alpha$, as well
as the required absolute continuity of the distribution of the
random parameters $\eta_\alpha$. Detailed statements and proofs of
these results are given in Sections \ref{sec:fraclemmas} -
\ref{sec:technical}. In the following, these preliminary bounds
serve as worst-case estimates, somewhat reminiscent of the role of
Wegner estimates in multi-scale analysis.

Let $r_0$ be the independence length introduced below next to the
assumption ${\mathcal IAD}$. Also define the boundary layer of a
set $\Lambda$ to be the (open) set
\begin{equation}
\delta \Lambda := \{ q \ : \ r<\dist(q, \Lambda^c) < 23r\},
\end{equation}
where the choice of the depth is somewhat arbitrary, but
convenient for our argument.

\begin{thm}\label{thm:1}
   Let $H$ be a random Schr\"odinger operator
   which satisfies the assumptions ${\mathcal A}$ as well as
   ${\mathcal IAD}$.  For each
   $s \in (0,1/3)$, $\lambda>0$ and $E\in \R$, there exists $M(s,\lambda,E) < \infty$,
   such that if for some $L > r_0 + 23r$,
   \begin{eqnarray}\label{eq:criterion}
     \e^{-\gamma} & := & M(s,\lambda,E)
     (1 + L)^{2(d-1)} \ \sup_{\alpha \in \I} \limsup_{\eps \downarrow 0}
     \E \left ( \norm{\chi_{\alpha}\,
     \frac{1}{H^{(B_{\alpha}^L)} -E-\im \eps} \, \1_{\delta
     B^L_\alpha} }^s
     \right ) \nonumber \\ & < & 1 \; ,
   \end{eqnarray}
   then there exists $A(s,\lambda,E)$ such that for any open $\Omega \subset \R^d$
   and any $x,y \in \Omega$
   \begin{equation}\label{eq:thm1conclusion}
     \limsup_{\eps \downarrow 0} \E \left (
       \norm{\chi_x \frac{1}{H^{(\Omega)} -E - \im \eps}
     \chi_{y}}^s \right )
     \ \le \e^\gamma A(s,\lambda,E) \, \e^{- \gamma \dist_\Omega(x,y) / 2 L}
     \; .
   \end{equation}

   Here, the constants $M(s,\lambda,E)$, $A(s,\lambda,E)$ can be chosen
   polynomially bounded: in $E$, in $1/\lambda$ as $\lambda \rightarrow 0$,
   and in $\lambda$ as $\lambda \rightarrow \infty$; if $H$
   satisfies also ${\mathcal A}3'$  then they can be chosen uniformly bounded
   in $\lambda>1$.  Furthermore, for any bounded region
   $\Omega$ \eq{eq:thm1conclusion} holds also with $\eps =0$.
\end{thm}

The above result serves as a finite-volume criterion for
localization. Eq.~(\ref{eq:thm1conclusion}) reflects the fact that
the scale $L$ at which \eqref{eq:criterion} is verified indeed
determines the localization length. The locally uniform $E$-bound
of the constant in \eqref{eq:thm1conclusion} allows one to deduce
integral bounds as required in \eqref{eq:corassump} of
Theorem~\ref{cor}. Theorem~\ref{thm:1} is proven in
Section~\ref{sec:fin-vol-criteria} with methods similar to those
developed in \cite{ASFH} in the proof of a corresponding result
for lattice operators. This uses what is frequently called the
{\em geometric resolvent identity}
(Lemma~\ref{lem:resolvent_expansion}) as well as decoupling and
re-sampling arguments to factorize expectations.

\subsection{Applications} \label{sec:appl}

In Section~\ref{sec:applications} we show how the general
framework provided here can be used to prove localization in
specific disorder regimes. This includes the familiar large
disorder (Theorem~\ref{thm:largedisorder}) and band edge or
Lifshitz tail regimes. For the latter we provide two results, one
based on smallness of the finite volume density of states
(Theorem~\ref{thm:bandedge}) and another
--- less traditional --- result using smallness of the infinite
volume density of states (Theorem~\ref{thm:lowdensity}). We also
show in Theorem~\ref{thm:msaappl} that the ``output'' of a
multi-scale analysis can be used to provide the ``input'' for
Theorem~\ref{thm:1}, thus proving that the stronger results found
by our methods hold throughout the multi-scale analysis regime. A
useful technical result (Lemma~\ref{lem:holdercont}) is a
continuity property of fractional resolvent moments. It shows that
in applications it suffices to check the bound
\eqref{eq:criterion} at a single energy.

This observation also leads to a proof of the following
complementary criterion which rounds off our discussion.

\begin{thm}\label{thm:necessary}
Let $H$ be a random Schr\"odinger operator which satisfies
${\mathcal A}$ and ${\mathcal IAD}$. Suppose that for some
$A<\infty$, $\mu>0$ and $E \in \R$
\begin{equation} \label{eq:greenloc}
\limsup_{\varepsilon \downarrow 0} \E
     \left ( \|\chi_\alpha \frac{1}{H - E -\im \varepsilon}
         \chi_\beta \|^s \right ) \le A {e}^{-\mu |\alpha-\beta|}
\end{equation}
for all $\alpha, \beta \in \I$. Then, for sufficiently large $L$,
\eq{eq:criterion} is satisfied uniformly for all $E'$ in an open
neighborhood of $E$.
\end{thm}

Combined with Theorems~\ref{cor} and \ref{thm:1} this shows that
exponential decay of the Green function as in \eq{eq:greenloc}
provides a necessary and sufficient condition for
\eq{eq:thm1conclusion}, and thus \eq{eq:corresult}, to hold in a
neighborhood of an energy $E$. Thus, in principle the entire
regime of localization in the sense of \eq{eq:thm1conclusion} may
be mapped out using the criterion provided by Theorem~\ref{thm:1}.

The applications of Section~\ref{sec:applications} are important
to tie our general method with concrete examples. But we stress
that these examples are somewhat secondary to the main goal of
this work, which is the outline of a general framework for
studying localization properties consisting of the three
interrelated Theorems~\ref{cor}, \ref{thm:1}, and
\ref{thm:necessary} and based on the preliminary bounds obtained
in Section~\ref{sec:s-moments}.

As such, we do not attempt to give an exhaustive list of
applications here, but rather try to illustrate how known methods
may be combined with the arguments developed in this paper.
Further developments based on the new techniques will be left to
future work. For example, the work \cite{BdMNSt2} will use a
variant of the techniques used here to study continuum random
surface models.

\subsection{Relation with past works} \label{sec:rel}

Mathematical analysis of localization for random operators has
been a very active field. The continuum operators have analogs in
the discrete setting,  obtained by replacing
  \eqref{eq:RSO} with analogous operators
on the $\ell^{2}$ space  of a graph such as $\Z^{d}$; i.e.,
replacing  the differential operator $\p_{\A}\cdot \p_{\A}$ by a
``hopping matrix''  and the potential $V_\omega(q)$ by a
multiplication operator with $\{ V_\omega(x)\}_{x\in \Z^d}$  iid
random variables. Localization phenomena are rather similar in the
two setups, in broad terms, but the discrete case is simplified by
the fact that the random coefficients affect terms of finite rank,
often just rank-$1$ or rank-$2$. Naturally, various analytical
arguments were initially developed in that setup, although this
was not true for the analysis in one dimension which provided the
first rigorous results, initiated in \cite{GMP}, and where special
tools are available.

The existing results on localization in the continuum of
dimension larger than one have been based on the multiscale
analysis, first obtained for discrete operators
\cite{Frolich/Spencer, Delyon/Levy/Souillard,
Frolich/Martinelli/Scoppola/Spencer, Simon/Wolff,
vonDreifus/Klein} and then extended to the continuum
\cite{Kotani/Simon, Combes/Hislop:JFA, Germinet/Klein:Bootstrap}.
We do not attempt to give an exhaustive survey of the related
literature. The reader is referred to the recent
book of P. Stollmann \cite{Stollmann:caught} for a review of the
history of the subject and a gentle introduction to the
multi-scale analysis\textemdash which is not used here. Our work
presents a continuum version of the fractional-moment method which
was developed for discrete systems in
\cite{Aizenman/Molchanov,A2,Aizenman/Graf,ASFH}. The
fractional-moment method was applied already to certain continuum
models in which some crucial features from the discrete case
persist, in particular the single site perturbations are rank
one~\cite{Hislop/Kirsch/Krishna, Dorlas/Macris/Pule}.

It seems appropriate to make a brief comparison of the two
approaches which have been developed to handle multidimensional
localization,  the multi-scale analysis (MSA) and the fractional
moment method (FMM). Both have now been found to apply to discrete
as well as continuum models. They lead to similar results:
spectral and dynamical localization, though expressed through
somewhat different estimates, and apply to essentially the same
disorder and energy regimes.

Two significant differences lie in: {\em i)\/}  the iterative
schemes which are used in the two methods for the derivation of
results for the infinite volume from finite volume characteristics
of localization, and {\em ii)\/}   the tools used to express
localization. MSA uses a KAM-type strategy through an infinite
collection of scales, whereas  FMM is a {\em single scale} method
- that of the localization length.  In MSA, the quantity to be
controlled is the probability of rare events, whereby the random
configuration locally manifests traits which may inhibit
localization.   In FMM the localization is expressed through rapid
decay of   the Green function's suitable (fractional) moments.

While MSA is a multiscale method, in FMM once a scale is reached
at which the finite volume localization criterion is met, all
spatial correlators are shown to decay exponentially on that
length scale, including in the mean, i.e., error estimates are
also exponentially small. The technical parts of the proof become
more involved in the continuum, as is also true for other methods,
but the basic mechanism of working with just one scale remains. A
particular consequence of this is that FMM yields exponential
decay of the Green function fractional moments in
\eqref{eq:thm1conclusion} and, subsequently, also in the dynamical
localization bound \eqref{eq:corresult}. This is a bit better than
the best result of this kind obtained through MSA, which
  is a bound of the form $C_{\zeta}\exp(-|x-y|^{\zeta})$ for any $\zeta<1$, see
\cite{Germinet/Klein:Transition}.  The limiting factor there is
the estimate for the probability of \emph{a-typical}
configurations, for which the multiscale scheme yields a fast, yet
suboptimal decay rate.

Possible directions for the extension of the analysis presented
here, which does not cover all the results which were derived
using MSA, include: removal of the condition that the random
potential bumps fully cover the space, which is required in
\eqref{eq:cover} (this will be addressed in \cite{BdMNSt2}), and
the relaxation of the regularity of the distribution of the random
parameters $\eta_{\alpha}$. The case which may be well beyond the
reach of the averaging methods used here, even under some natural
improvements, is that of $\eta_\alpha$ having the discrete
``Bernoulli'' distribution.

\subsection{Assumptions} \label{sec:model}

Following are the regularity conditions required for our results.
The condition ${\mathcal A}$ refers to the collection ${\mathcal
A}1$-${\mathcal A}3$.
\begin{enumerate}
\item[${\mathcal A}1$] The components of the vector potential
$\A$, its first derivatives $\partial_i \A$ for $i=1,...,d$, and
$V_{o,+}$ (the positive part of $V_0$) are locally bounded on
$\R^d$. The negative part $V_{o,-}$ of $V_0$  is bounded.

\item[${\mathcal A}2$] Each function $U_\alpha$ is bounded,
   non-negative, supported in a ball of radius $r$ around $\alpha$ for fixed
   $r >0$ and some $\alpha \in \I$, with $\I$ a discrete set of points
in $\R^d$.
   The number of points  $\alpha$  falling within any unit cube is
   uniformly bounded by some $N<\infty$,
   and the function
   \begin{equation}
     F(q) \ := \ \sum_{\alpha \in \I} U_\alpha(q) \;
   \end{equation}
   satisfies uniform bounds
   \begin{equation} \label{eq:cover}
     1 \ \le \ \inf_q F(q) \ \le  \ \sup_q F(q) \ =: \  b_+ \ < \
     \infty \; .
   \end{equation}
   Moreover,  $|\partial (\supp U_{\alpha})| =0$.

\item[${\mathcal A}3$]  The random variables $\eta_\alpha , \,
\alpha \in
       \I$ take values in $[0,1]$ and the conditional distribution of
       $\eta_\alpha$ at specified values of $\{\eta_\zeta\}_{\zeta \neq
       \alpha}$  has a density, denoted $\rho_\alpha(\eta \, | \, \omega)$,
       which is uniformly bounded: \begin{equation}\label{eq:D} D \ := \
       \sup_{\alpha} \norm{\rho_\alpha(\cdot | \cdot)}_\infty \  < \
       \infty \end{equation} where $\norm { \cdot }_\infty$ indicates the
       essential supremum over $\eta$ and $\omega$.

\end{enumerate}

Assumption ${\mathcal A}$ is sufficiently general to cover many
important examples, such as the two-dimensional Landau hamiltonian
with random potential, where $\A(q_1,q_2) = \frac{1}{2} (-Bq_2,
Bq_1)$. For both $\A$ and $V_0$ one could allow suitable,
dimension-dependent, $L^p$-type singularities, but we prefer to
avoid the additional technicalities which are caused by this.

A key requirement for Theorem \ref{thm:1} is ``independence at a
distance,'' namely the random functions obtained by restricting
$V_\omega$ to well separated regions are pairwise independent.
\begin{enumerate}
\item[${\mathcal IAD}$:] There exists $r_0>0$ such that
     if $\Lambda,\Lambda' \subset \R^d$ with $\dist(\Lambda,\Lambda')>r_0$ then
     the collections of random variables
     $\eta_\Lambda := \{\eta_\zeta : \zeta \in \Lambda\cap \I\}$ and
     $\eta_{\Lambda'} := \{\eta_\zeta : \zeta \in \Lambda' \cap \I \}$
     are independent , i.e.,
     \begin{equation}
     \Ev{f(\eta_\Lambda) g(\eta_{\Lambda'})} \ = \
     \Ev{f(\eta_\Lambda)} \Ev{g(\eta_{\Lambda'})} \; ,
     \end{equation}
     for arbitrary bounded measurable functions $f,g$ on
     $\R^{\Lambda \cap \I}$ and $\R^{\Lambda' \cap \I}$, respectively.
Without loss
     of generality we assume that $r_0 \ge 2r$.
\end{enumerate}
We note that this assumption is not required for the proof of  the
boundedness of the fractional moments in
Section~\ref{sec:s-moments} and also not for the derivation of
localization from the Green function decay (Theorem~\ref{cor}).

The restriction of  $\eta_{\alpha}$ to range over $[0,1]$ is not
essential; through the adjustment of the background potential and
the disorder parameter that range can be replaced by any other
bounded interval. However, the normalization becomes relevant when
one considers the strong disorder regime.

Some of the bounds derived below\textemdash c.f.,
Lemmas~\ref{lem:finite_smoment} and \ref{lem:decouple}\textemdash
exhibit coefficients which grow with increasing $\lambda$.  For
applications of these bounds to the ``large disorder regime''
($\lambda >> 1$) it is useful to break the coupling $\lambda
\eta_\alpha$ into a sum of variables $\eta_{1;\lambda} +
\eta_{2;\lambda}$ such that $\eta_{1;\lambda}$ is of order one and
obeys ${\mathcal A}3$. This could be accomplished in a number of
ways\textemdash e.g., let $\eta_{1;\lambda}$ be the fractional
part of $\lambda \eta_\alpha$.  However, without an additional
assumption these decompositions might become more and more
singular as $\lambda$ increases. Following is a useful notion.
\begin{defn}
   A real valued random variable $X$, with an absolutely continuous
   probability measure of density
   $\rho(\cdot)$ on $\R$, is  {\em blow-up regular}
   if there exist two sequences of real valued random variables, $\{ X^{(n)} : n
   \ge 1\}$ and $\{ Y^{(n)} : n \ge 1 \}$ such that
   \begin{enumerate}
     \item $X^{(n)}$ takes values in $[0,1]$.
     \item For each $n \ge 1$,
     \begin{equation}\label{eq:decompose}
       n X \ = \ Y^{(n)} \ + \ X^{(n)} \; .
     \end{equation}
     \item The conditional distribution of $X^{(n)}$,
       at a specified value $Y^{(n)} = y$, has a bounded density, $\rho_n(\cdot
       | y)$, and
       \begin{equation}\label{eq:finite}
      D\equiv \sup_{n\ge1; \, x,y\in \R}\abs{\rho_n(x| y)} \  <  \  \infty  \; .
       \end{equation}
    \end{enumerate}
The {\em blow-up
   norm} of  the probability density $\rho$, denoted $D_\rho$, is the  infimum
  of the above quantity $D$ taken over all sequences
  $\set{ X^{(n)},Y^{(n)} : n\ge 1}$ satisfying (1) and (2).
\end{defn}

The following assumption, in lieu of ${\mathcal A}3$, allows
better bounds for the strong disorder regime.
\begin{enumerate}
\item[${\mathcal A}3'$] The random variables $\eta_\alpha , \,
\alpha \in
     \I$ take values in $[0,1]$, and for each $\alpha$
     the conditional distribution of $\eta_\alpha$ at
     specified values of $\{ \eta_\zeta\} _{\zeta \neq \alpha}$ is blow-up
     regular, with the  blow-up norms bounded by a common
      $D < \infty$.
\end{enumerate}

The above condition is satisfied for independent uniform in
$[0,1]$ random variables, and also for i.i.d.\ variables with a
common density $\rho$ provided $\ln \rho$ is Lipshitz-continuous
in $[0,1]$ (see Appendix~\ref{sect:technicomm}).

\subsection{Outline of contents} \label{sec:contents}

The contents of Sections~\ref{sec:proofofthm2},
\ref{sec:s-moments}, \ref{sec:fin-vol-criteria} and
\ref{sec:applications} were discussed in Sections~\ref{sec:1.2},
\ref{sec:1.3}, \ref{sec:1.4} and \ref{sec:appl}, in that order.
Thus we focus here on briefly describing the contents of the four
Appendices of this paper:

The assumptions and results outlined above deserve a number of
more technical comments. In order to keep the introduction
relatively free of technicalities, further discussion of these is
postponed to Appendix~\ref{sect:technicomm}.

Appendix~\ref{sec:tools} contains a short description of the
Birman-Schwinger relation, which is used throughout this work.

The weak $L^1$ bound \eqref{eq:weakL1}, central to the extension
of the fractional moment method to the continuum, has not been
used previously in the literature on random operators. We thus
present a self-contained proof based on properties of the
vector-valued Hilbert transform in Appendix~\ref{sec:Weakbound}.
Much of the main argument is taken from \cite{Naboko}.

In Appendix~\ref{sect:app-spect-shift} we show how the methods of
Section~\ref{sec:s-moments} can be used to derive a bound for the
disorder averaged spectral shift which is locally uniform in
energy. As discussed in Section~\ref{sec:finitemoments}, such
bounds do not hold without averaging over the disorder. We do not
use this result in our main argument, however it may be of
independent interest.

\newpage

\section{From resolvent bounds to eigenfunction correlators and
dynamical localization} \label{sec:proofofthm2}

In this section we prove Theorem~\ref{cor}.  For discrete models
such results were derived in refs.\ \cite{A2,ASFH} employing the
observation that the eigenfunctions of the operator $H$ play the
role of Green functions for a re-sampled operator $\hat H$, at
other values of the coupling variables.  Key in that analysis were
properties of rank one perturbations.  Use was also made of a very
convenient interpolation argument which permits to extract bounds
on the off-diagonal matrix elements of the spectral projections of
$H$ from fractional moment bounds on the Green function of $\hat
H$.

We find that the approach of refs.\ \cite{A2,ASFH} can also be
applied to continuum operators, with the arguments which were
based on rank one perturbation replaced, or generalized, by
considerations of the Birman-Schwinger operator.

\subsection{Correlators\textemdash eigenfunction and other}
In the discussion of dynamical localization for discrete random
operators in ref.~\cite{ASFH}, estimates were developed for the
spectral measures $\mu_{x,y}$ which are defined through the Riesz
theorem by
\begin{equation}
\int f(E) \mu_{x,y}(\di E) \ = \  \langle x | f(H) | y \rangle \,
, \qquad f \in C_0(\R)
\end{equation}
Exponential decay (in $|x-y|$)  of the {\em total variation} of
these measures provided a strong description of dynamical
localization.  For analogous bounds in the present context, it is
convenient to work with the ``operator valued measures,''
\begin{equation}
   f \mapsto \chi_x f(H^{(\Omega)}) \chi_y \qquad f \in C_c(\R) \;
   .
\end{equation}
We introduce also the ``total variation'' of these measures,
\begin{equation} \label{eq:totalvar}
   Y_\Omega(\J;x,y) \ := \ \sup_{\substack{f \in C_c(\J) \\ \norm{f}_\infty \le
   1}} \norm{ \chi_x f(H^{(\Omega)}) \chi_y} \; ,
\end{equation}
defined for bounded open intervals $\J$ where $C_c(\J)$ denotes
the continuous functions compactly supported inside $\J$.

For a finite region $\Lambda$ and fixed $\alpha \in \I$, we use
the Birman-Schwinger relation  of Appendix~\ref{sec:tools} to study
the dependence of $H^{(\Lambda)}$ on the single random parameter
$\eta_{\alpha}$, keeping $\{ \eta_{\beta}\}_{\beta \not= \alpha}$
fixed. We express $H_{\eta_{\alpha}} \equiv H^{(\Lambda)}$ in
terms of a re-sampled reference operator $H_{\hat{\eta}_{\alpha}}$
as
\begin{equation} \label{eq:hatoperator}
H_{\eta_{\alpha}} = H_{\hat{\eta}_{\alpha}} - \lambda
(\hat{\eta}_{\alpha} - \eta_{\alpha}) U_{\alpha},
\end{equation}
where we will take the re-sampled variable $\hat{\eta}_{\alpha}$
to have the same conditional distribution as $\eta_{\alpha}$. The
family $H_{\eta_\alpha}$ has the form of the one-parameter family
\eq{eq:oneparSch} of Appendix~\ref{sec:tools}, with $H_0 =
H_{\hat{\eta}_{\alpha}}$, $V=U_{\alpha}$ and $\xi = \lambda
(\hat{\eta}_{\alpha} - \eta_{\alpha})$.

For  $0\le v \le 2$ we define the fractional ``eigenfunction
correlators'' as
\begin{equation}
     Q_v(\J; x, \alpha) \ = \ \sum_{n:\, E_n\in \J}
     \langle \chi_x \psi_n, \psi_n \rangle^{v/2} \
    \langle U_{\alpha} \psi_n, \psi_n \rangle^{1-v/2} \; ,
     \label{eq:Q}
\end{equation}
where $n$ labels  the eigenvalues $E_n = E_n(\xi)$ and the
corresponding orthonormal eigenfunctions $\psi_n = \psi_n(\xi)$ of
$H_{\eta_{\alpha}}$, choosing the labeling so that these are
holomorphic in $\xi$, as in Appendix~\ref{sec:tools}.

For $v=1$ the eigenfunction correlators provide bounds for
$Y_{\Lambda}(\J;x,y)$: If $f\in C_c(\J)$, then $f(H^{(\Lambda)}) =
\sum_{n:E_n\in \J} f(E_n) P_{\psi_n}$, where $P_{\psi_n}$ is the
orthogonal projector onto $\psi_n$. Thus, by the ``covering
condition'' \dn{eq:cover},
\begin{equation} \label{eq:YQbound}
\begin{aligned}
Y_{\Lambda}(\J;x,y) \le &\  \sum_{\substack{\alpha \in \I \\
\abs{y -
  \alpha} \le 2r}} \sup_{\substack{f\in C_c(\J) \\ |f|\le 1}}
  \norm{ \chi_x f(H^{(\Lambda)}) U_{\alpha}^{1/2}} \\ \le &\
  \sum_{\substack{\alpha \in \I \\  \abs{y -
  \alpha} \le 2r}} \sum_{n:E_n \in \J} \norm{ \chi_x P_{\psi_n}
  U_{\alpha}^{1/2}} \\ = &\  \sum_{\substack{\alpha \in \I \\  \abs{y -
  \alpha} \le 2r}} Q_1(\J;x,\alpha) \; .
\end{aligned}
\end{equation}
Here it was used that $\norm{\chi_x P_{\psi_n} U_{\alpha}^{1/2}} =
\norm{\chi_x \psi_n} \norm{ U_{\alpha}^{1/2} \psi_n}$.

For all $0< v < 2$ the quantity $Q_v(\J; x, \alpha)$ reflects the
overlap between the eigenfunctions at $x$ and $\alpha$.  That,
however, is not the case at the end-points $v = 0, 2$ for which
the corresponding values of $Q_v$ depend only on the density of
states:
  \begin{eqnarray}
     Q_0(\J; x, \alpha) & = & \tr U_{\alpha} \, P_{\J}(H^{(\Lambda)})
     \nonumber  \\
         Q_2(\J; x, \alpha) & = & \tr  \chi_x \,
P_{\J}(H^{(\Lambda)})
     \label{eq:Q01}
  \end{eqnarray}
with $P_{\J}(H^{(\Lambda)})$ the spectral projection operator. For
the Schr\"odinger operators considered here, both $Q_0(\J; x,
\alpha)$ and $Q_2(\J; x, \alpha)$ are of order one, for a finite
interval $\J$, in the sense that they are finite and do not decay
for increasing $\dist(x, \alpha)$. In particular, if $\J \subset
(-\infty, E)$ and $p>d/2$ we have
\begin{equation} \label{eq:density}
\begin{aligned}
Q_2(\J;x,\alpha)  \le &\ \Tr{\chi_x P_{\le E}(H^{(\Lambda)})}
\\ \le &\
(|E-E_0|+1)^{p} \Tr{\chi_x (H^{(\Lambda)} +E_0+1)^{-p}}\\
\le &\ C(|E-E_0|+1)^{p} \; ,
\end{aligned}
\end{equation}
where here and in the following we set $E_0 = \inf \sigma(H_0)$.
This bound is deterministic, and thus holds also for
$\E(Q_2(\J;x,\alpha))$.

\begin{lem} \label{lem:interpol}
$Q_v(\J; x, \alpha)$ is log convex in $v$, and  for any $v\in
(0,1)$:
\begin{equation}
\Ev{Q_1(\J)}  \ \le \ \Ev{Q_v(\J)}^{1/(2-v)}\
\Ev{Q_2(\J)}^{(1-v)/(2-v)}  \, . \label{eq:Einterpol}
\end{equation}
at any value of the (omitted) argument $(x, \alpha)$ of $Q_v$.
  \end{lem}

  \begin{proof}
The log convexity of $Q_v$ in $v$ is a standard observation for a
function of the form  $F(v) = \sum_n A_n \, B_n^v $. In
particular, for $ v < 1 < 2$, writing $1$ as a convex combination
of the other values: $1= a v+ (1-a) 2$,  one gets via the H\"older
inequality: $F(1) \le F(v)^a \, F(2)^{1-a}$ (with $a=1/(2-v)$). In
the present context that yields
\begin{equation}
   Q_1(\J)  \ \le \ Q_v(\J)^{1/(2-v)}\
   Q_2(\J)^{(1-v)/(2-v)} \, .
\label{eq:interpol}
\end{equation}
One more application of the H\"older inequality, this time to the
average   over the randomness ($\E(\cdot)$), yields
\eq{eq:Einterpol}.
  \end{proof}

\subsection{Eigenfunction correlators and resolvent moments}

Here we will relate the $Q_v$ to fractional resolvent moments by
applying the results of Lemma~\ref{lem2.2} to the family
\eq{eq:hatoperator}. By \eq{eq:FeynHell} and \eq{eq:FHconseq} we
have
\begin{equation}
\frac{d}{d\xi} E_n = - \langle U_{\alpha} \psi_n, \psi_n \rangle
\end{equation}
and for $\Gamma_n(E)$, the inverse function of $E_n$,
\begin{equation} \label{eq:gammaderiv}
\frac{d}{dE} \Gamma_n(E) = - \frac{1}{\langle U_{\alpha}
\psi_n(\Gamma_n(E)), \psi_n(\Gamma_n(E)) \rangle} \; .
\end{equation}

For $E \not\in \sigma(H_{\eta_{\alpha}})$ define
$K_{\eta_{\alpha},E} := U_{\alpha}^{1/2}
(H_{\eta_{\alpha}}-E)^{-1} U_{\alpha}^{1/2}$. If $E\not\in
\sigma(H_{\hat{\eta}_{\alpha}})$ then, by Lemma~\ref{lem2.2},
$\{\Gamma_n(E)\}$ are the repeated eigenvalues and $\phi_n(E) :=
U_{\alpha}^{1/2} \psi_n(\Gamma_n(E))$ corresponding complete
(non-normalized) eigenvectors for the unbounded self-adjoint
operator $K_{\hat{\eta}_{\alpha},E}^{-1}$. With this notation we
have

\begin{thm}
If $\eta_{\alpha} \not= \hat{\eta}_{\alpha}$ and
$\sigma(H_{\eta_{\alpha}}) \cap \sigma(H_{\hat{\eta}_{\alpha}})
\cap \J = \emptyset$, then the eigenfunction correlator
$Q_v(\J;x,\alpha)$ for $H^{(\Lambda)}$ admits the following
representation:
\begin{multline} \label{eq:Qrep}
     Q_v(\J; x, \alpha) \ = \ \sum_n \int_{\J} dE  \
     \delta( \Gamma_n(E) + \lambda( \eta_\alpha -\hat \eta_\alpha) ) \
      |\Gamma_n(E)|^{v} \\ \times
     \norm{\chi_x \, (H_{\hat{\eta}_{\alpha}} - E)^{-1}\,   U_{\alpha}^{1/2}
     \, \phi _n(E) }^{v} \big/ \norm{ \phi_n(E) }^v \: .
\end{multline}
Furthermore, for any $E$ and $a < b$ such that $E$ is not an
eigenvalue of $H_a$ or $H_b$,
\begin{equation}  \label{eq:count}
\int_a^b \di \eta_{\alpha}  \  \sum_{n}  \
     \delta( \Gamma_n(E) + \lambda( \eta_{\alpha} -\hat{\eta}_{\alpha}) ) \
        = \ \left[ \tr P_{\le E}(H_a) \ - \
      \tr P_{\le E}(H_b) \right] \, /
     \,  \lambda \; .
\end{equation}
  \end{thm}
\noindent{\em Remark:} Using Lemma~\ref{lem2.2} it is easy to see
that the condition $\sigma(H_{\eta_{\alpha}}) \cap
\sigma(H_{\hat{\eta}_{\alpha}}) \cap \J = \emptyset$ holds for
Lebesgue almost every $\eta_\alpha$.
  \begin{proof}
As $\Gamma_n(E) = E_n^{-1}(E)$ we have for arbitrary
$\eta_{\alpha}$ that $E\in \sigma(H_{\eta_{\alpha}})$ if and only
if $\Gamma_n(E) = \xi=\lambda (\hat{\eta}_{\alpha} -
\eta_{\alpha})$ for some $n$. Thus one may express sums over
eigenvalues as integrals with suitably weighted
$\delta$-functions:
\begin{equation}
\sum_{n:\,E_n \in \J} \, \ldots \ = \ \int_{\J} \di E \, \sum_n
\delta\left( \Gamma_n(E) + \lambda \, \Delta \eta_\alpha \right)
\,  \left| \frac{\di}{\di E} \Gamma_n(E)
  \right| \, \ldots   \, ,
\end{equation}
where $\Delta \eta_{\alpha} = \eta_{\alpha} -
\hat{\eta}_{\alpha}$. In particular, eq.~(\ref{eq:gammaderiv})
implies that
\begin{equation}
\begin{aligned}
     Q_0(\J; x, \alpha) \ = &\ \sum_{n:\, E_n\in \J}
       \langle U_{\alpha} \psi_n (\Gamma_n(E)), \psi_n (\Gamma_n(E)) \rangle \\
= &\ \int_{\J} \di E \, \sum_n \delta\left( \Gamma_n(E) + \lambda
\, \Delta \eta_\alpha \right) \   ,
\end{aligned}
     \label{eq:intQo}
\end{equation}
and for other values of $v$:
\begin{equation}
\begin{aligned}
     Q_v(\J; x, \alpha) \ = &\ \sum_{n:\, E_n\in \J}
\langle U_{\alpha} \psi_n, \psi_n \rangle \left( \frac{ \langle
\chi_x \psi_n, \psi_n \rangle }{ \langle U_{\alpha} \psi_n, \psi_n
\rangle } \right)^{v/2} \\
= &\ \int_{\J} \di E \, \sum_n \delta\left( \Gamma_n(E) + \lambda
\, \Delta \eta_\alpha \right) \ \frac{ \langle \chi_x \psi_n,
\psi_n \rangle^{v/2} }{ \norm{ \phi_n(E) }^v} \: .
     \label{eq:intQr}
\end{aligned}
\end{equation}
For $\Gamma_n(E) = -\lambda \Delta \eta_{\alpha} = \xi$, i.e.\
$E=E_n(\xi)$, it follows that
\begin{equation}\psi_n(\xi) = \xi
     (H_{\hat{\eta}_{\alpha}} - E)^{-1} U_{\alpha}^{1/2}
     \phi_n(E) \; ,
\end{equation}
since $K_{\hat{\eta}_{\alpha}}^{-1} \phi_n(E) = \Gamma_n(E)
U_{\alpha}^{1/2} \psi_n(\xi)$ . Thus eq.~(\ref{eq:Qrep}) follows
from eq.~(\ref{eq:intQr}).

For fixed $E$, the left-hand side of eq.~(\ref{eq:count}), up to a
factor $1/\lambda$, counts the number of $n$ for which
$\Gamma_n(E) = \lambda (\hat{\eta}_{\alpha} - \eta_{\alpha})$ has
a solution with $a < \eta_{\alpha} < b$. This number is exactly
the decrease in the number of eigenvalues of $H_{\eta_{\alpha}}$
below $E$ as $\eta_{\alpha}$ is moved from $a$ to $b$. That yields
\eq{eq:count}.
  \end{proof}

\subsection{Spectral shift  bounds}
In applying the interpolation argument seen in
Lem\-ma~\ref{lem:interpol}, we shall need bounds on the {\it
spectral shift function}, defined by
\begin{equation}
S_{\alpha,\lambda}(H^{(\Lambda)};E) \ := \ \tr \,  \left[ P_{\le E
}(H_{\eta_{\alpha}=0}^{(\Lambda)}) - P_{\le E
}(H_{\eta_{\alpha}=1}^{(\Lambda)}) \right]  \, ,
\end{equation}
and expressing how many energy levels are pushed over $E$  when
the value of the parameter $\eta_\alpha$ is increased by $1$. Note
that $S_{\alpha,\lambda}(H^{(\Lambda)};E)$ is non-negative and
bounded (since $|\Lambda| < \infty$):
\begin{equation}
0 \ \le \ S_{\alpha,\lambda}(H^{(\Lambda)};E) \ \le \ \tr P_{\le E
}(H_{\eta_{\alpha}=0}^{(\Lambda)}) \ \le \ C_p \, (1+|E-E_0|)^{p}
|\Lambda| \; ,
\end{equation}
for any $p>d/2$.

In the discrete setup, where the role of $U_\alpha$ is taken by a
rank-one operator, the shift is at most $1$.  For continuum
operators there is no such uniform bound independent of the volume
(see ref. \cite{Kirsch}), however
$S_{\alpha,\lambda}(H^{(\Lambda)};E)$ has locally  bounded $L^p$
norms as a function of the energy $E$, as was shown in
ref.~\cite{Combes/Hislop/Nakamura}.

\begin{lem}[$L^p$ boundedness of the spectral shift]
\label{lem:shiftbound} Let $1\le p < \infty$ and $n\in \N$ such
that $n> dp/2$. Then there exists a constant $C_{p,\lambda,n} <
\infty$ such that
\begin{equation} \label{eq:SpBound}
\int_{-\infty}^{E_+} S_{\alpha,\lambda}(H^{(\Lambda)};E)^p \di E \
\le \ C_{p,\lambda,n} (1+|E_+-E_0|)^{n+1}
\end{equation}
uniformly in the domain $\Lambda$ as well as in the choice of
$\alpha$ and the random parameters $\eta_{\beta}$ ($\beta \not=
\alpha$).
\end{lem}

\subsection{From fractional moment bounds to localization}

We shall now put together the elements introduced in the previous
sections and  prove that rapid decay of the resolvent moments
implies localization in its various manifestations.
\begin{thm} \label{thm:2}
Let $H_\omega$ be a random Schr\"odinger operator which satisfies
${\mathcal A}$.  Let $0<s\le 1$ and $\J \subset (-\infty, E_+]$ be
a bounded open interval. Suppose that for some $C<\infty$, $\mu>0$
and a bounded region $\Lambda$,
\begin{equation}
  \Ev { \int_{\J}\di E
  \norm{\chi_{x}\, \frac{1}{H^{(\Lambda)} -E}
    \, \chi_{y} }^s  }  \ \le C {e}^{-\mu \dist_\Lambda(x,y)} \,  ,
  \label{eq:cond1}
  \end{equation}
for all $x,y \in \Lambda$. Then, for any $v<1/(2-s)$, there exists
a volume independent constant $C_{s,v}(E_+,\lambda) < \infty$ such
that
\begin{equation}
  \Ev {\sup_{f\in C_c(\J):\, |f| \le 1} \norm{\chi_x \,
  f(H^{(\Lambda)}) \,
  \chi_y } } \
    \le   C_{s,v}(E_+,\lambda)\  {e}^{-v\mu \dist_\Lambda(x,y)} \; ,
\label{eq:concl1}
\end{equation}
for all $x,y \in \Lambda$.
\end{thm}
\noindent {\bf Remark:} The operators $H^{(\Lambda)}$ have finite
spectrum in $\J$. Thus, in \eq{eq:concl1} one could equivalently
take the supremum over {\it all} functions on $\J$ as long as they
pointwise  satisfy $|f|\le 1$. The only reason for stating
\eq{eq:concl1} in terms of functions in $C_c(\J)$ is to prepare
for a limiting argument in the proof of Thm~\ref{cor}.

\begin{proof}
Assume that for some open interval $\J$ the fractional moment
bound \eq{eq:cond1} holds. By \eq{eq:YQbound}, to prove
Theorem~\ref{thm:2} it suffices to establish a related bound on
$\E(Q_1(\J;x,\alpha))$ for $\alpha \in \I$ with $\dist(\alpha,y)
\le 2r$. This in turn is controlled through
Lemma~\ref{lem:interpol} by an interpolating product of
$\E(Q_2(\J;x,\alpha))$ and $\E(Q_v(\J;x,\alpha))$, with $v<1$.
Leaving room for one more interpolation, we choose $v<s$.

To estimate $\E(Q_v(\J;x,\alpha))$, we start from the
representation \eq{eq:Qrep} and average first over $\eta_{\alpha}$
at specified values of $\eta_{\zeta}$ for $\zeta \not= \alpha$.
For almost every choice of $\eta_{\zeta}$ ($\zeta \not= \alpha$),
$E$ is neither an eigenvalue of $H_{\eta_{\alpha}=0}^{(\Lambda)}$
nor of $H_{\eta_{\alpha}=1}^{(\Lambda)}$. Also, for fixed
$\hat{\eta}_{\alpha}$ and almost every $\eta_{\alpha}$ we have
$\sigma(H_{\eta_{\alpha}}^{(\Lambda)}) \cap
\sigma(H_{\hat{\eta}_{\alpha}}^{(\Lambda)}) \cap \J = \emptyset$.
Thus we can apply \eq{eq:Qrep} and \eq{eq:count} to conclude
\begin{multline} \label{eq:Qrest1}
\E (Q_v(\J;x,\alpha)) \\
\le \ 2^v {\lambda}^{v-1} D \; \Ev {\int_{\J} \di E \
   S_{\alpha,\lambda}(H^{(\Lambda)}; E) \, \| \chi_x (H^{(\Lambda)}-E)^{-1}
     U_{\alpha}^{1/2} \|^v } \; .
\end{multline}
Here we have used ${\mathcal A}3$, the bound $|\Gamma_n(E)| \le
2\lambda$ (imposed by the $\delta$ functions in \eq{eq:Qrep}), and
we chose the re-sampled variable $\hat{\eta}_{\alpha}$ to have the
same conditional distribution as $\eta_{\alpha}$.

Using H\"older's inequality we can further estimate the right hand
side of \eq{eq:Qrest1} to get
\begin{equation} \label{eq:Qrest2}
\begin{aligned}
\E(Q_v(\J;x,\alpha)) \ \le & \ 2^v {\lambda}^{v-1} D \left( \E
\int_{\J}
       \di E \; S_{\alpha,\lambda}(H^{(\Lambda)};E)^{\frac{s}{s-v}}
   \right)^{\frac{s-v}{s}} \\
& \qquad \times \left( \E \int_{\J} \di E \; \| \chi_x
   (H^{(\Lambda)}-E)^{-1} U_{\alpha}^{1/2} \|^s \right)^{v/s} \\
\le& \ C_{s,v,\lambda,n} (1+|E_+-E_0|)^{(n+1)(s-v)/s} {e}^{-\mu
   v\dist_\Lambda(x,\alpha)/s}
\end{aligned}
\end{equation}
for any integer $n> \frac{sd}{2(s-v)}$. In the final estimate we
have used the spectral shift bound from Lemma~\ref{lem:shiftbound}
and the assumption \eq{eq:cond1}.

Collecting all our bounds as well as the uniform bound for
$\E(Q_2(\J;x,\alpha))$ provided by eq.~(\ref{eq:density}), we
arrive at
\begin{equation} \label{eq:Ybound}
\E(Y_{\Lambda}(\J;x,y)) \le C_{v,s}(E_+,\lambda) {e}^{-\frac{\mu
     v\dist_\Lambda(x,\alpha)}{s(2-v)}} \; .
\end{equation}
After reorganizing the exponent this yields \eq{eq:concl1}.
\end{proof}

\subsection{Infinite volumes\textemdash proof of
Theorem~\ref{cor}}\label{sec:infinitevolumes} The eigenfunction
correlator methods used above are most easily implemented for the
finite volume operators $H^{(\Lambda)}$. However, it is important
to note that the bounds obtained in this way do not depend on the
size of the volume $\Lambda$ (except for the presence of the
modified distance $\dist_\Lambda(x,\alpha)$, which for fixed
$x,\alpha$ is equal to the usual distance $|x-\alpha|$ for
sufficiently large $\Lambda$). In this section we take the
infinite volume limit, proving Theorem~\ref{cor} and deduce the
results on spectral localization, as discussed in
Section~\ref{sec:1.2}. The fundamental analytic tools are 1)
strong resolvent convergence and 2) the RAGE theorem. The infinite
region $\Omega$ being fixed throughout this section, we write $H =
H^{(\Omega)}$.

\begin{proof}[Proof of Theorem~\ref{cor}:] For $g \in C_c(\J)$,
$g(H^{(\Lambda_n)})$ converges strongly to $g(H)$ since
$H^{(\Lambda_n)}$ converges to $H$ in the strong resolvent sense.
Thus
\begin{equation} \label{eq:semicont}
\| \chi_x g(H) \chi_y \| \le \liminf_{n\to\infty} \|\chi_x
g(H^{(\Lambda_n)}) \chi_y \| \: .
\end{equation}
Taking suprema and applying Fatou's lemma yields
\begin{equation} \label{eq:fintoinf}
\Ev {\sup_{g:\,|g|\le 1} \| \chi_x g(H) P_{\J}(H) \chi_y \|} \ \le
\ \liminf_{n\to \infty} \Ev { \sup_{f\in C_c(\J):\, |f|\le 1} \|
\chi_x
   f(H^{(\Lambda_n)}) \chi_y \|} \: .
\end{equation}

The supremum on the left-hand side is initially only taken over
$g\in C_c(\J)$, but can be extended to all Borel functions $g$
without changing its value. To see this, let $T_g := \chi_x g(H)
P_{\J}(H) \chi_y$ for a fixed Borel function $g$ with $|g|\le 1$.
For an orthonormal basis $(\phi_k)$, $k=1,2,\ldots$, set $P_N =
\sum_{k=1}^N \langle \cdot, \phi_k \rangle \phi_k$. As $T_g$ is
compact we have
\begin{equation} \label{eq:finiteapprox}
\lim_{N\to\infty} \norm{T_g P_N - T_g} =0 \; .
\end{equation}

The spectral measures $d\mu_k(\lambda) =
d\norm{E(\lambda)\chi_y\phi_k}^2$ associated with $H$ are Borel
measures on $\R$ and thus regular (e.g.\ \cite{Rudin}). Thus it
follows from elementary considerations that there exist $g_n \in
C_c(\J)$ with $|g_n| \le 1$ and $\int_{\J} |g_n-g|^2\, d\mu_k \to
0$ as $n\to\infty$ simultaneously for $k=1,\ldots, N$. We have
\begin{equation} \label{eq:normcalc}
\norm{T_{g_n} P_N - T_g P_N} \le \sum_{k=1}^N
\norm{(g_n(H)-g(H))P_{\J}(H) \chi_y \phi_k} \; .
\end{equation}
By the above, this implies that $T_{g_n} P_N \to T_g P_N$ as
$n\to\infty$. Together with \eqref{eq:finiteapprox} this implies
that for every $\varepsilon>0$ there exist $N$ and $n$ such that
$\norm{T_g} \le \norm{T_{g_n} P_N} + \varepsilon \le
\norm{T_{g_n}} + \varepsilon$. As $\varepsilon >0$ was arbitrary,
we conclude that the left-hand side of \eq{eq:fintoinf} does not
change if the supremum is taken over all Borel functions with
$|g|\le 1$.

Equation (\ref{eq:corresult}) now follows through the uniform
bounds \eq{eq:concl1} provided by Theorem~\ref{thm:2} under the
assumption \eq{eq:corassump}.
\end{proof}

We now turn to the spectral type of $H$, where $\Omega = \R^d$ is
assumed, and therefore $\dist_{\Omega}(x,y) = \dist(x,y)$. The
absence of continuous spectrum can be demonstrated from our
estimates using the RAGE theorem (e.g., see ref.~\cite{Cyconetal}
for discussion and references) which implies that the projection
$P_c(H)$ onto the continuous spectrum of $H$ satisfies
\begin{equation}
   \norm{P_c(H) \psi }^2 \ = \ \lim_{R
     \rightarrow \infty} \ \lim_{T \rightarrow \infty} \int_0^T \frac{\di t}{T}
     \ \norm{ \chi_{\set{\abs{x-x_0} \ge R}} \, \e^{-\im t H} \psi}^2
\end{equation}
As a consequence,
\begin{thm}[RAGE theorem for random operators]\label{thm:RAGE}
For the random operators considered here, if for some open
interval $\J$
\begin{equation}\label{eq:g(R)}
   \Ev{\norm{\chi_{\set{|x - x_0| \ge R}} \, {e}^{-\im t H} \,
P_{\J}(H) \, \chi_{x_0} }} \ \le \ g(R)
\end{equation}
with $g(R) \rightarrow 0$ as $R \rightarrow \infty$ uniformly in
$t$ and $x_0$, then $P_{c}(H) P_{\J}(H) = 0$ almost surely.
\end{thm}

\begin{proof} If $\phi \in L^2(\R^d)$ is compactly supported with
   $\supp \phi \subset B_{x_0}^r$ for some $x_0 \in \R^d$, then the
   RAGE theorem implies
\begin{equation} \label{eq:RAGEbound}
\|P_c(H) P_{\J}(H) \phi\|^2 \le \liminf_{R\to\infty}
\liminf_{T\to\infty} \int_0^T \frac{\di t}{T} \| \chi_{|x-x_0|\ge
R} {e}^{-\im tH} P_{\J}(H) \chi_{x_0}\|^2 \|\phi\|^2 \: .
\end{equation}

Upon taking expectations, Fatou's lemma and Fubini's theorem yield
\begin{equation} \label{eq:E=0}
\Ev { \|P_c(H) P_{\J}(H) \phi\|^2} \le \liminf_{R\to\infty}
\liminf_{T\to\infty} \int_0^T \frac{dt}{T} g(R) \|\phi\|^2 = 0 \:
.
\end{equation}
Thus $P_c(H) P_{\J}(H) \phi =0$ almost surely. This implies the
theorem since there exists a {\em countable} total set of $\phi$
with each $\phi$ supported in $B_{x_0}^r$ for suitable $x_0$.
\end{proof}

To complete the proof of pure point spectrum in $\J$ we note that
eq.~(\ref{eq:corresult}), proven above, yields the assumption of
Theorem~\ref{thm:RAGE}: Covering $\{|x-x_0| \ge R\}$ with balls
$B_{\alpha}^r$, $\alpha \in \I$, we get for every $\nu \in (0,
\mu/(2-s))$ that $\Ev {\| \chi_{\{|x-x_0|\ge R\}} {e}^{-\im tH}
   P_{\J}(H) \chi_{x_0} \|} \le C{e}^{-\nu R}$.

Another consequence of \eq{eq:corresult} is that
\begin{equation}
   \Ev{ \sum_{x,y} \frac{\e^{\nu | x-y|} }{1 + \abs{y}^{d+1}}
     \ \sup_{g : |g| \le 1} \norm{\chi_x g(H) P_{\J}(H) \chi_y} } \ < \
     \infty \; ,
\end{equation}
which implies
\begin{equation}\label{eq:eigprojdecay1}
   \sup_{g : |g| \le 1} \norm{\chi_x g(H) P_{\J}(H) \chi_y}  \ \le \
   \const \ (1 + \abs{y}^{d+1}) \
   \e^{-\nu |x-y|} \qquad \mbox{a.s.}.
\end{equation}
Since the Kronecker delta functions $\delta_E(x)$ are Borel
measurable, this implies that with probability one all
eigen-projections of $H$ satisfy \eqref{eq:eigprojdecay}.

Almost surely all eigenvalues of $H$ in $\J$ are finitely
degenerate. An argument for this (based on compactness
considerations and spectral averaging) which applies to our model
was provided in the proof of Theorem~3.2 of
ref.~\cite{Combes/Hislop:JFA}.

Given this and using \eqref{eq:eigprojdecay}, we can proceed as
follows.  We have, for almost every configuration $\omega$ and
$E\in \J$, ({\em i.}) The range of $\delta_E(H)$, denoted
$\mathcal{R} (\delta_E(H))$, is finite dimensional and ({\em ii.})
$\norm{\chi_x \delta_E(H) \chi_{|x|\le R}} \ = \ \mathrm{O} (e^{-
\nu |x|}) $ for every $x\in \R^d$ and $R>0$.

{F}rom ({\em i.}) it follows that $\mathcal{R} (\delta_E(H)) = \mathcal{R}
(\delta_E(H) \chi_{|x|\le R})$ for $R\ge R_0(\omega,E)$. Thus ({\em ii.})
implies $\|\chi_x \phi\| \ = \ \mathrm{O} (e^{- \nu |x|})$ for every $\phi$ in
the range of $\delta_E(H)$, i.e.\ all eigenfunctions. An $L^\infty$ version of
exponential decay follows from well know facts regarding the smoothness of
eigenfunctions of Schr\"odinger operators (``elliptic regularity'', e.g. see
ref.~ \cite{Simon:semigroups}).

This completes the proof of spectral localization properties.

\newpage

\section{Finiteness of the fractional-moments}
\label{sec:s-moments}

In this section we prove two technical results, Lemmas~\ref{lem:finite_smoment}
and~\ref{lem:decouple}, which permit to bound disorder averages of resolvent
norms (raised to a fractional power) at fixed energy.  
These are the analogues of Proposition~\ref{prop:sfinite} required in the proof
of Theorem~\ref{thm:1} in Section~\ref{sec:fin-vol-criteria}, where energy
averaging is replaced by disorder averaging in ``local environments.'' An
improved bound under the stronger assumption ${\mathcal A}3'$ is presented in
Proposition~\ref{prop:extend} below.

We begin with an overview of the argument and then present the two lemmas.
Finally we give a proof of the two main technical results.

\subsection{Why are disorder averages finite?} \label{sec:finitemoments}
The analysis presented below has its genesis in the discrete setup, where the
finiteness of disorder averages of the Green function is quickly implied by a
rank-one perturbation argument. However, for the continuum operators considered
here, that short argument requires thorough remaking since the local potential
term ($U_\alpha$) is now an operator of infinite rank.

To contrast the two cases, discrete and continuum, it is instructive to compare
them in a unified framework provided by the Birman-Schwinger relation (see
\eq{eq:BirSch}, \eq{eq:BirSchrel2}, and \eq{eq:BirSchrel3}):
\begin{equation}
   U^{1/2} \frac{1}{ \widehat H +  \eta U - E} U^{1/2} \ = \
   \left [ \left [ \widehat K_E \right ]^{-1}  \ + \ \eta \right ]^{-1} \; ,
\label{eq:BS2}
\end{equation}
with
\begin{equation}
   \widehat K_E
   \ = \   U^{1/2} \frac{1}{\widehat H - E} U^{1/2} \; .
\end{equation}
In the discrete case, with $U = \left | o \right > \left < o \right |$ a
rank-one operator, $\widehat K_E^{-1}$ is simply a complex number, and
\eq{eq:BS2} readily implies
\begin{equation}\label{eq:prelimweak11}
   \abs { \set{\eta \ : \ \abs{ \left < o \right | \frac{1}{ \widehat H +
       \eta
       \left | o \right > \left < o \right |
       - E} \left | o \right > } > t } } \ = \ \frac{2}{t} \; ,
\end{equation}
and thus finiteness of fractional $\eta$-moments by the ``layer-cake''
representation \eq{layercake}.

For general $U$, \eq{eq:BS2} implies the weak $L^1$ bound
\begin{equation}\label{eq:wk11xi}
   \abs{ \set{ \eta \in [0,1] \ : \
   \norm{U^{1/2} \frac{1}{\widehat H +  \lambda \eta U - E} U^{1/2}} >
t } } \ \le \
   \frac{2(1 + \xi_{E,\lambda})}{\lambda t} \;
\end{equation}
where $\xi_{E,\lambda}$ denotes the number of eigenvalues of $\widehat
K_E^{-1}$ in the interval $(-\lambda,0)$ (see \eq{eq:BirSchrel3}).  For trace
class perturbations $U$, the simple bound $\xi_{E,\lambda} \le \Tr \, U$ allows
the analysis to proceed from \eq{eq:wk11xi} much as in the rank-one case,
however for the Schr\"odinger operators considered here, with $U$ one of the
``bumps'' $U_\alpha$,  the perturbation $\lambda \eta U$ is not trace class.
Nonetheless, it is {\em relatively compact} with respect to $\widehat H$ due to
the kinetic term $\p_\A \cdot \p_\A $.  It follows that for $E \not \in
\sigma(\widehat H)$ the operator $\widehat K_E$ is {\em compact} and therefore
$\xi_{E,\lambda}$ is {\em finite.} However, there is no reason to expect a
bound on $\xi_{E,\lambda}$ which is uniform in $E$ and in the various
parameters implicit in $\widehat H$.

To examine the factor $\xi_{E,\lambda}$ more closely, note that $E$ is an
eigenvalue of  $H = \widehat H + \lambda \eta U$ precisely when $\lambda \eta$
is an eigenvalue of $(-\widehat K_E^{-1})$. Hence, we may equate
$\xi_{E,\lambda}$ with the number of times that $E$ becomes an eigenvalue of
$H$ as $\eta$ is moved from $0$ to $1$. By the monotonicity of $H$ in $\eta$
(implied by the positivity of $U$), we get
\begin{equation}
   \xi_{E,\lambda} \ \le \ \tr \, [P(\widehat H < E) - P( \widehat H +
\lambda U < E)] \; , \label{eq:shift} \end{equation} with equality
unless $E$ is a degenerate eigenvalue for some $\eta \in (0,1)$.

The quantity on the right side of \eq{eq:shift} is the `Krein spectral shift'
at energy $E$ between $\widehat H$ and $\widehat H + \lambda U$, and has
recently been the subject several of studies
\cite{Combes/Hislop/Nakamura,Hundertmark/Simon}. A key fact is that for
Schr\"odinger operators the Krein spectral shift is locally integrable as a
function of $E$\textemdash  indeed, it has been shown to be locally $L^p$ for
every $p \in [1,\infty)$ with explicit bounds \cite{Combes/Hislop/Nakamura}, a
fact which we used in Section~\ref{sec:proofofthm2} in the proof of
Theorem~\ref{cor}. However, in general the spectral shift is not locally
bounded; examples exist of Schr\"odinger operators for which it is arbitrarily
large at certain energies \cite{Kirsch}.

Thus, the moments of continuum Green functions differ from their
discrete counterparts in an essential way\textemdash after
averaging over a {\em single} coupling one still does not get a
uniform bound. Nevertheless, it is natural to guess that for
random operators the spectral shift has a finite expectation
value, since averaging over disorder may play a role somewhat
similar to averaging over energy.

In the derivation of our fractional moment bounds (see
Lemma~\ref{lem:finite_smoment} below), we do not consider directly
the average of the spectral shift.  Instead, we find it more
convenient to derive an analogue of \eq{eq:prelimweak11} for
continuum operators (from which fractional moment bounds follow
quite easily).  Nonetheless, the notion that a disorder averaged
spectral shift is bounded is one of the driving ideas behind this
work, and we find that the techniques developed in this section
imply a result of this type (Theorem~\ref{thm:spectral-shift},
stated and proved in Appendix~\ref{sect:app-spect-shift} below).
In the end however, the fractional moment bounds obtained here are
somewhat stronger than those which follow from
Theorem~\ref{thm:spectral-shift}.

\subsection{The $1/t$-tails}\label{tails}

A useful result for moment bounds is a general weak-$L_1$ type
bound for the boundary values of resolvents of maximally
dissipative operators. We state this as the second part of the
following Lemma, which is essentially a special case of results
proven in \cite{Naboko}. For completeness, we give a proof based
on properties of the Hilbert transform in
Appendix~\ref{sec:Weakbound}. The first part of the Lemma is a
well known result in scattering theory, e.g.\ \cite{deBranges}.

\begin{lem}[Weak $L_1$ bound]
\label{lem:HSweakbound} Let $\Hi$ and $\Hi_1$ be separable Hilbert
spaces, $A$ a maximally dissipative operator in $\Hi$, as well as
$M_1: \Hi \to \Hi_1$ and $M_2: \Hi_1 \to \Hi$ Hilbert-Schmidt
operators.  Then
\begin{enumerate}
\item The ``boundary values''
   \begin{equation}
     M_1 \frac{1}{A - v + \im 0} M_2  \ = \ \lim_{\epsilon \downarrow 0}
     M_1 \frac{1}{A - v + \im \epsilon} M_2
   \end{equation}
exist as Hilbert-Schmidt operators for almost every $v \in \R$,
with convergence in the Hilbert-Schmidt norm. \item There exists
$C_W < \infty$ (independent of $A$, $M_1$, $M_2$) such that
\begin{equation} \label{eq:weakbound}
   \abs{ \set{ v \ : \norm{ M_1 \frac{1}{A - v + \im 0} M_2 }_{HS} > t} }
   \ \le \ C_W \, \norm{M_1}_{HS} \, \norm{M_2}_{HS} \, \frac{1}{t} \; .
\end{equation}
\end{enumerate}
\end{lem}
\noindent {\bf Remark:} Recall that a densely defined operator $A$
is called {\em dissipative} if, for each $\varphi \in {\mathcal
D}(A)$, we have $\Im \langle \varphi, A \varphi \rangle \ge 0$. It
is called \emph{maximally dissipative} if it has no proper
dissipative extension, which is equivalent to contractivity of the
semi-group $e^{\im tA}$ generated by $\im A$. If a dissipative
operator $A$ has strictly positive imaginary part, such as a
Birman-Schwinger operator $A_{BS}$ of the type consider in
Lemma~\ref{lem2.1}, then the $i0$ in (\ref{eq:weakbound}) is not
needed since $(\xi-A_{BS})^{-1}$ is norm-continuous for $\xi$ in
the closed lower half plane.

In our applications of Lemma~\ref{lem:HSweakbound} we will use the
following ``off-diagonal'' version which follows from the lemma
via the Birman-Schwinger identity and a simple change of
variables. For ease of presentation, we state this version with
the additional assumption that the operator $A$ has strictly
positive imaginary part, thus avoiding the issue of whether
certain limits exist.
\begin{prop}\label{prop:offdiag}
Let $A$ be a maximally dissipative operator with strictly positive
imaginary part on a Hilbert space $\Hi$, let $M_1, M_2$ be
Hilbert-Schmidt operators, and let  $U_1$, $U_2$ be non-negative
operators. Then
\begin{multline}\label{eq:offdiag}
\left | \left \{ \langle v_1,v_2 \rangle \in [0,1]^2\ :
     \norm{M_1 U_1^{1/2} \frac{1}{A - v_1 U_1 - v_2 U_2} U_2^{1/2} M_2}
     > t \right \} \right |  \\
\le \ 2 \, C_W \norm{M_1}_{HS} \norm{M_2}_{HS}  \frac{1}{t} \; .
\end{multline}
\end{prop}
\begin{proof}
The key to the proof is the change of variables $v_{\pm} = \half
(v_1  \pm v_2)$, so that $ v_1 U_1 + v_2 U_2  =  v_+ (U_1 + U_2) +
v_- (U_1 - U_2)$. From the Birman-Schwinger identity,
\eq{eq:BirSchrel2},
\begin{equation}
\left [ (U_1 + U_2)^{1/2} \frac{1}{A - v_1 U_1 - v_2 U_2} (U_1 +
U_2)^{1/2} \right ] \ = \ \left [ \left [ K(v_-) \right ]^{-1} \ -
\ v_+  \right ]^{-1} \; ,
\end{equation}
where $[ \cdot ]$ denotes the restriction of an operator to $\ker
(U_1 + U_2)^\perp$ and
\begin{equation}
K(v_-) \ = \ (U_1 + U_2)^{1/2} \frac{1}{A - v_- (U_1 - U_2)} (U_1
+ U_2)^{1/2} \; .
\end{equation}
Thus,
\begin{equation}
M_1 U_1^{1/2} \frac{1}{A - v_1 U_1 - v_2  U_2} U_2^{1/2} M_2 \ = \
     \widetilde M_1 \left [\left [ K( v_-) \right ]^{-1}
     \ - \ v_+  \right ]^{-1}
\widetilde M_2
\end{equation}
with $\widetilde M_1 = M_1 U_1^{1/2} P [U_1 + U_2]^{-1/2} P$ and
$\widetilde M_2 \ = \ P [U_1 + U_2]^{-1/2} P U_2^{1/2} M_2$ where
$P$ denotes orthogonal projection onto $\ker (U_1 + U_2)^\perp$.

Since $U_1$ and $U_2$ are positive, $ \norm{U_j^{1/2} P [U_1 +
U_2]^{-1/2} P } \le 1 $ and therefore
\begin{equation}
\norm{\widetilde M_j}_{HS} \ \le \ \norm{M_j}_{HS} \; .
\end{equation}
Furthermore $[K(v_-)]^{-1}$ is maximally dissipative as shown in
Lemma~\ref{lem2.1}. Thus for each fixed $v_-$ we are in the
situation governed by Lemma~\ref{lem:HSweakbound}, and
Prop.~\ref{prop:offdiag} follows via the Fubini Theorem.  The
factor of $2$ on the right hand side results from the Jacobian of
the transformation $\langle v_1, v_2 \rangle \mapsto \langle v_+,
v_- \rangle$.
\end{proof}

We will use Prop.~\ref{prop:offdiag} to conclude that for the
random operator $H$,
\begin{multline}\label{eq:BSweak1-1}
   \abs{ \set{ \langle \eta_\alpha, \eta_\beta \rangle \in [0,1]^2 \ :
\norm{ M_1 U_\alpha^{1/2} \frac{1}{ H - z }   U_\beta^{1/2} M_2
}_{HS} > t} }
\\ \le \ 2 \ C_W \, \norm{M_1}_{HS} \, \norm{M_2}_{HS} \,
\frac{1}{\lambda t} \; .
\end{multline}
Several comments are in order:
\begin{enumerate}
\item When the energy $z$ lies in the lower half plane
\eqref{eq:BSweak1-1} follows directly from
Prop.~\ref{prop:offdiag} with $v_1 =\eta_\alpha, v_2 = \eta_\beta$
and $U_1 = \lambda U_\alpha, U_2 = \lambda U_\beta$. \item
Eq.~\eqref{eq:BSweak1-1} also holds for $z$ in the upper half
   plane, as can be seen by taking conjugates.
\item For real energies ($z = E \in \R$),
   \eq{eq:BSweak1-1} holds also in the
   limits $z \rightarrow E \pm \im 0$,
   as follows from the bound at complex energies and Fatou's lemma
   (compare the argument for proving Lemma~\ref{lem:HSweakbound}
   in Appendix~\ref{sec:Weakbound}).
\end{enumerate}

The above results display that once the resolvent of the continuum
operator is bracketed with a pair of Hilbert-Schmidt operators
$M_1$, $M_2$ its fractional moments can be handled similarly to
the discrete case. In the proofs of Lemmas
\ref{lem:finite_smoment} and \ref{lem:decouple} we use an argument
which shows that $\chi_x (H -z)^{-1} \chi_y$ can be presented as a
sum of a {\em bounded} operator and one  of the form
\begin{equation}
   M_x \1_{B_x} (H - z)^{-1} \1_{B_y} M_y \,  ,
\end{equation}
with $M_\sharp$ Hilbert-Schmidt operators and $B_x, B_y$ sets
somewhat larger than the balls of radius $r$ around $x$ and $y$.
This will allow us to  conclude finiteness of fractional moments
after ``averaging over the local environment,'' i.e., averaging
over all $\eta_\alpha$ with $U_\alpha$ non-zero in $B_x$ or $B_y$.

\subsection{A pair of fractional-moment lemmas}
\label{sec:fraclemmas}

We now present the two  basic technical results which give
finiteness of the fractional moments, as well as a ``decoupling
argument'' to be used in the proof of Theorem~\ref{thm:1}.

\begin{defn}
   For $\alpha \in \I$, we denote
   \begin{equation}
     \I_\alpha \ := \ \set{\zeta \in \I : \dist(\alpha, \zeta) < 3r}
   \end{equation}
   and set $\I_{\alpha, \beta} = \I_\alpha \cup \I_\beta$.
   Likewise, for any subset $\L \subset \I$,
   $\I_{\L} := \cup_{\alpha \in \L} \I_\alpha$.  By $\Fc_{\L}$ we denote the
   $\sigma$-algebra generated by all $\eta_\alpha$ with $\alpha$
     {\em not} in $\I_{\L}$.
   Thus, $\Ev{\cdot | \Fc_{\L}}$ represents averaging over
   the ``local environment'' of $\L$.
\end{defn}

Our first bound yields the finiteness of the $s$-moments after
averaging over the local-environment.
\begin{lem} \label{lem:finite_smoment}
   Let $H$ be a random Schr\"odinger operator which satisfies {$\mathcal
   A$} with disorder strength $\lambda>0$.
   Then there exists $C_\lambda \ < \infty$
   such that the restriction of $H$ to a region $\Omega$ obeys
   \begin{equation}\label{eq:weak1-1_1}
     \Pr \left (  \left . \norm{ U_\alpha
       \frac{1}{H^{(\Omega)} -E-\im \eps} U_\beta
       } > t  \right | {\mathcal F}_{\alpha, \beta}^c  \right ) \
     \le \ C_\lambda
     (1 + |E-E_0|)^{d+2}    \ \frac{D^2}{ t} \; ,
   \end{equation}
   for any $\alpha,\beta \in \I$, any $E \in \R$ and
   $\eps > 0$, where the coefficient $C_\lambda$ can be chosen such that
   \begin{equation}\label{eq:Clambdabound}
     C_\lambda \ \le \ \mathrm{const.} \ (1 + \lambda^{-1}) (1+\lambda)^{d+2} \; .
   \end{equation}
\end{lem}
\noindent {\bf Remarks:}
\begin{enumerate}
\item Recall that $E_0 = \inf \sigma(H_0)$ and that $D$ is a bound
on the conditional densities for
   $\eta_\alpha$\textemdash see eq.~\eqref{eq:D}.
\item In Prop.~\ref{prop:extend} below we show that if $H$
satisfies ${\mathcal A}3'$ then Lemma~\ref{lem:finite_smoment} can
be improved so that \eq{eq:weak1-1_1} holds with $\sup_{\lambda >1}
C_\lambda < \infty$. \item For fractional moments we use the
``layer-cake'' representation\textemdash $\Ev{X^s} = \int_0^\infty
\Pr(X > t^{1/s}) \di t$\textemdash to conclude that
\begin{equation}
   \label{eq:finite_smoment}
   \E \left (  \left . \norm{ U_\alpha \frac{1}{H^{(\Omega)} - E-\im
\varepsilon}
     U_\beta }^s  \right | {\mathcal F}^c_{\alpha, \beta}
   \right ) \ \le \  \frac{C_{\lambda}^s}{1-s}  \ (1 + |E-E_0|)^{s(d+2)} \ D^{2s}\; .
\end{equation}
Of course, this bound implies a similar estimate for the average
over all variables. \item  Using condition ${\mathcal A}2$, we
obtain the following bound from \eq{eq:finite_smoment}: For all
measurable $\Lambda, \Lambda' \subset \R^d$,
\begin{multline}\label{eq:generalfm}
   \E \left ( \left . \norm{ \1_\Lambda
       \frac{1}{H^{(\Omega)} - E - \im \varepsilon} \1_{\Lambda'}}^s
\right | {\mathcal
       F}^c_{\L(\Lambda \cup \Lambda')} \right )
   \\ \le \ \frac{C_{\lambda}^s}{1-s}
     b_-^{-2s} \ (1 + |E-E_0|)^{s(d + 2)} \ D^{2s} \ N_\Lambda \,
N_{\Lambda'}
\end{multline}
where $\L(\Lambda) := \{ \gamma\in \I: \1_{\Lambda} U_{\gamma}
\not= 0\}$ and $N_\Lambda$ indicates the number of points in
$\L(\Lambda)$.
\end{enumerate}

The second lemma (Lemma~\ref{lem:decouple}) focuses on the average
of $\norm{U_\alpha (H - z)^{-1} U_\beta}$ with respect to the
local environment of {\em one} of the sites $\alpha,\beta$. The
idea underlying this result is ``re-sampling:'' we compare the
distribution of $(H - z)^{-1}$ with that of a reference operator
$(\widehat H - z)^{-1}$, where $\widehat H$ is obtained from $H$
by redrawing the coupling variables $\eta_\zeta$ for $\zeta$ near
$\beta$. The basic result is an estimate of the form
\begin{equation}
\Ev{ \left . \norm{ U_\alpha \frac{1}{H_\omega -z}
       U_\beta }^s \right | \Fc_\beta } \ \le \ \const
     \norm{ \chi_\alpha \frac{1}{\widehat H_\omega -z} \1_{S} }^s \; ,
\end{equation}
with $S$ an appropriate neighborhood of $\beta$. However, the
result permits a number of variations, and the estimate
\eqref{eq:decouple} stated below is slightly complicated because
it is tailored to the required application in the proof of
Theorem~\ref{thm:1}.

For that application it is convenient to use a smooth function in
place of $U_\beta$.  Thus we fix a choice of a partition of unity,
$\{ \Theta_\alpha : \alpha \in \I\}$, with the following
properties:
\begin{enumerate}
\item Each function $\Theta_\alpha$ is non-negative and smooth
with compact support in the ball of radius $4r/3$ centered at
$\alpha$. \item The collection $\set{\Theta_\alpha}$ is a
partition of unity: $\sum_{\alpha} \Theta_\alpha(q) = 1$. \item
$\sup_\alpha (\norm{\grad \Theta_\alpha}_\infty , \norm{\Delta
     \Theta_\alpha}_\infty ) \ < \ \infty$.
\end{enumerate}
Any choice with these properties will do, although the constant
$\widetilde C$ in \eq{eq:decouple} below will depend on the
supremum in (3).

\begin{lem} \label{lem:decouple}
Let $H$ be a random Schr\"odinger operator which satisfies
${\mathcal A}$ with disorder strength $\lambda >0$. Then there
exists $\widetilde C_\lambda < \infty$ such that any restriction
of $H$ to a region $\Omega$ obeys
   \begin{multline}\label{eq:decouple}
     \Pr \left ( \left .  \norm{ \chi_x \frac{1}{H^{(\Omega)} -z}
       \Theta_{\beta}(1 + H_0-E_0)^{1/2} }  > t \right |
         {\mathcal F}^c_{\beta,\gamma} \right ) \\
      \le \ \widetilde C_\lambda  \ (1 + \abs{z-E_0} )^{(d+3)}  \
      \norm{
        \chi_x \frac{1}{\widehat H^{(\Omega)} - z}
     \1_{S_{\beta,\gamma}} } \ \frac{D^2}{t} \; ,
   \end{multline}
   for any $z \in \C \setminus \R$, $x\in \R^d$, and $\beta, \gamma \in \I$ with
   $\dist(x,\beta)$, $\dist(x,\gamma) > 6r$.  The coefficient
   $\widetilde C_\lambda$ obeys
   \begin{equation}
   \widetilde C_\lambda \ \le \ \mathrm{const.} \ (1+
   \lambda)^{d+4}\; .
   \end{equation}
   Here
   $S_{\beta,\gamma} = \{q : \dist(q,\{\beta,\gamma\}) \le 5r\}$ and
$\widehat H$
     is obtained from $H$ by replacing $\set{\eta_\zeta : \zeta \in
     \I_{\beta,\gamma}}$ with
   arbitrary values $\hat \eta_\zeta \in [0,1]$:
   \begin{equation} \label{eq:re-sample}
     \widehat H \ = \ H + \lambda \sum_{\zeta \in \I_{\beta,\gamma} } (\hat
     \eta_\zeta - \eta_\zeta) U_\zeta(q) \; .
   \end{equation}
\end{lem}
\noindent {\bf Remark:} {\em i. \/} Note that we have resampled
$H$ at a second site $\gamma \in \I$ as well as $\beta$, with the
result that $\Theta_\beta$ is replaced by the characteristic
function of a neighborhood of $\gamma$ and $\beta$.  This
additional resampling is required in the application of this lemma
in the proof of Theorem~\ref{thm:1}, but does not play a key role
in the present section. {\ \em ii. \/}  The factor
$(1+H_0-E_0)^{1/2}$ is included here to control various
commutators $\com{H}{\Theta}$ which appear in applications of the
lemma.  It is to bound this factor that we choose to work with the
smooth function $\Theta_\beta$.

In the applications of Lemma~\ref{lem:decouple} we will choose
$\widehat \eta_\zeta$ to have the same probability distribution as
$\eta_\zeta$.  This will provide us with a ``decoupling
argument,''  as follows: If $X$ is a quantity for which $\Ev{X^s|
\Fc_{\beta,\gamma}} \le A_s < \infty$, then \eq{eq:decouple} and
the Cauchy-Schwarz inequality imply that
\begin{multline}
   \Ev{ \left . X^{s/2} \, \norm{ \chi_x \frac{1}{H^{(\Omega)} - z}
       \Theta_{\beta}(1 + H_0-E_0)^{1/2} }^{s/2} \right | \Fc_{\beta,\gamma}  }
\\
\begin{aligned}
\le & \ A_s^{1/2} \ \Ev{ \left . \norm{ \chi_x
             \frac{1}{H^{(\Omega)} - z}
       \Theta_{\beta}(1 + H_0-E_0)^{1/2} }^s  \right |
             \Fc_{\beta,\gamma} }^{1/2} \\
\le&
       \  \const \ A_{s}^{1/2} \ \norm{
        \chi_x \frac{1}{\widehat H^{(\Omega)} - z} \1_S
        }^{s/2} \; .
\end{aligned}
\end{multline}
If $\widehat \eta_\zeta$ is distributed identically to
$\eta_\zeta$ then, upon taking expectations, we obtain
\begin{multline}
   \Ev{ X^{s/2} \, \norm{ \chi_x \frac{1}{H^{(\Omega)} - z}
       \Theta_{\beta}(1 + H_0-E_0)^{1/2} }^{s/2}   } \\ \le
       \  \const \ A_{s}^{1/2} \ \Ev{ \norm{
        \chi_x \frac{1}{H^{(\Omega)} - z } \1_S
        }^{s/2} } \; .
\end{multline}
In the final expression, we have replaced $\widehat H$ by $H$
since the two are identically distributed.

\subsection{Averaging over local
environments}\label{sec:technical}

We now derive the two technical lemmas stated above. This
subsection is essential for our analysis, but the reader may wish
to skip it at first reading, as the arguments here are not
required elsewhere in the article.

\begin{proof}[Proof of Lemma~\ref{lem:finite_smoment}]
We will write $z=E+i\varepsilon$ and assume without loss that
$\varepsilon <1$. Thus $1+|E-E_0| \sim 1+|z-E_0|$. Before applying
Prop.~\ref{prop:offdiag} and the associated
eq.~\eqref{eq:BSweak1-1}, we must introduce Hilbert-Schmidt
operators to the left and the right of $U_\alpha (H-z)^{-1}
U_\beta$. The procedure we use to insert these operators is
somewhat involved, so let us first outline the argument:
\begin{enumerate}
\item We replace $U_\alpha, \, U_\beta$ by the
   upper bounds $b_+ \Theta^2, \, b_+ \Psi^2$,
   \begin{equation}
     \norm{U_\alpha\frac{1}{H-z} U_\beta} \ \le \ b_+^2
     \norm{\Theta^2 \frac{1}{H-z} \Psi^2} ,
   \end{equation}
   where $\Theta, \Psi$ denote smoothed
   characteristic functions for the balls $B^r_\alpha$, $B^r_\beta$
respectively. \item The smoothness of $\Theta$ allows us to prove
an {\em identity}
   \begin{equation}\label{eq:HSidentity}
     \Theta^2 \frac{1}{H-z} \ = \ B \ + \ T \widetilde \Theta^2
     \frac{1}{H - z} \; ,
   \end{equation}
   where $\widetilde \Theta$ is a ``fattened'' version of $\Theta$
   and $T$ is Hilbert-Schmidt with HS-norm which
   is uniformly bounded with respect to the disorder.  The operator $B$
   is norm bounded (uniformly with respect to the
   disorder).
\item We repeat this procedure to the right of the resolvent to
obtain
   \begin{equation}
     \Theta^2 \frac{1}{H-z} \Psi^2 \ = \
     T_\alpha \widetilde \Theta^2
     \frac{1}{H - z} \widetilde \Psi^2 T_\beta \ + \ B' \; ,
   \end{equation}
   where $B'$ is norm bounded.
\item We introduce the partition of unity $\sum U_\alpha / F$
   between $\widetilde\Theta^2, \widetilde \Psi^2$ and the resolvent, and then
   apply Prop.~\ref{prop:offdiag} to each term in the resulting sum.
   (The actual argument is complicated somewhat by
   the fact that $T_\alpha,T_\beta$ carry some dependence on the randomness.)
\end{enumerate}

We now turn to specifics. A good deal rests on the proof of \eq{eq:HSidentity}.
This identity is a consequence of the following: Let $\Theta_j$ be any sequence
of smooth functions such that $\Theta_j$ takes value $1$ on the support of
$\Theta_{j-1}$. Then for each $n \ge 1$
\begin{equation}\label{repr1}
   \Theta_1^2  \frac{1}{H - z} \ = \ T_n
   \, \Theta_{n+1}^2 \frac{1}{H - z} \ + \ B_n
\end{equation}
where
\begin{equation}
   T_n \ = \ A_1 \, \cdots  \, A_{n}
\end{equation}
with
\begin{equation}\label{expan}
   A_j \ = \ \Theta_j \frac{1}{H_0 + a} \left ( \Theta_j (a + z -
\lambda V_\omega)
   \ - \ \com{\Theta_j}{H_0} \right ) \; ,
\end{equation}
and
\begin{equation}\label{b}
   B_n \ = \  \Theta_1 \frac{1}{H_0 + a} \Theta_1 \ + \ T_1 \Theta_2
   \frac{1}{H_0 + a} \Theta_2 \ + \ \cdots \ + \ T_{n-1} \Theta_n
   \frac{1}{H_0 + a} \Theta_n \; .
\end{equation}
Here $a := 1-E_0 $, so $H_0 + a \ge 1$.

To verify \eqref{repr1}, use induction on $n$ with the induction
step provided by the following ``commutator argument:''
\begin{equation}
   \begin{aligned}
   \Theta^2_j \frac{1}{H - z} \ =& \ \Theta_j
   \frac{1}{H_0 + a} \, ( H_0 + a)
   \Theta_j \frac{1}{H - z} \\
   =& \ \Theta_j
   \frac{1}{H_0 + a} \Theta_j (H_0 + a) \frac{1}{H-z} \ - \
   \Theta_j \frac{1}{H_0 + a} \com{\Theta_j}{H_0} \frac{1}{H - z} \\
   =& \Theta_j \frac{1}{H_0 + a} \Theta_j  \\ & \qquad  + \
   \Theta_j \frac{1}{H_0 + a} \left ( \Theta_j (a + z - \lambda V_\omega)
   \ - \ \com{\Theta_j}{H_0} \right ) \frac{1}{H - z} \; .
\end{aligned}
\end{equation}
The crucial point here is that
\begin{equation}
     \Theta_{j} (a + z - \lambda V_\omega)
     \ - \ \com{\Theta_j}{H_0} \ = \ \left ( \Theta_j (a + z -
\lambda V_\omega)
       \ - \ \com{\Theta_j}{H_0} \right ) \ \Theta_{j+1}^2
      \; ,
\end{equation}
since $\Theta_{j+1}$ is identically one throughout the support of
$\Theta_j$ and $\Theta_j (-a + z - \lambda V_\omega) \ - \
\com{\Theta_j}{H_0}$ is a differential\textemdash hence
local\textemdash operator.

The representation provided by \eq{repr1} has two key features:
\begin{enumerate}
\item The norm of $B_n$ is bounded uniformly in $\omega$ and
locally uniformly in energy. \item $T_n$ is in the Schatten class
$\I_p$ (see Remark (2) in Appendix A) for arbitrary $p>d/n$ with
$\|\cdot\|_p$-norm bounded uniformly in $\omega$ and locally
uniformly in energy.
\end{enumerate}

To see this we first observe that $A_j \in \I_p$ for $p>d$ with
uniform $\|\cdot\|_p$-bounds. This follows since
$\theta_j(H_0+a)^{-1} \in \I_p$ for $p>d/2$, $\theta_j
(H_0+a)^{-1/2} \in \I_p$ for $p>d$ and $(H_0+a)^{-1/2} [\theta_j,
H_0] = (H_0+a)^{-1/2} (-2\im {\p}_{\A} \cdot (\nabla \theta_j) -
(\Delta \theta_j))$ is bounded. Thus (2) follows from the H\"older
property of Schatten classes and (1) from

\begin{equation} \label{eq:b_iterate}
B_n = B_{n-1} + T_{n-1} \theta_n \frac{1}{H_0+a} \theta_n.
\end{equation}

Once $n > d/2$, property (2) above implies that the operator $T_n$
is Hilbert-Schmidt. Henceforth, we fix $n$ to be the least integer
greater than $d/2$\textemdash in any case $n \le (d +2)/2$. This
provides the representation \eq{eq:HSidentity} with $\widetilde
\Theta = \Theta_n$.

There is much flexibility in the definition of $\Theta_j$ which,
in turn, affects the norm estimates for $T_n$ and $B_n$. For our
purposes, it is sufficient to let $\Theta_j$ be supported in
$\set{q : \dist(q,\alpha) \ \le \
   r + [j/n] r}$ so that $\widetilde \Theta = \Theta_n$ is
supported in $\set{q : \dist(q,\alpha) \ \le \ 2 r}$.  The
specific choice of $\Theta_j$ is not so important.  We note,
however, that the choice may be made so that
\begin{equation}
   \norm{\grad \Theta_j}_\infty \ \le \ {\mathrm O}(d) \; , \
   \norm{\Delta \Theta_j}_\infty \ \le \ {\mathrm O}(d^2) \; ,
\end{equation}
since the gradient of $\Theta_j$ is supported on a set of width
$r/n \approx 2r/d$.

Having fixed the sequence $\Theta_j$ we obtain a Hilbert-Schmidt
operator $T_n$. More precisely, $T_n$ is a {\em random}
Hilbert-Schmidt operator since $A_j$ depend on the random
potential through the terms $\Theta_j V_\omega$.  However, $T_n$
is a product of $n$ terms each of which is linear in $\Theta_j
V_\omega$. Since
\begin{equation}
   \Theta_j V_\omega = \sum_{\zeta \in \I_\alpha} \eta_\zeta \Theta_j
   U_\zeta \;,
\end{equation}
we conclude that $T_n$ is a polynomial of degree $n$ in the
variables $\{ \eta_\zeta : \zeta \in \I_\alpha \}$ with
Hilbert-Schmidt valued coefficients. Recall that $\I_\alpha$ is
the set of lattice sites within distance $3r$ of $\alpha$.

We now repeat this procedure to the right of the resolvent to
obtain the two sided representation:
\begin{equation}\label{repr2}
   \Theta \frac{1}{H - z} \Psi \ = \ B \ + \ T_\alpha \widetilde
   \Theta^2 \frac{1}{H - z} \widetilde \Psi^2
   T_\beta \; .
\end{equation}
The above discussion shows that we may obtain the following
properties for the terms in this identity:
\begin{enumerate}
\item There is $C_d < \infty$, which depends on the
   parameters of the model and the choice of $\Theta_j$ (but not
   on the coupling variables $\eta_\zeta$), such that
   \begin{equation}\label{eq:Bzbound}
     \begin{aligned}
     \norm{B} \ \le& \ C_d \ ( 1 + \lambda + |z-E_0|)^{2 n} \\ \le& \
     C_d \ (1+\lambda)^{2n} ( 1 + |z-E_0|)^{2 n} \; .
     \end{aligned}
   \end{equation}
   \item Each $T_\sharp$ is a polynomial of degree $n \le (d+2)/2$
   in the variables $ \eta_\zeta$ for $\zeta \in \I_\sharp$ with
   coefficients which are (non-random) Hilbert-Schmidt operators:
   \begin{equation}\label{eq:coeffbound}
     T_\sharp \ = \ \sum_{(k)} T_{\sharp}^{(k)} \lambda^{|k|} \prod_{\zeta \in
         \I_\sharp}  \eta_\zeta^{k_\zeta} \; ,
   \end{equation}
   where the summation is over multi-indices $(k) \in {\mathbb N}^{\I_\sharp}$
   with $|k|:=\sum_\zeta k_\zeta \le n$.
\item There is $\widetilde C_d < \infty$ such that
   \begin{equation} \label{eq:HSbound}
   \norm{T_{\sharp}^{(k)}}_{HS} \ \le \
     \widetilde C_d ( 1 + \abs{z-E_0})^n \; ,
   \end{equation}
   for $\sharp = \alpha,\beta$ and each $(k) \in {\mathbb
   N}^{\I_\sharp}$ with $|k| \le n$.
\item $\widetilde \Theta$ and $\widetilde \Psi$ are
   bounded by one and supported in $\set{q : \dist(q,\alpha) < 2 r}$
   and $\set{q : \dist(q,\beta) < 2 r}$ respectively.
\end{enumerate}

With this representation in hand we insert the partition of unity
$\sum U_\zeta/F$ between each factor $\widetilde \Theta$,
$\widetilde \Psi$ and the resolvent $(H-z)^{-1}$.  Upon taking
norms and applying the triangle inequality this yields
\begin{equation}\label{eq:boundandsplit}
   \norm{ U_\alpha \frac{1}{H - z} U_\beta}
  \le  \ \norm{B} \ + \  \sum_{\substack{ \zeta \in \I_\alpha \\ \zeta' \in
     \I_\beta}} \sum_{\substack{(k) \in \N^{\I_\alpha} \\ (l) \in
\N^{\I_\beta}}}
     \lambda^{|k| + |l|}
     \norm{T_\alpha^{(k)} \widetilde \Theta^2 \frac{U_\zeta}{F} \frac{1}{H - z}
     \frac{U_{\zeta'}}{ F} \widetilde  \Psi^2 T_\beta^{(l)}} \; ,
\end{equation}
where we have used that $\abs{\eta_\zeta} \le 1$.

Consider now the probability, conditioned on $\Fc_{\alpha,\beta}$,
that $\norm{U_\alpha \frac{1}{H - z} U_\beta} > t$.  This may be
bounded from above by the probability that {\em one} of the terms
on the right hand side of \eq{eq:boundandsplit} is greater than
$t/M$, where $M$ is the number of terms. In turn this may be
bounded by the sum of the individual probabilities:
\begin{multline}
\Pr \left ( \left . \norm{ U_\alpha \frac{1}{H - z} U_\beta} 
> t \right |
         \Fc_{\alpha,\beta} \right ) \
     \le \ \Pr \left ( \left .\norm{B} > t/M \right |
         \Fc_{\alpha,\beta} \right ) \\ + \
     \sum_{\substack{ \zeta \in \I_\alpha \\ \zeta' \in
         \I_\beta}} \sum_{\substack{(k) \in \N^{\I_\alpha} \\
         (l) \in \N^{\I_\beta}}}
     \Pr \left ( \left .\lambda^{|k| + |l|}
         \norm{T_\alpha^{(k)} \widetilde \Theta
         \frac{U_\zeta}{F} \frac{1}{H - z} \frac{U_{\zeta'}}{ F}
         \widetilde  \Psi T_\beta^{(l)}} > t/M \right |
         \Fc_{\alpha,\beta} \right ) \; .
\end{multline}
Applying Prop.~\ref{prop:offdiag}\textemdash via
\eq{eq:BSweak1-1}\textemdash and the bound on
$\norm{T_\sharp^{(\cdot)}}_{HS}$ provided by \eq{eq:HSbound}, we
see that each term of the summation is bounded:
\begin{multline}\label{eq:onelittlepiece}
\Pr \left ( \left .\lambda^{|k| + |l|}\norm{T_\alpha^{(k)}
\widetilde \Theta
     \frac{U_\zeta}{F} \frac{1}{H - z} \frac{U_{\zeta'}}{ F}
     \widetilde  \Psi T_\beta^{(l)}} > t/M \right |
     \Fc_{\alpha,\beta} \right ) \\
\le \ 2 \, C_W \, b_+ \,
     \widetilde C_d^2   (1 + |z-E_0|)^{2n} D^2
         \frac{(1+\lambda)^{2n} M}{\lambda t} \; ,
\end{multline}
while by \eq{eq:Bzbound}
\begin{equation}
\Pr \left ( \left .\norm{B} > t/M \right |
         \Fc_{\alpha,\beta} \right ) \
         \le \ C_d \ (1+\lambda)^{2n}( 1+ \abs{z-E_0})^{2n} \,
     \frac{M}{t} \; .
\end{equation}
The factor $D^2$ in \eq{eq:onelittlepiece} is an upper bound for
the joint density of $\eta_\zeta$ and $\eta_{\zeta'}$.

Putting this all together we find that
\begin{multline}
\Pr \left ( \left . \norm{ U_\alpha \frac{1}{H - z} U_\beta}
     > t \right | \Fc_{\alpha,\beta} \right ) \\
     \begin{aligned}
     \le & \ M \ \left (
         C_d \ + \ (M - 1) \times \left [ 2 \,
         C_W \, \widetilde C_d^2 \, b_+\right ]
         \times \frac{D^2}{\lambda}
     \right ) \\
     & \qquad \qquad \times (1+\lambda)^{2n} \ (1 + \abs{z-E_0})^{2n}
\ \frac{1}{t}
     \\ \le & \ M \ \left (
         C_d \ + \ (M - 1) \times \left [ 2 \,
         C_W \,  \widetilde C_d^2 \, b_+ \right ]
     \right ) \\
     & \qquad \qquad \times (1+\lambda)^{2n} \ ( 1 + \abs{z-E_0})^{2n} \
         (1+\lambda^{-1}) \ \frac{D^2}{t} \; ,
     \end{aligned}
\end{multline}
since $D \ge 1$.  By ${\mathcal A}2$ the number of terms $M$ is
bounded independent of $\alpha, \beta$.  This completes the proof
of Lemma~\ref{lem:finite_smoment}.
\end{proof}

\begin{proof}[Proof of Lemma~\ref{lem:decouple}]
This proof follows rather closely that of
Lemma~\ref{lem:finite_smoment}. The main new ingredients are (1)
controlling the (un-bounded) factor $(a + H_0)^{1/2}$ and (2) the
``re-sampling.''

The unbounded factor is controlled by a one-step ``commutator
argument,'' similar to that used to generate the Hilbert-Schmidt
operators in the proof of Lemma~\ref{lem:finite_smoment}. The key
is the following identity
\begin{multline}
\chi_x \frac{1}{H - z} \Theta_\beta (a + H_0)^{1/2}
\\  = \ \chi_x
\frac{1}{H-z} \ \widetilde \Theta_\beta^2 \
     \left ( (z + a - \lambda V_\omega) \Theta_\beta +
\com{\Theta_\beta}{H_0} \right ) \
     \frac{1}{ (a + H_0)^{1/2}} \; ,
\end{multline}
where we have used that $\chi_x \Theta_\beta = 0$. Here
$\widetilde \Theta_\beta$ may be any function which is one
throughout the support of $\Theta_\beta$.  As a result,
\begin{equation}
\norm{ \chi_x \frac{1}{H - z} \Theta_\beta (a + H_0)^{1/2} } \ \le
\ c \ (1+\lambda)(1 +\abs{z-E_0}) \ \norm{\chi_x \frac{1}{H-z} \
\widetilde \Theta_\beta^2} \; ,
\end{equation}
with $c < \infty$ (depending on the size of $\grad \Theta_\beta$).
We choose $\widetilde \Theta_\beta$ to have support in the ball of
radius $ 5r/3$ centered at $\beta$.

Thus it suffices to prove \eq{eq:decouple} with $\Theta_\beta (a +
H_0)^{1/2}$ replaced by $\widetilde \Theta_\beta^2$ provided the
power of $(1 + \abs{z-E_0})$ on the right hand side and the
permitted power growth of $\widetilde C_\lambda$ are each reduced
by one, i.e., we must show
\begin{multline}
     \Pr \left ( \left .  \norm{ \chi_x \frac{1}{H^{(\Omega)} -z}
     \widetilde \Theta_\beta^2}  > t \right |
         {\mathcal F}^c_{\beta,\gamma} \right ) \\
     \le \ C (1+\lambda)^{d+3} \ (1 + \abs{z-E_0} )^{(d+2)}  \
     \norm{
         \chi_x \frac{1}{\widehat H^{(\Omega)} - z}
     \1_{S_{\beta,\gamma}} } \ \frac{D^2}{t} \; .
\end{multline}

  We want to average over $\zeta
\in \I_{\beta,\gamma}$, using the Birman Schwinger identity
together with the weak $L^1$ inequality, and compare the result to
the ``re-sampled'' operator:
\begin{equation}
\chi_x \frac{1}{H -z} \widetilde \Theta_{\beta}^2
     \ = \ \chi_x \frac{1}{\widehat H - z} \widetilde
     \Theta_{\beta}^2
  \ + \  \chi_x \frac{1}{\widehat H - z}
  (\widehat H \ - \ H)   \frac{1}{H - z}
     \widetilde \Theta_{\beta}^2 \; ,
\end{equation}
where
\begin{equation}
     \widehat H \ = \ H \ + \   \lambda \sum_{\zeta \in
     \I_{\beta,\gamma} } (\widehat \eta_\zeta - \eta_\zeta) U_\zeta(q) \,.
\end{equation}
We use the construction presented in the proof of
Lemma~\ref{lem:finite_smoment}\textemdash see
\eq{repr1}\textemdash to introduce the necessary Hilbert-Schmidt
operators.

We begin by noting that
\begin{equation}
\chi_x \frac{1}{\widehat H - z} (\widehat H - H ) \ = \ \chi_x
\frac{1}{\widehat H - z} \Psi^2 (\widehat H - H)
\end{equation}
where $\Psi$ is identically one throughout $
\{ q  : \dist(q,\{\beta,\gamma\}) < 4 r \}$ and smooth.  Thus the commutator
argument in the proof of Lemma~\ref{lem:finite_smoment} yields the
identity
\begin{equation}
   \chi_x \frac{1}{\widehat H - z} (\widehat H - H ) \ = \ \chi_x
   \frac{1}{\widehat H - z} \widetilde \Psi^2 \widehat T (\widehat H - H)
\end{equation}
where $\widetilde \Psi$ is one on the support of $\Psi$ and
supported in the set $\set{q :  \dist(q,\{\beta,\gamma\}) < 5r}$,
and $\widehat T$ is Hilbert-Schmidt with uniformly bounded norm:
$\|\widehat T\|_2 \le \widetilde C_d (1+\lambda)^n(1+ |z-E_0|)^n$
($n$ the smallest integer greater than $d/2$). The bounded term,
``$B$'', drops out of this representation because $\chi_x
\widetilde \Psi^2 = 0$ and multipliers of support no larger than
that of $\widetilde \Psi$ appear on the left of the terms of $B$.
Note that $\widehat T$ is independent of $\{\eta_\zeta : \zeta \in
\I_{\beta,\gamma} \}$ since it is constructed from $\widehat H$.

Next, we apply the commutator argument to the left of $(H-z)^{-1}$
to obtain
\begin{equation}
   \frac{1}{H-z} \widetilde \Theta_\beta^2  \ = \
   \widehat \Theta_\beta B \ + \ \frac{1}{H-z} \widehat \Theta_\beta^2 T
\end{equation}
with $B$ bounded, $ \norm{B} \le C_d (1 +
\lambda+\abs{z-E_0})^{2n}$, and $T$ a polynomial in $\{\eta_\zeta
: \zeta  \in \I_\beta\}$,
\begin{equation}
T \ = \ \sum_{\substack{ (k) \in \N^{\I_\beta} \\
             \abs{k} \le n}} \lambda^{|k|}
     T^{(k)} \prod_{\zeta \in \I_\beta} \eta_\zeta^{k_\zeta} \; ,
\end{equation}
with Hilbert-Schmidt coefficients,
\begin{equation}
\norm{T^{(k)}}_{HS} \ \le \ \widetilde C_d ( 1 +  \abs{z-E_0})^{n}
\; .
\end{equation}
Here the function $\widehat \Theta_{\beta}$ is bounded by one and
supported in $\set{q : \dist(q,\beta) < 2r}$. Note that $U_\zeta
\widehat \Theta_{\beta} = 0$ for $\zeta \not \in \I_\beta$.

Putting this all together, we obtain
\begin{multline}
     \norm{ \chi_x \frac{1}{H - z} \widetilde \Theta_{\beta}^2 } \le \  \norm{
         \chi_x \frac{1}{\widehat H - z} \widetilde \Psi } \times
     \left ( 1 \ + \ \norm{(\widehat H - H) \widehat \Theta_\beta B}
         \phantom{\sum_{\N^{\I_\beta}}\norm{ \frac{1}{H - z}}}
     \right . \\
     + \ \left . \sum_{(k) \in \N^{\I_\beta}} \lambda^{|k|}
         \norm{\widetilde \Psi \, \widehat T \,(\widehat H \ - \ H)
         \frac{1}{H - z} \widehat \Theta_{\beta}^2 T^{(k)}}
     \right )\; .
\end{multline}
Thus,
\begin{multline}\label{eq:banana}
     \norm{ \chi_x \frac{1}{H - z} \widetilde \Theta_{\beta}^2 }
     \le \ \  \norm{ \chi_x \frac{1}{\widehat H - z}
             \widetilde \Psi } \times
     \left (1\ + \ 2 \, b_+ \lambda \norm{B} \ +
         \phantom{\sum_{\substack{ \I_{\beta,\gamma}\\
         \N^{\I_\beta}}} \norm{ \frac{1}{H - z} }}
     \right . \\
     \left . + \sum_{\substack{ \zeta \in \I_{\beta,\gamma}, \zeta' \in
                 \I_{\beta} \\ (k) \in \N^{\I_\beta}}}
     2 \lambda^{1+|k|}
     \norm{\widetilde \Psi \, \widehat T U_\zeta
         \frac{1}{H - z} \frac{U_{\zeta'}}{F}
         \widehat \Theta_{\beta}^2 T^{(k)} }
     \right )\; .
\end{multline}
Based on \eqref{eq:banana}, using Prop.~\ref{prop:offdiag} and the
arguments from the proof of Lemma~\ref{lem:finite_smoment}, it
follows that
\begin{multline}
\Pr \left ( \left .  \norm{ \chi_x \frac{1}{H -z}
     \widetilde\Theta_{\beta}^2}  > t \right |
         {\mathcal F}^c_{\beta,\gamma} \right ) \\
     \begin{aligned} \le& \ \norm{ \chi_x \frac{1}{\widehat H - z}
         \widetilde \Psi } \times
         \left ( 1 \ + \ 2 b_+  C_d \lambda (1+\lambda)^{2n}
         (1 + \abs{z-E_0})^{2n}
         \phantom{\frac{b_+}{b_-}}\right . \\ & \quad \quad \quad \left . + \
         (M - 2) \, 4 C_W \widetilde C_d^2  (1+\lambda)^{2n}(1 +
\abs{z-E_0})^{2n}
         \, b_+ \, D^2 \right ) \ \frac{M}{t} \\
         \le& \ \norm{ \chi_x \frac{1}{\widehat H - z}
         \widetilde \Psi } \times M
         \left ( 1 \ + \ 2 b_+ C_d
         (M - 2) \, 4 C_W \widetilde C_d^2  \right )
     \end{aligned} \\
         \times (1+\lambda)^{2n+1}
         (1 + \abs{z-E_0})^{2n} \frac{D^2}{t} \; ,
\end{multline}
where $M$ is the number of terms between the brackets in
\eqref{eq:banana}. By ${\mathcal A}2$, $M$ is bounded uniformly in
$x,\beta,\gamma$.  This completes the proof of
Lemma~\ref{lem:decouple}.
\end{proof}

\newpage

\section{Finite-volume criteria}
\label{sec:fin-vol-criteria}

\subsection{The finite-volume inequality}
With the bounds provided by Lemmas~\ref{lem:finite_smoment} and
\ref{lem:decouple}, the proof of Theorem~\ref{thm:1} proceeds
according to a set of arguments familiar from the fractional
moment method for discrete operators \cite{ASFH}, and related to
the multi-scale analysis of random Schr\"odinger operators, e.g.
\cite{vonDreifus/Klein,Combes/Hislop:JFA}, as well as the analysis
of a number of lattice models in statistical mechanics, e.g.
\cite{Hammersley, Simon:correlationinequalities, Lieb,
Aizenman/Simon:Wardidentities, Aizenman/Newman,
Dobrushin/Shlosman}. We begin by proving a ``correlation
inequality''\textemdash \eq{eq:simonlieb} below\textemdash and
then iterate this inequality to obtain exponential decay of the
bulk Green-function.

\begin{lem}
   \label{lem:simon-lieb}
   Let $H$ be a random Schr\"odinger operator with disorder parameter
$\lambda > 0$
   which obeys ${\mathcal A}$ and ${\mathcal IAD}$. Then, for each $s < 1/3$ and $z
   \in \C \setminus \R$
   there exists $\widetilde C_{\lambda,s,z} <\infty$
   such that if we define, for $L > 23 r$,
   \begin{equation}
     a(x;L)  \ := \   L^{d-1} \,
     \sum_{\zeta \in
       {\mathcal S}_{x,L}}  \Ev{ \norm{ \chi_x \frac{1}{H^{(B^L_x)} - z }
     \chi_\zeta }^s }
   \end{equation}
   where ${\mathcal S_{x,L}}  =  \set{\zeta \in \I : L - 23r  < \abs{\zeta -x}
     < L-3r}$, then for any region $\Omega \supset B^L_x$,
   \begin{equation}
     \label{eq:simonlieb}
     \Ev{ \norm{ \chi_x \frac{1}{H^{(\Omega)} - z }
     \chi_y }^s }
     \le  \  \widetilde C_{\lambda,s,z} \
     a(x;L) \ \sum_{\zeta \in {\mathcal S'_{x,L}}}  \, \E
     \left ( \norm{ \chi_{\zeta}
     \frac{1}{H^{(\Omega )} -
       z} \chi_y}^s \right )
   \end{equation}
   for any $y$ with $\dist(x,y) > L + r_0 + 23 r$,
   where ${\mathcal S'_{x,L}} =  \{ \zeta \in \I : L + r_0 - 13r < |x-\zeta| <
   L+ r_0 + 23 r\}$ (recall that $r_0$ is
   the length scale appearing in ${\mathcal IAD}$), with
   \begin{equation}\label{eq:slcoeff}
     \widetilde C_{\lambda,s,z} \ \le \ \frac{\mathrm{Const.}}{1-3s}
     (1+\lambda^{-1})^{2s}(1 + \lambda)^{5s (d+4)} (1 + |z-E_0|)^{5s(d+2)} D^{10s} \; .
   \end{equation}
   If $H$ satisfies the stronger condition ${\mathcal A}3'$ then
   \begin{equation}\label{eq:brslcoeff}
     \widetilde C_{\lambda,s,z} \ \le \ \frac{\mathrm{Const.}}{1-3s}
     (1+\lambda^{-1})^{2s} (1 + |z-E_0|)^{5s(d+2)} D^{10s} \; .
   \end{equation}
\end{lem}

The main tools in the proof are the moment bounds presented above
and a well known analog of the resolvent expansion commonly used
for discrete Schr\"odinger operators, sometimes called the {\it
geometric resolvent identity}.
\begin{lem}
   \label{lem:resolvent_expansion}
   Let $H$ be a Schr\"odinger operator.
   Consider a sequence of three open sets $\Lambda_0 \subset \Lambda \subset
   \Omega$ with $\dist(\Lambda_0, \Lambda^c) > 0$
   and let $\Theta$ be a smooth function
   which is identically $1$ in a neighborhood of $\Lambda_0$ and identically
   zero in a neighborhood of $\Lambda^c$.  Given any restrictions $H^{(\Omega)}$
   and $H^{(\Lambda)}$ of $H$ to $\Omega$ and $\Lambda$ respectively,
   \begin{equation}
     \label{eq:resolvent_expansion1}
     \1_{\Lambda_0} \frac{1}{ H^{(\Omega)} - z} \ = \ \1_{\Lambda_0} \,
     \frac{1}{H^{(\Lambda)} - z} \Theta \ + \ \1_{\Lambda_0}
\frac{1}{H^{(\Lambda)} -
     z}  \com{H}{\Theta} \frac{1}{H^{(\Omega)} - z}
   \end{equation}
   for any $z$ at which both resolvents exist.

   If furthermore, $\Lambda'_0 \subset \Lambda' \subset \Omega$
   with $\dist(\Lambda'_0, {\Lambda'}^c) > 0$ and $\dist(\Lambda', \Lambda) >
   0$,
   and $\Theta'$ is a function which is identically $1$ in a neighborhood of
   $\Lambda'_0$ and identically $0$ in a neighborhood of ${\Lambda'}^c$ then
   \begin{equation}\label{eq:resolvent_expansion2}
     \1_{\Lambda_0} \frac{1}{H^{(\Omega)} -z} \1_{{\Lambda'_0}}
     \ = \ - \1_{\Lambda_0} \frac{1}{H^{(\Lambda)} - z}
     \com{H}{\Theta} \frac{1}{H^{(\Omega)} -z} \com{H}{\Theta'}
     \frac{1}{H^{(\Lambda')} -z} \1_{{\Lambda'_0}} \; ,
   \end{equation}
   with $H^{(\Lambda')}$ any restriction of $H$ to $\Lambda'$ and
   $z$ such that all three inverses exist.
\end{lem}
\begin{proof}
   The second identity, eq.~\eqref{eq:resolvent_expansion2},
   is a consequence of the first,
   eq.~\eqref{eq:resolvent_expansion1}, and its transpose
   (eq.~\eqref{eq:resolventatconjugatez} below). A number of terms
   drop out because $\Lambda \cap \Lambda' = \emptyset$.  To verify
   eq.~\eqref{eq:resolvent_expansion1} use the
   identity
   \begin{equation}
     \com{H}{\Theta} \ = \ (H^{(\Lambda)} - z) \Theta  - \Theta
     (H^{(\Omega)} - z)
   \end{equation}
   on ${\mathcal D}(H^{(\Omega)})$, which follows from the fact that
   $H^{(\Lambda)}f = H^{(\Omega)}f$ if the
   support of $f$ is strictly contained in $\Lambda$.  Multiplying on the left
   by $\1_{\Lambda_0} (H^{(\Lambda)} - z)^{-1} $ and on the right by
   $(H^{(\Omega)} - z)^{-1}$  yields eq.~\eqref{eq:resolvent_expansion1}.
\end{proof}

\noindent {\bf Remarks:}
\begin{enumerate}
\item  The identity,
   \begin{equation}\label{eq:resolventatconjugatez}
     \frac{1}{ H^{(\Omega)} - z}  \1_{\Lambda_0} \ = \  \,
     \Theta \frac{1}{H^{(\Lambda)} - z} \1_{\Lambda_0} \ - \
\frac{1}{H^{(\Omega)} -
       z}  \com{H}{\Theta} \frac{1}{H^{(\Lambda)} - z} \1_{\Lambda_0} \; ,
   \end{equation}
   follows from the transpose of
   eq.~\eqref{eq:resolvent_expansion1}  (at conjugate $z$).
\item Eq.~\eqref{eq:resolvent_expansion2} holds in a number of
other contexts.
   In  particular, it is true for discrete Schr\"odinger operators
   and in that case gives the usual geometric resolvent expansion (see
   \cite[eq.~(2.16)]{ASFH}).
\end{enumerate}

\begin{proof}[Proof of  Lemma~\ref{lem:simon-lieb}]
We shall assume ${\mathcal A}3$.  For the extension of the
argument to couplings which satisfy ${\mathcal A}3'$ see
Prop.~\ref{prop:extendsimonlieb} below.

We start by using the geometric resolvent identity
\eq{eq:resolvent_expansion2} of
   Lemma~\ref{lem:resolvent_expansion} with the sets
   \begin{equation}
     \begin{array}{ll}
       \Lambda_0  \ = \  B_x^{L -11r} \; , &
       \Lambda \ = \ B_x^L \; ,\\
       \Lambda'_0 \ = \ \Omega \setminus B_x^{L + r_0+ 12 r} \; , &
       \Lambda' \ = \ \Omega \setminus B_x^{L + r_0} \; ,
     \end{array}
   \end{equation}
   and $\grad \Theta, ~ \grad \Theta'$ supported in
   $\{ q : L - 11r < |q-x| < L - 10r\}$, $\{q : L + r_0+11 r < | q -
x| < L + r_0+12 r\}$
   respectively.

   The operators $\com{H}{\Theta^\sharp}$ are {\em
     local} and supported in
   the set where $\grad \Theta$ is non-zero. A
   particular consequence of this observation is that
   \begin{equation}
     \begin{aligned}
       \com{H}{\Theta} \ =& \ \sum_{\zeta_1 \in {\mathcal S}_0}
       \sum_{\substack{ \zeta_2 \\
       U_{\zeta_2} \Theta_{\zeta_1} \not \equiv 0 }}
       \Theta_{\zeta_1} \com{H}{\Theta} \frac{U_{\zeta_2}}{F}
       \\ \com{H}{\Theta'} \ =& \ \sum_{\zeta_1' \in {\mathcal S}_0'}
       \sum_{\substack{ \zeta_2' \\
       U_{\zeta_2'} \Theta_{\zeta_1'} \not \equiv 0 }}
        \frac{U_{\zeta_2'}}{F} \com{H}{\Theta'} \Theta_{\zeta_1'}
       \; ,
     \end{aligned}
   \end{equation}
   with $ {\mathcal S}_0 = \set{ q \in \I \ : \ L - 13 r < |q -x| < L
- 8 r }$ and
   ${\mathcal S}_0' =  \{ q \in \I \ : \  L +  r_0 + 9 r < |q - x| < L
+r_0 + 14 r \}$.
   Here $\Theta_\zeta$ is the smooth partition of unity used in
   Lemma~\ref{lem:decouple}.

   Recall that
   \begin{equation}
     \norm{\frac{1}{(a + H_0)^{1/2}} \com{H}{\Theta^\sharp} } \ \le \
     (\norm{\Delta \Theta^\sharp}_\infty \ + \ \sqrt{2} \norm{\grad
       \Theta^\sharp}_\infty ) \; ,
   \end{equation}
   which is bounded for smooth $\Theta$.  We may
   choose $\Theta$ so that the right hand side is bounded uniformly in
$L$, being no
   larger than $(\const) \, d^2/r^2$.

   With these observations, we obtain from \eq{eq:resolvent_expansion2} the
   inequality:
   \begin{multline}
     \label{eq:splitit3ways}
     \norm{ \chi_x \frac{1}{H^{(\Omega)} - z} \chi_y}
     \\ \le \ (\const) \ \sum_{ \substack{\zeta_1 \in {\mathcal
     S}_0 \\
     \zeta_2 : \abs{\zeta_2 - \zeta_1} < 3r }}
     \sum_{ \substack{\zeta_1' \in {\mathcal S}_0' \\ \zeta_2' :
     \abs{\zeta_2' - \zeta_1'} < 3r}}
     \norm{ \chi_x \frac{1}{H^{(B_x^L)} - z}
       \Theta_{\zeta_1} (a + H_0)^{1/2}} \\ \times \norm{U_{\zeta_2}
       \frac{1}{H^{(\Omega)} - z}  U_{\zeta_2'} }
     \\ \times \norm{ (a + H_0)^{1/2} \Theta_{\zeta_1'} \frac{1}{H^{(\Omega
       \setminus B_x^{L+r_0})} - z} \chi_y} \; .
   \end{multline}
   We now raise \eq{eq:splitit3ways} to the power $s < 1/3$ and take
   expectation values, using the inequality $(\sum a_n)^s \le \sum a_n^s$.
   Consider each term on the right-hand side
   separately: first estimate the expectation conditioned on
   $\Fc_{\zeta_1,\zeta_2,\zeta_1',\zeta_2'}$, using
   the H\"older inequality to separate factors. The central factor may be
   estimated with Lemma~\ref{lem:finite_smoment} (see \eq{eq:finite_smoment}),
   to yield
   \begin{multline}
     \Ev{ \norm{ \chi_x \frac{1}{H^{(\Omega)} - z} \chi_y}^s} \
     \le  \ (\const) \ \frac{C_\lambda^s}{(1-3s)^{1/3}} \
     (1 + |z-E_0|)^{s(d+2)} \ D^{2s}  \\
     \begin{aligned}
       & \times \sum_{ \substack{ \left < \zeta_1, \zeta_2 \right > \\
       \left < \zeta_1', \zeta_2' \right >}}
       \E \left [  \Ev{ \left .
         \norm{ \chi_x \frac{1}{H^{(B_x^L)} - z}
           \Theta_{\zeta_1} (a + H_0)^{1/2}}^{3s} \right |
       \Fc_{\zeta_1,\zeta_2} } ^{1/3} \right .
       \\
       & \left . \ \times \   \Ev{ \left .
         \norm{ (a + H_0)^{1/2} \Theta_{\zeta_1'}  \frac{1}{H^{(\Omega
           \setminus B_x^{L+r_0} )} - z} \chi_y}^{3s}
       \right | \Fc_{\zeta_1',\zeta_2'} }^{1/3} \right ]\; .
     \end{aligned}
   \end{multline}
   \noindent Here we have noted that $H^{(B_x^L)}$ ($H^{(\Omega
   \setminus B_x^{L+r_0})}$) does not depend on the variables
   $\eta_{\alpha}$ with $\alpha \in \I_{\zeta_1',\zeta_2'}$
   ($\alpha \in \I_{\zeta_1,\zeta_2}$).

   The remaining two factors in each term are i) independent by ${\mathcal IAD}$
   and ii) of the correct form to be estimated using Lemma~\ref{lem:decouple}.
   Thus,
   \begin{multline} \label{eq:resamplebound}
     \Ev{ \norm{ \chi_x \frac{1}{H^{(\Omega)} - z} \chi_y}^s} \\
     \le  \ (\const) \left [ \frac{C_\lambda^s}{(1-3s)^{1/3}}
     (1 + |z-E_0|)^{s(d+2)} \ D^{2s} \right ] \\
     \times \left [ \frac{\widetilde C_\lambda^s}{(1 - 3s)^{1/3}}
         (1 + |z-E_0|)^{s (d + 3)} D^{2s} \right ]^2 \\
     \times \sum_{ \substack{ \left < \zeta_1, \zeta_2 \right > \\
             \left < \zeta_1', \zeta_2' \right >}}
     \Ev{ \norm{ \chi_x \frac{1}{\widehat H_{\zeta_1,\zeta_2}^{(B_x^L)} - z}
         \1_{S_{\zeta_1,\zeta_2}} }^s } \\
     \times \Ev{ \norm{ \1_{S_{\zeta_1',\zeta_2'}} \frac{1}{\widehat
     H_{\zeta_1',\zeta_2'}^{(\Omega \setminus B_x^{L+r_0})} - z} \chi_y }^s }
     \; .
     \end{multline}
   Here $\widehat H_{\zeta_1,\zeta_2}$ ($\widehat
   H_{\zeta_1',\zeta_2'}$) denotes a version of $H$ which is
   re-sampled over $\zeta \in \I_{\zeta_1,\zeta_2}$ ($\zeta \in
   \I_{\zeta_1',\zeta_2'}$) in the sense of \eq{eq:re-sample}.
   The assumptions $L>23r$ and $\dist (x,y) > L+r_0+23r$ in
   Lemma~\ref{lem:simon-lieb} are used to satisfy the distance
   requirements of Lemma~\ref{lem:decouple}.

   We now pick the re-sampled variables with the same distribution
   as the $\{\eta_{\zeta}\}$, and include averaging over these
   variables in the expectations. Since $S_{\zeta_1,\zeta_2}
   \subset B_{\zeta_1}^{8r}$, $S_{\zeta_1', \zeta_2'} \subset
   B_{\zeta_1'}^{8r}$ and for each $\zeta_1$, $\zeta_1'$ there is
   only a fixed finite number of values for $\zeta_2$, $\zeta_2'$ (by
${\mathcal A}2$),
   we infer from \eq{eq:resamplebound}, after adjusting the
   constant, that
   \begin{multline} \label{eq:secondresamplebound}
     \Ev{ \norm{ \chi_x \frac{1}{H^{(\Omega)} - z} \chi_y}^s}
       \le  \ (\const) \ \frac{C_\lambda^s \widetilde C_\lambda^{2s}}{1-3s}
     (1 + |z-E_0|)^{3s(d+3)} \ D^{6s} \\
       \times \sum_{ \substack{ \zeta_1 \in S_0 \\
                 \zeta_1' \in S_0'}}
       \Ev{ \norm{ \chi_x \frac{1}{H^{(B_x^L)} - z}
       \1_{B_{\zeta_1}^{8r}} }^s } \\
       \times \Ev{ \norm{ \1_{B_{\zeta_1'}^{8r}}
       \frac{1}{ H^{(\Omega \setminus B_x^{L+r_0})} - z} \chi_y }^s } \; .
   \end{multline}

   To complete the proof of \eq{eq:simonlieb} we need a bound for
   the resolvent of $H^{(\Omega \setminus B_x^{L+r_0})}$ in terms of
   the resolvent of $H^{(\Omega)}$. For this we will apply
   eq.~(\ref{eq:resolvent_expansion1}) of Lemma~\ref{lem:resolvent_expansion}
   with
   \begin{equation}
   \Lambda = \Omega \setminus B_x^{L+r_0}, \qquad \Lambda_0 = \Omega
   \setminus B_x^{L+r_0+r}
   \end{equation}
   and a smooth function $\Theta''$ such that $\nabla \Theta''$ is
   supported in $\{q: L+r_0 < |q-x| < L+r_0 + r \}$. Note that $\Theta''
   \chi_y = \chi_y$ and $B_{\zeta_1'}^{8r} \subset \Lambda_0$ for
   every $\zeta_1' \in {\mathcal S}_0'$. In place of (\ref{eq:splitit3ways})
   we obtain
   \begin{multline}
     \label{eq:splitit2ways}
     \norm{ \1_{B_{\zeta_1'}^{8r}} \frac{1}{H^{(\Omega
     \setminus B_x^{L+r_0})} - z} \chi_y}
     \le \norm{ \1_{B_{\zeta_1'}^{8r}} \frac{1}{H^{(\Omega)} - z} \chi_y}
     \\ \ + \ (\const) \ \sum_{ \substack{\zeta_1'' \in {\mathcal S}_0'' \\
     \zeta_2'' : \abs{\zeta_2'' - \zeta_1''} < 3r }}
     \norm{ \1_{B_{\zeta_1'}^{8r}}  \frac{1}{H^{(\Omega \setminus
B_x^{L+r_0})} - z}
       U_{\zeta_2''}}
     \\ \times \norm{(a+H_0)^{1/2} \Theta_{\zeta_1''}
       \frac{1}{H^{(\Omega)} - z}  \chi_y } \; ,
   \end{multline}
   where ${\mathcal S}_0'' := \{q \in \I: \; L + r_0 - 2r < |q-x| <
L+r_0 + 3r\}$.

   This yields the expectation bound
   \begin{multline}
     \label{eq:expect2ways}
     \Ev{ \norm{ \1_{B_{\zeta_1'}^{8r}} \frac{1}{H^{(\Omega
     \setminus B_x^{L+r_0})} - z} \chi_y}^s}
     \le \Ev{ \norm{ \1_{B_{\zeta_1'}^{8r}} \frac{1}{H^{(\Omega)} - z}
     \chi_y}^s}
     \\ \ + \ (\const) \ \sum_{ \left< \zeta_1'', \zeta_2'' \right> }
     \sum_{\substack{ \zeta\in \I \\ \abs{\zeta - \zeta_1'} < 9r}}
     \E \left( \norm{ U_{\zeta}  \frac{1}{H^{(\Omega \setminus
B_x^{L+r_0})} - z}
       U_{\zeta_2''}}^s \right.
     \\ \left. \times \norm{(a+H_0)^{1/2} \Theta_{\zeta_1''}
       \frac{1}{H^{(\Omega)} - z}  \chi_y }^s \right) \; .
   \end{multline}
   For the terms in the sum of \eq{eq:expect2ways} we first take
   the expectation conditioned on $\Fc_{\zeta_1'', \zeta_2'',
   \zeta}$ and use the Cauchy-Schwarz inequality on the product.
   The resulting term
   \begin{equation}
   \Ev{ \left. \norm{ U_{\zeta} \frac{1}{H^{(\Omega \setminus
   B_x^{L+r_0})} -z} U_{\zeta_2''}}^{2s} \right| \Fc_{\zeta_1'',
   \zeta_2'', \zeta}} \; ,
   \end{equation}
   is treated by \eq{eq:finite_smoment}, while
   \begin{equation}
   \Ev{ \left. \norm{ (a+H_0)^{1/2} \Theta_{\zeta_1''}
   \frac{1}{H^{(\Omega)}-z} \chi_y}^{2s} \right| \Fc_{\zeta_1'',
   \zeta_2'', \zeta}}
   \end{equation}
   can be estimated through a straightforward variant of
   Lemma~\ref{lem:decouple}, with re-sampling done over all variables
   in $\I_{\zeta_1'',\zeta_2'',\zeta}$.

   Averaging over the re-sampled variables yields
   \begin{multline} \label{eq:tralala}
     \Ev{ \norm{ \1_{B_{\zeta_1'}^{8r}} \frac{1}{H^{(\Omega
     \setminus B_x^{L+r_0})} - z} \chi_y}^s}
     \le \Ev{ \norm{ \1_{B_{\zeta_1'}^{8r}} \frac{1}{H^{(\Omega)} - z}
     \chi_y}^s}
     \\ \ + \ (\const) \frac{C_\lambda^s \widetilde C_\lambda^s}{1-2s}
     (1+|z-E_0|)^{2s(d+3)} D^{4s}
     \\ \times \sum_{ \left< \zeta_1'', \zeta_2'', \zeta \right> }
     \Ev{ \norm{ \1_{S_{\zeta_1'', \zeta_2'', \zeta}}
\frac{1}{H^{(\Omega)} - z}
     \chi_y}^s} \; .
   \end{multline}
   After inserting this into \eq{eq:secondresamplebound} we arrive at
   \eq{eq:simonlieb} by covering all the sets
   $B_{\zeta_1}^{8r}$, $B_{\zeta_1'}^{8r}$ and $S_{\zeta_1'',
   \zeta_2'', \zeta}$ with balls $B_{\alpha}^r$, $\alpha \in \I$.
   The factor $L^{d-1}$ appears in this calculation since by
   ${\mathcal A}2$ the sets $S_0'$ and $S_0''$ contain $CL^{d-1}$ points in
   $\I$. This completes the proof of \eq{eq:simonlieb}.

   The resulting bound for the coefficient $\widetilde
   C_{s,\lambda}$ is
   \begin{equation}
     \widetilde C_{s,\lambda,z} \ \le \ (\const) \frac{C_\lambda^{2s}
     \widetilde C_{\lambda}^{3s}}{(1-2s)(1-3s)} (1 + |z-E_0|)^{5s(d+2)} D^{10s} \; ,
   \end{equation}
   and the growth bounds for $C_\lambda$ and $\widetilde C_\lambda$
   provided by Lemmas~\ref{lem:finite_smoment} and \ref{lem:decouple}
easily yield
   \eq{eq:slcoeff}.  For the proof of \eq{eq:brslcoeff}, we refer
   to Prop.~\ref{prop:extendsimonlieb} in the following section.
\end{proof}

\subsection{Large disorder and blow-up regularity}\label{sec:largedisorder}
In this section we consider the large disorder regime {($\lambda >1$)} and
discuss the use of assumption ${\mathcal A}3'$ to improve the bounds provided
by Lems.~\ref{lem:finite_smoment} and \ref{lem:simon-lieb}\textemdash i.e.,
\eq{eq:finite_smoment} and \eq{eq:simonlieb}.  The basic idea is to apply
Lemmas~\ref{lem:finite_smoment} and \ref{lem:decouple} to fluctuations of
$\lambda \eta_\zeta$ which are of order one.

It is instructive first to consider independent identically distributed random
couplings $\eta_\zeta$. Given {$\lambda >1$}, we decompose the interval $[0,1]$
as a union of a finite number of intervals of length less than $1/\lambda$ with
disjoint interiors, $I_j = [a_j, a_{j+1}]$, and consider $\eta_\zeta$ as a
super-position of its ``integer'' and ``fractional'' parts with respect to this
decomposition:
\begin{equation}
   \lambda \eta_\zeta \ = \ \lambda a_\zeta \ + \ f_\zeta \; .
\end{equation}
Here the random variable $a_\zeta$ takes value $a_j$ when
$\eta_\zeta$ falls in the interval $I_j$, and $f_\zeta = \lambda
(\eta_\zeta - a_\zeta)$ is a random variable which takes values in
the interval $[0,1]$. The conditional distribution of
$f=f_\alpha$, at a specified value of $a_\alpha$, is given by the
following expression:
\begin{equation}
   \frac{\rho(f/\lambda + a_\alpha)}{\lambda \int_{I_j}
     \rho(\eta)
   \di \eta } \ \di f \; ,
\end{equation}
where $\rho$ is the common density for $\eta_\zeta$. The
denominator is just the probability that $\eta_\alpha$ falls in
the interval $I_j$.

To apply Lemmas~\ref{lem:finite_smoment} and \ref{lem:decouple} to
conditional averages with respect to $f_\zeta$, we would need to
use the following value for $D$:
\begin{equation}
   D \ = D(\lambda) = \
     \sup_j \mathop{\operatorname{ess-sup}}_{f
     \in I_j  }
   \frac{\rho(f/\lambda +a_j)}{\lambda
     \int_{I_j} \rho(\eta)
   \di \eta} \; .
\end{equation}
For fixed $\lambda > 1$, this is certainly finite. However,
$D(\lambda)$ depends on $\lambda$ as well as the choice of $I_j$.
Blow-up regularity of the distribution $\rho$, which is guaranteed
by assumption ${\mathcal A}3'$, is precisely the requirement that
$I_j=I_j(\lambda)$ may be chosen so that $D(\lambda)$ remains
bounded for large $\lambda$.

The situation for general couplings $\{\eta_\zeta\}$ obeying
${\mathcal A}3'$ is somewhat more complicated since we must admit
{\em random} decompositions of the interval $[0,1]$. Specifically,
${\mathcal A}3'$  guarantees that the conditional distributions
$\rho_\alpha(\eta_\zeta|\omega)$ of $\eta_\alpha$ at specified
values of the remaining couplings are almost-surely blow-up
regular with uniformly bounded blow-up norm: $D_{\rho_\alpha(\cdot
| \omega)} \le D$.  Thus for each $\alpha$ we may choose
$I_j^\alpha =
I_j^\alpha(\omega)=[a_j^\alpha(\omega),a_{j+1}^\alpha(\omega)]$
which are measurable with respect to $\Fc_{\alpha}$, satisfy
$|I_j^\alpha| \le 1/\lambda$, and such that
\begin{equation}
   \sup_\alpha \, \mathop{\operatorname{ess-sup}}_{\omega} \, \max_{j} \,
     \mathop{\operatorname{ess-sup}}_{f \in I_j^\alpha(\omega)  }
   \frac{\rho_\alpha(f/\lambda + a_j^\alpha(\omega) | \omega )}{\lambda
     \int_{I_j^\alpha(\omega)} \rho_\alpha(\eta | \omega )
   \di \eta} \ \le \ D \; .
\end{equation}
We now define $a_\zeta$ and $f_\zeta$ as above: $a_\zeta =
a_j^\alpha(\omega)$ if $\eta_\zeta$ falls in $I_j^\alpha(\omega)$
and $f_\zeta = \lambda \eta_\zeta - \lambda a_\zeta$.  The density
of the distribution of $f=f_\alpha$ conditioned on $a_\alpha$ and
$\Fc_\alpha$ is
\begin{equation}
   \frac{\rho_\alpha( f/\lambda + a_\alpha|\omega )}{\lambda
\int_{I_j^\alpha(\omega)}
     \rho_\alpha(\eta |\omega)
   \di \eta } \; ,
\end{equation}
which is a density supported in $[0,1]$ and bounded by $D$.

With these notions, it is an easy exercise in conditional
expectations to prove the following extension of
Lemma~\ref{lem:finite_smoment}:
\begin{prop}\label{prop:extend}
Let $H$ be a random Schr\"odinger operator which satisfies
${\mathcal A}1$, ${\mathcal A}2$, and ${\mathcal A}3'$. Then
\eq{eq:weak1-1_1} of Lemma~\ref{lem:finite_smoment} holds with a
coefficient $C_\lambda$ which satisfies $\sup_{\lambda \ge 1}
C_\lambda < \infty$, provided we use for $D$ the constant
appearing in ${\mathcal A}3'$\textemdash i.e, the blow-up norm of
the distribution.
\end{prop}

Turning now to Lemma~\ref{lem:decouple}, we note that the disorder
strength plays a role in the re-sampling procedure in addition to
the averaging.  Hence, it is most natural to use the lemma as
stated and re-sample only the variables $f_\zeta$ with the
$a_\zeta$ fixed. An argument along these lines will be used to
complete the proof of \eq{eq:brslcoeff} from
Lemma~\ref{lem:simon-lieb}:
\begin{prop}\label{prop:extendsimonlieb}
Let $H$ be a random Schr\"odinger operator which satisfies
${\mathcal IAD}$, ${\mathcal A}1$, ${\mathcal A}2$, and ${\mathcal
A}3'$. Then \eq{eq:simonlieb} of Lemma~\ref{lem:simon-lieb} holds
with a coefficient $\widetilde C_{s,\lambda}$ which obeys
\eq{eq:brslcoeff}.
\end{prop}
\begin{proof}
The proof follows closely the derivation of \eq{eq:simonlieb}
given above with a few modifications which we now indicate.

Let ${\mathcal A}$ denote the sigma-algebra generated by
$a_\zeta$, $\zeta \in \I$. Conditioning on ${\mathcal A}$, we
obtain a random Schr\"odinger operator with $\lambda =1$ and
couplings $f_\zeta$, $\zeta \in \I$ which obey ${\mathcal IAD}$
and ${\mathcal A}3$. Thus, following the proof of
Lemma~\ref{lem:simon-lieb} we obtain the following analog of
\eq{eq:resamplebound}:
\begin{multline} \label{eq:conditionalresamplebound}
     \Ev{ \left . \norm{ \chi_x \frac{1}{H^{(\Omega)} - z} \chi_y}^s
     \right | \mathcal A}
       \le  \ (\const) \ \frac{1}{1-3s}
     (1 + |z-E_0|)^{3s(d+3)} \ D^{6s} \\
       \times \sum_{ \substack{ \zeta_1 \in S_0 \\
                 \zeta_1' \in S_0'}}
       \Ev{ \left .\norm{ \chi_x \frac{1}{H^{(B_x^L)} - z}
       \1_{B_{\zeta_1}^{8r}} }^s \right | \mathcal A} \\
       \times \Ev{ \left . \norm{ \1_{B_{\zeta_1'}^{8r}}
       \frac{1}{ H^{(\Omega \setminus B_x^{L+r_0})} - z} \chi_y }^s
       \right | \mathcal A} \; .
   \end{multline}
We now average with respect to ${\mathcal A}$, noting that
$H^{(B_x^L)}$ and $H^{(\Omega \setminus B_x^{L+r_0})}$ are
independent by ${\mathcal IAD}$, and this independence is
inherited by conditional averages over $\mathcal A$. The resulting
bound is identical to \eq{eq:resamplebound} with $\lambda=1$. To
complete the proof, use a similar argument to prove the analog of
\eq{eq:tralala}, first conditioning on $\mathcal A$ to obtain
coefficients with $\lambda=1$.
\end{proof}

\subsection{The criteria\textemdash proof of Theorem~\ref{thm:1}.} Let
\begin{equation}
  M(s,\lambda,E+\im \delta) \ := \ \sup_{0\le\eps \le\delta}
  \widetilde C_{\lambda,s,E+\im \eps}
  \ \sup_{\substack{\alpha \in \I \\ L >
23 r}} \frac{ \# S_{\alpha,L}'}{L^{d-1}} \; ,
\end{equation}
with the sets ${\mathcal S}_{\alpha, L}'$  as in
Lemma~\ref{lem:simon-lieb}, and suppose \eq{eq:criterion} holds
for some $s \in (0,1/3)$ and $E\in \R$.

For each open set $\Omega \subset \R^d$ and $\delta > 0$, we
define
\begin{equation}
G_\Omega^\delta(\alpha,\beta) := \sup_{|\eps| < \delta} \Ev{
\norm{\chi_{\alpha}
   \frac{1}{H^{(\Omega)}  - E-\im \eps} \chi_{\beta}}^s } \;
\end{equation}
with $\alpha,\beta \in \I \cap \Omega$. Whenever $\alpha,\beta \in
\I \cap\Omega$ with $B^L_\alpha \subset \Omega$ and
$|\alpha-\beta| > L + r_0 + 23 r$, Lemma~\ref{lem:simon-lieb}
implies that
\begin{equation}\label{eq:basicineq}
   G_\Omega^\delta(\alpha,\beta) \ \le
   \ \e^{-\gamma(\delta)} \frac{1}{\# S_{\alpha,L}'} \sum_{\zeta \in
S_{\alpha,L}'}
   G_\Omega^\delta(\zeta,\beta) \; ,
\end{equation}
with
\begin{multline}
\e^{-\gamma(\delta)} \ := \ M(s,\lambda, E + \im \delta)
\sup_{\eps < \delta} \sup_{\alpha \in \I}
     \E \left ( \norm{\chi_{\alpha}\,
     (H^{(B_{\alpha}^L)} -E-i\eps)^{-1} \, \1_{\delta
     B^L_\alpha} }^s
     \right ) \; .
\end{multline}
Note that $\gamma(\delta)$ is an increasing function of $\delta$
and the assumed finite volume bound \eq{eq:criterion} shows that
$\lim_{\delta \rightarrow 0} \gamma(\delta) > 0$. In particular,
there is $\delta_0 > 0$ such that $\gamma(\delta) > 0$ for $\delta
< \delta_0$.

Let us define,
\begin{equation}\label{eq:FOmegadefn}
   F_\Omega^\delta(\nu) \ := \ \sup_{\alpha, \beta \in \I \cap \Omega} \e^{\nu
   \dist_\Omega(\alpha,\beta) /2L}  G_\Omega^\delta(\alpha,\beta)\; .
\end{equation}
The proof will proceed as follows.  First, for $\nu <
\gamma(\delta)$ and bounded $\Omega$, we derive an $\Omega$
independent bound on $F_\Omega^\delta(\nu)$. Second, unbounded
$\Omega$ are handled via finite volume approximations. Finally, we
pass to the limits $\delta \rightarrow 0$ and $\nu \rightarrow
\gamma$.

Now, consider a bounded region $\Omega$ and $\alpha,\beta \in \I \cap \Omega$
with {$\dist_\Omega(\alpha,\beta)>2L$.} Observe that either \eq{eq:basicineq}
or its conjugate holds for this pair, since $|\alpha-\beta|> 2L > L + r_0 +
23r$ and one of $B^L_\alpha$ or $B^L_\beta$ is contained entirely in $\Omega$.
If, say $B^L_\alpha \subset \Omega$, then \eq{eq:basicineq} implies
\begin{equation}
\begin{aligned}
\e^{\nu \dist_\Omega(\alpha,\beta)
     /2L}G_\Omega^\delta(\alpha,\beta) \
     \le& \ \e^{-\gamma(\delta)} \frac{1}{\# S_{\alpha,L}'} \sum_{\zeta \in
     S_{\alpha,L}'} \e^{\nu \dist_\Omega(\alpha,\beta)/2L}
     G_\Omega^\delta(\zeta,\beta) \\
     \le & \e^{\nu} \e^{-\gamma(\delta)} \frac{1}{\# S_{\alpha,L}'}
     \sum_{\zeta \in S_{\alpha,L}'} \e^{\nu \dist_\Omega(\zeta,\beta)/2L}
     G_\Omega^\delta(\zeta,\beta) \; ,
\end{aligned}
\end{equation}
where we have used the triangle inequality for $\dist_\Omega$ and
observed that $\dist_\Omega(\alpha, \zeta) \le 2L$ for $\zeta \in
S_{\alpha,L}'$.   If instead $B^L_\beta \subset \Omega$, the
conjugate version of \eq{eq:basicineq} implies the conjugate
version of this bound.  Either way, the end result is that
\begin{equation}\label{eq:largedistbound}
\e^{\nu \dist_\Omega(\alpha,\beta)
     /2L}G_\Omega^\delta(\alpha,\beta) \ \le \ \e^{\nu} \e^{-\gamma(\delta)}
     F_\Omega^\delta(\nu) \; .
\end{equation}

If $\nu < \gamma(\delta)$, \eq{eq:largedistbound} implies that
$F_\Omega^\delta(\nu)$ may be found by restricting the supremum in
\eq{eq:FOmegadefn} to ``nearby'' pairs (here we use that
$F_\Omega^\delta(\nu)$ is finite due to boundedness of $\Omega$):
\begin{equation}\label{eq:FOmegabound}
F_\Omega^\delta(\nu) \ = \ \sup_{\alpha, \beta :
\dist_\Omega(\alpha,\beta) < 2L} \e^{\nu
   \dist_\Omega(\alpha,\beta) /2L}  G_\Omega^\delta(\alpha,\beta) \ \le \
   \e^{\nu} A(s,\lambda,E) \; ,
\end{equation}
where $A(s,\lambda,E)<\infty$ is the \textit{a priori} bound on
fractional moments provided by Lemma~\ref{lem:finite_smoment}, so
\begin{equation}
A(s,\lambda,E) \ \le \ \frac{C_{\lambda}^s}{1-s} \ (1 +
|E-E_0|)^{s(d+2)} \ D^{2s}
\end{equation}
by \eq{eq:finite_smoment}.

To complete the proof, we must extend \eq{eq:FOmegabound} to
unbounded regions.  For this purpose, fix an open set $\Omega
\subset \R_d$ and let $\Omega_j = [-j,j]^d \cap \Omega$. One may
verify, for example using the geometric resolvent identity
\eq{eq:resolvent_expansion1} and the Combes-Thomas estimate
\cite{Combes/Thomas}, that for any $\alpha, \beta \in \I \cap
\Omega$ and $z \in \C \setminus \R$
\begin{equation}
\lim_{j \rightarrow \infty} \norm{ \chi_{\alpha}
\frac{1}{H^{(\Omega_j)} - z} \chi_{\beta} - \chi_{\alpha}
\frac{1}{H^{(\Omega)} - z} \chi_{\beta}} \ = \ 0 \; .
\end{equation}
Thus
\begin{equation}
\sup_{|\eps| \le \delta} \norm{\chi_{\alpha} \frac{1}{H^{(\Omega)}
- E - \im \eps} \chi_{\beta}} \ \le \ \liminf_{j \rightarrow
\infty} \sup_{|\eps| \le \delta} \norm{\chi_{\alpha}
\frac{1}{H^{(\Omega_j)} - E - \im \eps} \chi_{\beta}} \; ,
\end{equation}
and by Fatou's Lemma,
\begin{equation}
   G_\Omega^\delta(\alpha,\beta) \ \le \ \liminf_{j \rightarrow \infty}
   G_{\Omega_j}^\delta(\alpha,\beta) \; ,
\end{equation}
Since $\dist_{\Omega_j}(\alpha,\beta) \rightarrow
\dist_\Omega(\alpha,\beta)$, this implies \eq{eq:FOmegabound}
holds for $\Omega$ when $\nu < \gamma(\delta)$.

In the limit $\delta \rightarrow 0$, \eq{eq:FOmegabound} implies
that
\begin{equation}
\limsup_{\eps \rightarrow 0} \Ev{ \norm{\chi_{\alpha}
   \frac{1}{H^{(\Omega)}  - E-\im \eps} \chi_{\beta}}^s } \ \le \
   \e^{\nu} A(s,\lambda,E) \e^{-\nu
   \dist_\Omega(\alpha,\beta) /2L}
\end{equation}
for any $\nu < \gamma$.  Finally, taking $\nu \rightarrow \gamma$,
we obtain \eq{eq:thm1conclusion}. \qed

\newpage

\section{Applications: results for distinct energy
ranges}\label{sec:applications} To apply the above results on
localization we need to verify the sufficiency criteria of
Theorem~\ref{thm:1} in specific disorder regimes and energy
ranges.  In this section we present several such results for
regimes of interest, including the familiar large disorder and
band edge (Lifshitz tail) regimes. The mechanisms which allow us
to check \eq{eq:criterion} are essentially those used to verify
the initial length scale estimates for a multiscale analysis. As
such, we do not attempt to give an exhaustive list of applications
here, but rather try to illustrate how known methods may be
combined with the arguments developed in this paper.

A useful feature of our results is that it suffices to check
\eq{eq:criterion} at a single energy $E \in \R$, since the
following continuity result allows us to extend the obtained
bounds onto an interval containing $E$ (as well as off the real
axis).

\begin{lem} \label{lem:holdercont}
Let $H$ satisfy ${\mathcal A}$. Fix $\Lambda$, and subsets $\Lambda_1$ and
$\Lambda_2$ of $\Lambda$. Let $G(z) := \1_{\Lambda_1} (H^{(\Lambda)}-z)^{-1}
\1_{\Lambda_2}$. Then, for any $0<s \le 1/2$, $z \mapsto \Ev {\|G(z)\|^s}$ is
locally H\"older-continuous with exponent $s$ for all $z\in \C$. More
precisely, there exists a constant $A_{s,\lambda} < \infty$ such that
\begin{multline} \label{eq:holderbound}
\left| \Ev { \|G(z)\|^s} - \Ev {\|G(w)\|^s} \right| \le
A_{s,\lambda} N_{\Lambda} N_{\Lambda_1}^{1/2} N_{\Lambda_2}^{1/2}
\\ \times (1+|z-E_0|)^{s(d+2)} (1+|w-E_0|)^{s(d+2)} |z-w|^s
\end{multline}
for all $z, w \in \C$. Here $N_{\Lambda}=\#(\I\cap\Lambda)$ is the
``volume'' defined in the discussion following
Lemma~\ref{lem:finite_smoment} and
\begin{equation}
A_{s,\lambda} \ \le \ \mathrm{const.} \ \frac{(1+
1/\lambda)^{2s}}{1-2s} \times \begin{cases}
(1+\lambda)^{2s(d+2)} & \text{ if ${\mathcal A}3$ holds,} \\
1 & \text{ if ${\mathcal A}3'$ holds.}
\end{cases}
\end{equation}
\end{lem}

\begin{proof} By the resolvent identity, noting that $\1_\Lambda =
\1$ on $L^2(\Lambda)$,
\begin{equation} \label{eq:resid}
G(z)-G(w) = (z-w) \1_{\Lambda_1} (H^{(\Lambda)}-z)^{-1}
\1_{\Lambda} (H^{(\Lambda)}-w)^{-1} \1_{\Lambda_2} \; .
\end{equation}
Thus
\begin{multline} \label{eq:Gcalc}
\left| \Ev { \|G(z)\|^s} - \Ev {\|G(w)\|^s} \right| \le \Ev
{\|G(z) -
G(w)\|^s} \\
\begin{aligned}
\ \le& \ |z-w|^s \Ev { \|\1_{\Lambda_1} \frac{1}{H^{\Lambda}-z}
\1_{\Lambda} \frac{1}{H^{(\Lambda)}-w} \1_{\Lambda_2} \|^s} \\
\ \le& \ |z-w|^s \Ev { \|\1_{\Lambda_1} \frac{1}{H^{(\Lambda)}-z}
\1_{\Lambda} \|^{2s}}^{1/2} \Ev { \|\1_{\Lambda}
\frac{1}{H^{(\Lambda)}-w} \1_{\Lambda_2}\|^{2s}}^{1/2} \: .
\end{aligned}
\end{multline}
The result now follows from
Lemma~\ref{lem:finite_smoment}\textemdash see
\eq{eq:generalfm}\textemdash combined with
Prop.~\ref{prop:extend}.
\end{proof}

A preliminary application of the continuity provided by
Lemma~\ref{lem:holdercont} is the proof of
Theorem~\ref{thm:necessary}.

\begin{proof}
Due to Lemma~\ref{lem:holdercont}, we need only verify
\eq{eq:criterion} for the energy $E$ appearing in
\eq{eq:greenloc}.  For this we argue along the lines of the proof
of Lemma~\ref{lem:simon-lieb}, using the geometric resolvent
identity \eq{eq:resolvent_expansion1} to obtain for $L > 23 r$,
\begin{equation}
\chi_\alpha \frac{1}{H^{(B^L_\alpha)} - z} \1_{\delta B^L_\alpha}
\ = \ \chi_\alpha \frac{1}{H - z} \1_{\delta B^L_\alpha}
+\chi_\alpha \frac{1}{H - z} \com{H}{\Theta_L}
\frac{1}{H^{(B^L_\alpha)} - z} \1_{\delta B^L_\alpha}
\end{equation}
where $\Theta_L$ is any smooth function equal to one on
$B^{L-r}_\alpha$ and zero on $\R^d \setminus
B^L_\alpha$\textemdash recall that $\delta B^L_\alpha = \{q \, :
\, r < \dist(q,\partial \Lambda) < 23 r \} \subset
B^{L-r}_\alpha$. From this we conclude, as in the proof of
\eq{eq:tralala}, that
\begin{equation}
\Ev{\norm{\chi_\alpha \frac{1}{H^{(B^L_\alpha)} - z} \1_{\delta
B^L_\alpha}}^s} \ \le \ \const \ L^{2(d-1)} \sum_{\substack{\zeta
\\ \dist(\zeta , \delta B^L_\alpha) \, <\, r}} \Ev{\norm{\chi_\alpha
\frac{1}{H - z} \chi_\zeta}^s} \; .
\end{equation}
By assumption the right hand side is $\mathcal{O} (L^{3(d-1)}
\e^{-\mu L})$ uniformly in $\alpha$.  Thus \eq{eq:criterion} is
satisfied for sufficiently large $L$ and the theorem follows.
\end{proof}

\subsection{The multi-scale analysis regime}
The region in which a multiscale analysis may be carried out
provides another characterization of the localization regime. An
important observation is that the ``output'' of the multiscale
analysis implies the ``input'' for Theorem~\ref{thm:1}, i.e.,
\eq{eq:criterion}. Thus for the operators considered here the
stronger results proved by our methods hold throughout the
multiscale regime.  A precise formulation of this statement is the
following.
\begin{thm} \label{thm:msaappl}
Let $H$ be a random Schr\"odinger operator which satisfies ${\mathcal A}$ and
${\mathcal IAD}$. Suppose that for some $A<\infty$, $\mu>0$, $\xi >2(d-1)$,
$C<\infty$ and $E \in \R$ it holds for $L$ sufficiently large that
\begin{equation} \label{eq:probbad}
\sup_{\alpha \in \I} \Pr \left [ \norm{ \chi_\alpha
\frac{1}{H^{(B_\alpha^L)}-E} \1_{\delta
     B_0^L} } > A{e}^{-\mu L} \right ] \le CL^{-\xi} \;.
\end{equation}
Then, there exist $0<s<1/3$, $A'<\infty$, $\mu'>0$ and an open interval
$\mathcal{J}$ containing $E$ such that
\begin{equation} \label{eq:msacons}
\limsup_{\varepsilon \downarrow 0} \E \left ( \norm{\chi_x
\frac{1}{H^{(\Omega)} -
   E' -\im \varepsilon} \chi_y }^s \right ) \le A' {e}^{-\mu'
   \dist_\Omega(x,y)}
\end{equation}
for all open sets $\Omega \subset \R^d$, $x,y \in \Omega$ and $E'
\in \mathcal{J}$.
\end{thm}

Bounds on the probability of exceptional behavior of the type of
\eq{eq:probbad} are a characterization of the localized regime
mapped by the multi-scale analysis, where $\xi$ can be made
arbitrarily large, e.g.\ \cite{Combes/Hislop:JFA}. Using a
``bootstrap'' approach in which the output of one multi-scale
analysis serves as the input for another, Germinet and Klein have
shown that this probability is $\mathcal{O}(\e^{-L^\alpha})$ with
any $\alpha < 1$ throughout the multi-scale regime
\cite{Germinet/Klein:Bootstrap}. {\it A posteriori}, we conclude
from Theorem~\ref{thm:1} that this bound may be replaced by
$\mathcal{O}(\e^{-\nu L})$ with $\nu > 0$ for the operators
considered here.

For ergodic random Schr\"odinger operators (see ${\mathcal A}4$ in
Section~\ref{sec:applications}), there is a notion of  ``strong
localization region,'' introduced in
\cite{Germinet/Klein:Transition}, analogous to Dobrushin and
Shlosman's regime of complete analyticity in statistical mechanics
\cite{Dobrushin/Shlosman}. The strong localization region may be
characterized by any of a number of criteria, one of which is the
applicability of the multi-scale analysis, and throughout this
region all of those criteria hold (see
\cite[Theorem~4.4]{Germinet/Klein:Transition}). In the words of
the authors of \cite{Germinet/Klein:Transition}, it is a region
``possessing every possible virtue we can imagine!'' For the
random operators considered here, one may add to that list of
virtues exponential localization of Green function moments and
dynamical localization in the (stronger) sense of
Corollary~\ref{cor}.

\begin{proof} Following the argument for the lattice case from Section~4.4
of ref.~\cite{ASFH}, we define complementary ``good'' and ``bad''
subsets of the probability space $\Omega$ by
\begin{equation} \label{eq:goodset}
\Omega_{G} \ := \ \left\{ \omega \,| \; \| \chi_\alpha
   (H^{(B_\alpha^L)}-E)^{-1} \1_{\delta B_\alpha^L} \| \le A
   {e}^{-\mu L} \right\}
\end{equation}
and $\Omega_B := \Omega_G^c$. Then
\begin{multline} \label{eq:goodbad}
\E \left ( \|\chi_\alpha (H^{(B_\alpha^L)}-E)^{-1} \1_{\delta
     B_\alpha^L} \|^s \right ) \\
= \ \E \left ( \|\chi_\alpha (H^{(B_\alpha^L)}-E)^{-1} \1_{\delta
     B_\alpha^L} \|^s I[\omega \in \Omega_G] \right) \\
\  + \E \left ( \|\chi_\alpha (H^{(B_{\alpha}^L)}-E)^{-1}
\1_{\delta
     B_\alpha^L} \|^s I[\omega \in \Omega_B] \right) \;.
\end{multline}
The first term is bounded by ${A}^s {e}^{-s\mu L}$, while we may
apply the H\"{o}lder inequality to bound the second term for any
$s<t<1$ by
\begin{equation} \label{eq:holderbad}
\E \left ( \|\chi_\alpha (H^{(B_\alpha^L)}-E)^{-1} \1_{\delta
     B_\alpha^L} \|^t \right)^{s/t} \, \E \left ( I[\omega \in
     \Omega_B] \right)^{1-s/t} \;.
\end{equation}
We may further estimate this by $C(s,E,d) L^{(s(d-1)-\xi
(t-s))/t}$ using (1) the fractional moment bound \eq{eq:generalfm}
and (2) the assumed bound \eq{eq:probbad} on $\Pr(\Omega_B)$. Here
we have used that $\delta B_0^L$ contains $\mathcal{O}(L^{d-1})$
points in $\I$.

In summary, we get the bound
\begin{equation} \label{eq:Lbound}
\E \left ( \|\chi_\alpha (H^{(B_\alpha^L)}-E)^{-1} \1_{\delta
     B_\alpha^L}\|^s \right ) \le {A}^s {e}^{-s\mu L} +
     C(t,E,\lambda)^{s/t} \,
     L^{(s(d-1)-\xi(t-s))/t} \; ,
\end{equation}
uniformly in $\alpha$. Since $\xi >2(d-1)$, we can choose $s$ sufficiently
close to $0$ and $t$ close to $1$ to guarantee
$(\xi(t-s) -s(d-1))/t>2(d-1)$.
Thus
\begin{equation}
\limsup_{L \rightarrow \infty} \, L^{2(d-1)} \,
     \sup_{\alpha \in \I}
     \Ev{\| \chi_\alpha (H^{(B_\alpha^L)}-E)^{-1}
     \1_{\delta B_\alpha^L} \|^s} \ = \ 0 \; ,
\end{equation}
and we may choose $L$ large enough that \eq{eq:criterion} holds at
$E$.

Working in the finite volumes $B^L_\alpha$, we can use the
continuity given by Lemma~\ref{lem:holdercont} and the continuity
of $b_s(\lambda,E)$ in $E$ to conclude the existence of a complex
neighborhood $\mathcal U$ of $E$ such that \eq{eq:criterion} holds
for every $E'+i\varepsilon \in {\mathcal U}$.  Then
Theorem~\ref{thm:1} applies at all real $E' \in {\mathcal U}$,
which concludes the proof by Theorem~\ref{thm:msaappl}.
\end{proof}

\subsection{Large disorder}
Perhaps the easiest localization regime to understand is that
induced at the bottom of the spectrum by large disorder. If we fix
an energy $E > E_0$ and a length scale $L$ we may adjust $\lambda$
so that for any $x \in \R^d$ it is overwhelmingly likely that $E$
is far below the bottom of the spectrum of the ``local
Hamiltonian,'' i.e., $H^{(B_x^L)}$. Heuristically, this suggests
that $E$ lies in the localization regime, since the resolvent
$(H^{(B_x^L)} - E)^{-1}$ is typically bounded with small norm.

In the discrete setting, the bound $\E (|G(x,x)|^s) \lesssim 1/\lambda^s$
provides the basis for the `single-site' criterion of
\cite{Aizenman/Molchanov}.  For the operators considered here, by taking
$\lambda$ large enough one may directly verify the localization condition
\eq{eq:criterion} at any fixed finite scale $L$ allowed in Theorem~\ref{thm:1}.
In fact, \eq{eq:criterion} may be satisfied uniformly for all energies in an
arbitrary finite interval. For this result, we assume blow-up regularity
${\mathcal A}3'$ to ensure that the constants which appear in \eq{eq:criterion}
remain bounded as $\lambda$ increases.

\begin{thm} \label{thm:largedisorder}
Let $H$ satisfy ${\mathcal IAD}$, ${\mathcal A}1$, ${\mathcal A}2$ and
${\mathcal A}3'$. Then to every $E_1 \in \R$ and $0<s<1/3$ there exists
$\lambda_0 = \lambda_0(E_1,s)$ such that for every $\lambda>\lambda_0$ there
are constants $A<\infty$ and $\mu>0$ such that for any open set $\Omega \subset
\R^d$
\begin{equation} \label{eq:ldcons}
\limsup_{\varepsilon \downarrow 0} \E \left ( \norm{\chi_x
\frac{1}{H^{(\Omega)}
   -E-\im \varepsilon} \chi_y }^s \right ) \le A {e}^{-\mu
   \dist_{\Omega}(x,y)}
\end{equation}
for all $E \in [E_0, E_1]$ and $x,y \in \Omega$.
\end{thm}
\noindent {\bf Remark:}  The proof will show that we can take
$\lambda_0 \propto (1 + |E_1-E_0|)^{2+3d/2}$.
\begin{proof}
We cannot use \eq{eq:finite_smoment} to verify \eq{eq:criterion},
since the r.h.s.\ does not approach $0$ as $\lambda \to \infty$,
even assuming $\mathcal A 3'$. Instead we fix any $L>23r$ and use
the method discussed in Section~3.1 to show that $\E ( \|
\chi_\alpha (H^{(B_\alpha^L)}-E)^{-1} \1_{\delta B_\alpha^L} \|^s
)$ approaches $0$ as $\lambda \to \infty$. This method gives
volume dependent bounds on fractional moments, but as $L$ is fixed
in the present argument that is of no importance.

By the covering condition in ${\mathcal A}2$ it suffices to show
that
\begin{equation} \label{eq:fmto0}
\Ev {\norm{ U_\beta \frac{1}{H^{(B_\alpha^L)}-E} U_\zeta }^s } \to
0 \quad \mbox{as $\lambda \to \infty$}
\end{equation}
for all $\beta,\zeta \in \I$. Convergence needs to be shown
uniformly with respect to $\alpha,\beta,\zeta \in \I$ and $E \in
[E_0, E_1]$. One may then switch to complex energy using
Lemma~\ref{lem:holdercont} and complete the proof with
Theorem~\ref{thm:1}.

To derive \eq{eq:fmto0}, we write $H = H^{(B_\alpha^L)}$,
understanding henceforth that all operators are restricted to
$L^2(B_\alpha^L)$. Let $\widehat{H} = H - \lambda \eta_\beta
U_\beta - \lambda \eta_\zeta U_\zeta$ and define $\eta_{\pm} =
\frac{1}{2}(\eta_\beta \pm \eta_\zeta)$, so that $\eta_\beta
U_\beta + \eta_\zeta U_\zeta = \eta_+ (U_\beta+U_\zeta) + \eta_-
(U_\beta - U_\zeta)$. Assumption ${\mathcal A}2$ and the
Birman-Schwinger argument yield
\begin{equation} \label{eq:BSnondiag}
\begin{aligned}
\|U_\beta \frac{1}{H-E} U_\zeta \| \ \le& \ b_+ \|
(U_\beta+U_\zeta)^{1/2}
\frac{1}{H-E} (U_\beta+U_\zeta)^{1/2} \| \\
=& \ b_+ \left\| [ \widehat{K}_{E,\eta_-}^{-1} + \lambda
\eta_+]^{-1} \right\| \;,
\end{aligned}
\end{equation}
where $\widehat{K}_{E,\eta_-} = (U_\beta+U_\zeta)^{1/2} (\hat{H}+
\lambda \eta_- (U_\beta-U_\zeta) -E)^{-1} (U_\beta+U_\zeta)^{1/2}$
in $L^2((\supp U_\beta \cup \supp U_\zeta) \cap B_\alpha^L)$.

Now $\eta_+ \in [0,\infty)$ and with $\eta_-$ fixed one can use
the argument described in Section~3.1 to show that
\begin{equation} \label{eq:ldweak1-1}
\left| \left\{ \eta_+ \in [0,\infty) \;: \;
\|[\widehat{K}_{E,\eta_-}^{-1} + \lambda \eta_+]^{-1}
     \| > t \right\} \right| \le \frac{(1+\xi_{E,\eta_-})}{\lambda t}
\;,
\end{equation}
where $\xi_{E,\eta_-}$ may be expressed as a spectral shift
function:
\begin{multline} \label{eq:ldshift}
\xi_{E,\eta_-} \ \le \ \tr \left[ P(\widehat{H} + \lambda \eta_-
     (U_x-U_y) < E) \right ] \\
     \ - \tr \left[ P(\widehat{H} +
     \lambda \eta_- (U_x-U_y) + \lambda (U_x+U_y) < E ) \right] \;.
\end{multline}
Since both Schr\"{o}dinger operators appearing in \eq{eq:ldshift}
are positive perturbations of $H_0$, we may bound each trace by a
Weyl-type bound (see \eq{eq:density}) to yield
\begin{equation} \label{eq:Weylbound}
\xi_{E,\eta_-} \ \le \ \tr P(H_0 < E) \ \le \  \const \
(1+|E-E_0|)^{d/2} L^d \;.
\end{equation}

Averaging over $\eta_+,\eta_-$ as in the proof of
Prop.~\ref{prop:offdiag} we obtain the following bound:
\begin{equation}
\Pr \left ( \norm{ U_\beta \frac{1}{H^{(B_\alpha^L)}-E} U_\zeta}>t |
\Fc_{\beta,\zeta} \right ) \ \le \ \const \ D^2 \ L^d \
\frac{(1+|E_1-E_0|)^{d/2} }{\lambda t} \; ,
\end{equation}
for any $\alpha,\beta,\zeta \in \I$ and $E \in [E_0,E_1]$, from
which \eq{eq:fmto0} follows.
\end{proof}

\subsection{Localization via density of states bounds}\label{sec:smalldensity}
To state the bounds in this section, which are based on the notion
of the {\em density of states}, it is necessary to assume that the
random operator $H$ is {\em ergodic}, i.e. its distribution is
invariant under a sufficiently large group of translations. In
particular this implies that the spectrum of $H$ is a non-random
set, c.f. \cite{Pastur/Figotin,Stollmann:caught}. Hence,
throughout we make the additional assumption
\begin{enumerate}
\item[$\mathcal A4$:] $\I$ is a lattice containing the origin,
$U_{\alpha} = U(\cdot -\alpha)$, and $H$ is ergodic with respect
to shifts in $\I$.
\end{enumerate}
We will drop the supremum over $\alpha$ in \eq{eq:criterion} and
work with $\alpha=0$, since if ${\mathcal A}4$ holds
\eq{eq:criterion} is equivalent to
\begin{equation}\label{eq:ergodiccriterion}
     M(s,\lambda,E)
     (1 + L)^{2(d-1)} \limsup_{\eps \downarrow 0}
     \E \left ( \norm{\chi_{0}\,
     \frac{1}{H^{(B_{0}^L)} -E-i\eps} \, \1_{\delta
     B^L_0} }^s
     \right ) \ < \ 1 \; .
\end{equation}

For simplicity, we shall assume that the lattice $\I = \Z^d$ and
work throughout with the `$\ell^\infty$-norm' on $\R^d$: $|x| =
\max_j |x_j|$. Thus balls are cubes with sides parallel to the
co-ordinate axes and unit balls are fundamental cells for the
lattice $\I$.  The reader should have no problem extending the
results below to more general lattices, in which case it is
natural to work with the `$\ell^\infty$-norm' induced on $\R^d$ by
the decomposition of a vector into components parallel to lattice
generators.

The density of states measure for an ergodic random Schr\"odinger
operator is a Borel measure $\kappa$ on the real line defined
through a limiting procedure via its action on compactly supported
continuous functions:
\begin{equation}
\int f(t) \kappa(dt) := \lim_{L\rightarrow \infty}
\frac{1}{(2L)^d} \tr f(H^{(\Lambda_L)})
\end{equation}
where $\Lambda_L = B_0^L = [-L,L]^d$.  For the operators considered here,
$\mathcal A4$ implies that this limit exists almost surely, is non-random
\cite{Pastur/Figotin}, and is equal to
\begin{equation}
\int f(t) \kappa(\di t) \ = \ \lim_{L\rightarrow \infty}
\frac{1}{(2L)^d} \tr \Ev{f(H^{(\Lambda_L)})} \ = \
\frac{1}{|\mathcal{C}|}\Ev{\tr \1_{\mathcal C} f(H)}\; ,
\end{equation}
where $\mathcal{C}$ is any unit cell of the lattice $\I$, e.g.,
$[0,1]^d$. In addition, the random operators studied here satisfy
a Wegner estimate which shows that the density of states measure
is absolutely continuous with a density which is in $L^p_{loc}$
for all finite $p$ \cite{Combes/Hislop/Nakamura}.

We also introduce the finite volume measures
\begin{equation}
\int_A \kappa_L(\di t) \ = \ \Ev{\tr P_A(H^{(\Lambda_L)})} \; .
\end{equation}
Again $\kappa_L$ is absolutely continuous with a density in
$L^p_{loc}$ \cite{Combes/Hislop/Nakamura}. Note that $\lim_L
\kappa_L = \kappa$ in the sense of weak convergence in the dual
space of the compactly supported continuous functions.

If $E$ is a band edge of the almost sure spectrum of $H$, e.g. if
$E = \inf \sigma(H)$, then it is generally expected that
$\kappa[E,E+\Delta E]$ vanishes to very  high order in $\Delta E$.
This phenomenon was first observed by I. M. Lifshitz
\cite{Lifshitz} and the regions where such estimates hold are
called `Lifshitz tails'.

In the Lifshitz tail regime one expects it is very rare to find an
eigenvalue of a finite volume operator near the band edge $E$.
With favorable estimates this suggests localization should hold
near $E$ since the Combes-Thomas estimate
\cite{Combes/Thomas,Barbaroux/Combes/Hislop} may typically be used
to obtain resolvent decay across finite volumes.

One result in this vein shows that smallness bounds for the finite
volume measures $\kappa_L$ may be used to verify the localization
criterion in Theorem~\ref{thm:1} in a similar way as done for
initial length scale estimates in proofs of band edge localization
via multiscale analysis.
\begin{thm} \label{thm:bandedge}
Let $H$ satisfy ${\mathcal IAD}$, ${\mathcal A}1-4$. Suppose that for some
$\beta \in (0,2)$, $\xi >2(d-1)$, $C_1>0$, $C_2>0$, and $E \in \R$ it holds for
sufficiently large $L$ that
\begin{equation} \label{eq:ilse}
\kappa_L\left ( [E - C_1 L^{-\beta}, E + C_1 L^{-\beta}] \right )
\ < \ C_2 L^{-\xi-d} \: .
\end{equation}
Then the conclusion of Theorem~\ref{thm:msaappl} holds.
\end{thm}

\begin{proof}
The proof is very similar to the proof of
Theorem~\ref{thm:msaappl}. This time, define $\Omega_G := \{
\omega: \, \dist(\sigma(H^{(\Lambda_L)}), E) > C_1 L^{-\beta}\}$
and $\Omega_B = \Omega_G^c$.  Then, using \eq{eq:ilse}
\begin{equation}
\begin{aligned}
\Pr(\Omega_B) \ \le& \ \Ev{\tr P_{[E - C_1 L^{-\beta}, E + C_1
L^{-\beta}]}(H^{(\Lambda_L)})} \\ =& \ (2L)^d \kappa_L\left ( [E -
C_1 L^{-\beta}, E + C_1 L^{-\beta}] \right ) \ \le \ 2^d C_2
L^{-\xi}
\end{aligned}
\end{equation}
The improved Combes-Thomas estimate due to
ref.~\cite{Barbaroux/Combes/Hislop} implies that there are
$\eta>0$ and $C<\infty$ such that for $\omega \in \Omega_G$,
\begin{equation} \label{eq:combesthomas}
\norm{\chi_0 \frac{1}{H^{(\Lambda_L)}-E} \1_{\delta \Lambda_L}}
\le CL^{\beta} {e}^{-\eta L^{1-\beta/2}} \: .
\end{equation}
This gives a bound for the analogue of the first term on the right
hand side of \eq{eq:goodbad}. From here on the proof is identical
to the proof of Theorem~\ref{thm:msaappl}.
\end{proof}

There are two well-known situations in which bounds of the form
(\ref{eq:ilse}) can be derived. One is the aforementioned
Lifshitz-tail regime, which generally holds if the background
operator $H_0$ is $\I$-periodic and $E$ is the infimum of the
almost sure spectrum. In this case one gets \eq{eq:ilse} for
arbitrary $\xi>0$, see e.g.\ \cite{Combes/Hislop:JFA, KSS1}.
Conditions on the periodic background operator for the appearance
of this regime at more general band edges were given by F.~Klopp
\cite{Klopp}.

It is not known if the conditions used in ref.~\cite{Klopp} hold for general
periodic $H_0$. But one always has the second option of ``forcing'' a bound
like \eq{eq:ilse} to hold by assuming the distribution density $\rho$ of the
random coupling constants $\eta_{\alpha}$ has small tails near the boundary of
their support. For example, one gets \eq{eq:ilse} with $\xi \in (0, 2\tau -d)$
at a lower band edge $E_0$ of the almost sure spectrum if $\int_0^h
\rho(q)\,\di q = {\mathcal O}(h^{\tau})$ for $h\to 0$, see
\cite[Prop.~4.1]{KSS1}. Thus Theorem~\ref{thm:bandedge} is applicable if $\tau
>(3d-2)/2$.

To complement the above discussion, it is interesting to note that
one may prove localization directly from an estimate on the
infinite volume density of states.
\begin{thm} \label{thm:lowdensity}
Let $H$ satisfy ${\mathcal IAD}$, ${\mathcal A}1-4$. For each
$\xi >  3d-2$ and $E \in \R$, there exists $C = C(E,\xi,\lambda) >0$
and  $\delta_0= \delta(E,\xi,\lambda)$ such that if for some
$\delta < \delta_0 $
\begin{equation}\label{eq:densitycriterion}
\kappa([E-\delta ,E+\delta]) \ \le \ C \, \delta^\xi \; ,
\end{equation}
then there exist $s \in (0,1)$, $A < \infty$, and $\mu > 0$ such
that for any open set $\Omega \subset \R^d$
\begin{equation}
     \limsup_{\eps \downarrow 0} \E \left (
       \norm{\chi_x \frac{1}{H^{(\Omega)} -E' - \im \eps}
     \chi_{y}}^s \right )
     \ \le \ A \, \e^{- \mu \dist_\Omega(x,y)}
     \; ,
\end{equation}
for every $E' \in [E - \delta/4, E+\delta/4]$.
\end{thm}

\begin{proof} Klopp has shown that the densities of states for periodic
approximations to a random Schr\"odinger operator approach the
infinite volume density of states extremely rapidly, c.f. refs.~
\cite{Klopp,Klopp:periodic}.  We shall adapt his argument to use
\eq{eq:densitycriterion} to bound the probability that a certain
finite volume operator with random quasi-periodic boundary
conditions has spectrum in the interval $[E-\delta/2,E+\delta/2]$.
Together with arguments in the proof of Thms.~\ref{thm:msaappl}
and \ref{thm:bandedge} this will imply the theorem via an analogue
of Theorem~\ref{thm:1} in which the smallness criterion is
satisfied for a quasi-periodic finite volume Hamiltonian.  That
result may be proved in exactly the same way as
Theorem~\ref{thm:1} since, as remarked in the introduction,
quasi-periodic boundary conditions preserve all the properties
implied by ${\mathcal A}1$ which were needed in the proof.

We begin by introducing a sequence of periodic approximations to
$H$. For each $\ell \in \N$ greater than $r_0$, define a random
potential $V_\ell^P$ which is periodic under translations in $\ell
\I$:
\begin{equation}
V_\ell^P(q) \ := \ \sum_{\alpha \in \I \cap \Lambda_{\ell/2}} \
\eta_\alpha \sum_{\zeta \in \ell \I} U(q - \alpha -\zeta) \; ,
\end{equation}
and note that for any set  $\Lambda \subset \R^d$ with diameter
less than $\ell-r_0$ the random functions $V_\ell^P(q)
\1_\Lambda(q)$ and $V(q) \1_\Lambda(q)$ are identically
distributed.  We define
\begin{equation}
H_\ell^P  \ := \ H_0 \ + \ V_\ell^P \; .
\end{equation}
Then $H_\ell^P$ is periodic under shifts in $\ell \I$  and its
distribution is invariant under shifts in $\I$.

The averaged density of states measure for $H_\ell^P$, denoted
$\kappa_\ell^P$, is defined to be
\begin{equation}
\kappa_\ell^P(A) \ := \ \frac{1}{\ell^d} \Ev{\tr
\1_{\Lambda_{\ell/2}} P_A(H_\ell^P)} \; .
\end{equation}
By translation invariance of the distribution of $H_\ell^P$, this
is also given by
\begin{equation}
\kappa_\ell^P(A) \ = \  \Ev{\tr \1_{\mathcal{C}} P_A(H_\ell^P)} \;
,
\end{equation}
with $\mathcal{C} = [0,1]^d$.

In the present situation, the arguments of
ref.~\cite{Klopp:periodic} may be adapted to show:

{\it For each $n > 0$ there exists $C_n = C_n(E)$ such that
\begin{equation}\label{eq:Kloppbound}
\kappa_\ell^P([E-\delta/2,E+\delta/2]) \ \le \
\kappa([E-\delta,E+\delta]) \ + \ C_n \, \ell^{d+1-n} \delta^{-n}
\; .
\end{equation} }

Since our assumptions differ somewhat from those of ref.\
\cite{Klopp:periodic}, let us describe the proof of
\eq{eq:Kloppbound}.  In the following we use $C(E)$ to denote a
generic energy and dimension dependent parameter whose value does
not depend on the length scale $\ell$ but may change from line to
line. Choose a $C^\infty$ function $f$ with $\1_{\J_{\delta/2}}
\le f \le \1_{\J_\delta}$ where $\J_t := [E-t,E+t]$.  Then
\begin{equation}
\kappa_\ell^P(\J_{\delta/2}) \ \le \ \Ev{ \tr \1_{\mathcal{C}}
f(H_\ell^P)} \ \le \ \kappa(\J_\delta) +
\Ev{\tr\1_{\mathcal{C}}(f(H_\ell^P) - f(H))\1_{\mathcal{C}}} \; .
\end{equation}
Using the Helffer-Sj\"ostrand formula
\cite{Helffer/Sjostrand,Davies}, write
\begin{multline}\label{eq:HS}
\1_{\mathcal{C}}(f(H_\ell^P) - f(H))\1_{\mathcal{C}} \\ = \
\int_{\J_\delta \times [-1,1]} \left (
\partial_{\bar z} \widetilde f_n(z) \right ) \, \1_{\mathcal{C}}\left (
\frac{1}{z - H_\ell^P} - \frac{1}{z - H} \right )\1_{\mathcal{C}}
\, \di x \di y \; ,
\end{multline}
where $z=x + \im y$ and $\widetilde f_n$ is an ``almost-analytic''
extension of $f$ which vanishes to order $n$ at $y=0$:
\begin{equation}
\widetilde f_n(x+\im y) \ = \ \left [ \sum_{j=0}^n \frac{1}{j!}
f^{(j)}(x) (\im y)^j \right ] \sigma(y) \; .
\end{equation}
Here $\sigma$ is a fixed cut-off function supported in $[-1,1]$
and identically one in a neighborhood of zero.  The key property
here is that $\partial_{\bar z} \widetilde f(x+ \im y) =
\mathcal{O}\left ( |y|^n/\delta^{n+1} \right )$ as may be verified
directly.

The difference of resolvents appearing in \eq{eq:HS} can be
expressed in terms of the geometric resolvent identity
\eq{eq:resolvent_expansion1}:
\begin{equation}
\1_{\mathcal{C}}\left ( \frac{1}{z- H_\ell^P} - \frac{1}{z - H}
\right )\1_{\mathcal{C}} \ = \ \1_{\mathcal{C}}\frac{1}{z-
H_\ell^P} \com{H}{\Theta} \frac{1}{z - H}\1_{\mathcal{C}}
\end{equation}
with $\Theta$ any function which is identically one in a
neighborhood of $\mathcal{C}$ and supported inside
$\Lambda_{\ell/2-r_0}$. We choose $\Theta$ with $\grad \Theta$
supported in a strip of width one near the boundary of
$\Lambda_{\ell/2-r_0}$.

Using a ``commutator argument'' similar to that in the proof of
Lemmas~\ref{lem:finite_smoment} and \ref{lem:decouple}, we may
show that for $z \in \mathcal{J_\delta} \times \left \{
[-1,0)\cup(0, 1] \right \}$,
\begin{equation}
\com{H}{\Theta} \frac{1}{z - H}\1_{\mathcal{C}} \ = \ T \Psi
\frac{1}{z - H}\1_{\mathcal{C}} \; ,
\end{equation}
where $T$ is a trace class operator with
\begin{equation}
\tr |T| \ \le \ C(E) \ \ell^{d-1}
\end{equation}
and $\Psi$ is a function which is identically one on the support
of $\grad \Theta$ and also supported in a strip of width of order
one near the boundary of $\Lambda_{\ell/2-r_0}$.  Therefore
\begin{equation}
\begin{aligned}
\abs{\tr  \1_{\mathcal{C}}\frac{1}{z- H_\ell^P} \com{H}{\Theta}
     \frac{1}{z - H}\1_{\mathcal{C}} } \le& \ C(E) \ell^{d-1} \
     \norm{\1_{\mathcal{C}}\frac{1}{z- H_\ell^P} \Psi} \
     \norm{\Psi \frac{1}{z - H}\1_{\mathcal{C}} } \\
     \le & \ C(E) \ \ell^{d-1} \ \frac{1}{y^2} \e^{-\mu(E)
     y \ell}\; ,
\end{aligned}
\end{equation}
where we have used the Combes-Thomas bound
\cite{Combes/Thomas,Barbaroux/Combes/Hislop} to estimate the
resolvent norms.

Putting this into \eq{eq:HS} we find that
\begin{equation}
\begin{aligned}
\abs{\Tr \1_{\mathcal{C}}(f(H_\ell^P) - f(H))\1_{\mathcal{C}}} \
\le & \ C(E) \int_{\J_\delta \times [-1,1]}
\frac{y^n}{\delta^{n+1}}
\ell^{d-1} \ \frac{1}{y^2} \e^{-\mu(E) y \ell} \, \di x \di y \\
\le & \ C(E,n) \, \ell^{d-n} \delta^{-n} \; ,
\end{aligned}
\end{equation}
which completes the proof of \eq{eq:Kloppbound}.

We now recall some standard facts from the Bloch/Floquet theory of
periodic operators, c.f.  \cite{Reed/Simon:vol4}.  For each
$\mathbf{k} \in BZ$ with $BZ=[0,2 \pi]^d$ define the restriction
of $H_\ell^P$ to $\Lambda_{\ell/2}$ with quasi-periodic boundary
conditions at quasi-momentum $\mathbf{k}/\ell$ to be the
self-adjoint operator $H_{\ell;\mathbf{k}}$ on
$L^2(\Lambda_{\ell/2})$ that agrees with $H$ when applied to
functions compactly supported in the interior of
$\Lambda_{\ell/2}$ and whose domain includes all functions on
$\Lambda_{\ell/2}$ of the form $\e^{\im \mathbf{k} \cdot
x/\ell}\phi(x)$ with $\phi(x)$ smooth and periodic.

It is well known that the periodic density of states
$\kappa_\ell^P$ may be obtained as an average of the densities for
$H_{\ell;\mathbf{k}}$:
\begin{equation}\label{eq:kaverage}
\kappa_\ell^P(\J_{\delta/2}) \ = \ \frac{1}{(2 \pi)^d}\int_{BZ}
\kappa_{\ell;\mathbf{k}}(\J_{\delta/2}) \, \di \mathbf{k} \; ,
\end{equation}
where the quasi-periodic densities $\kappa_{\ell;\mathbf{k}}$ are
defined by
\begin{equation}
\kappa_{\ell;\mathbf{k}}(A) \ := \ \frac{1}{\ell^d}\Ev{\tr
P_A(H_{\ell;\mathbf{k}})} \; .
\end{equation}

Now consider the probability space $\Omega' = \Omega \times BZ$ with associated
measure $\Pr' = \Pr \ \times \ \di \mathbf{k} / (2 \pi)^d$, and let $H_\ell=
H_{\ell;\omega'}$ be a random Schr\"odinger operator with
$\omega'=(\omega,\mathbf{k}) \in \Omega'$ distributed according to $\Pr'$. As
in the proof of Thms.~\ref{thm:msaappl} and \ref{thm:bandedge}, we define
complementary good and bad sets, $\Omega_G := \{\omega' :
\dist(\sigma(H_{\ell;\omega'}), E)>\delta/2 \}$ and $\Omega_B=\Omega' \setminus
\Omega_G$.

Using \eq{eq:kaverage} and \eq{eq:Kloppbound} we see that
\begin{equation}
\begin{aligned}
\Pr' ( \Omega_B ) \ \le \ \E' \left ( \tr
P_{\J_{\delta/2}}(H_{\ell;\omega'}) \right ) \ =& \ \ell^d
\kappa_\ell^P(\J_{\delta/2}) \\ \le& \ \ell^d
\kappa(\mathcal{J_\delta}) \ + \ C_n \ell^{2d-n} \delta^{-n} \; .
\end{aligned}
\end{equation}
On the other hand, the improved Combes-Thomas bound
\cite{Barbaroux/Combes/Hislop} shows that there are $A < \infty$
and $\mu>0$ such that for $\omega' \in \Omega_G$ and $E' \in
\J_{\delta/4}$
\begin{equation}
\norm{ \chi_0 \frac{1}{H_{\ell;\omega'} - E'} \delta \Lambda_\ell}
\ \le \ A \, \delta^{-1} \, \e^{-\mu \delta^{1/2} \ell} \; .
\end{equation}
Arguing as in the proof of Theorem~\ref{thm:msaappl} this implies
\begin{multline}
\E' \left ( \norm{ \chi_0 \frac{1}{H_{\ell;\omega'} - E'} \delta
\Lambda_\ell}^s \right ) \ \le \ A^s \, \delta^{-s}  \, \e^{-s\mu
\delta^{1/2} \ell} \\ + \ C(t,\lambda,E)^{s/t} \ell^{(2+s)(d-1)}
\left ( \ell^d \kappa(\mathcal{J_\delta}) \ + \ C_n \ell^{2d-n}
\delta^{-n} \right )^{1-s/t} \; ,
\end{multline}
for any $s < t< 1$.  Thus
\begin{equation}
\ell^{2(d-1)} \E' \left ( \norm{ \chi_0 \frac{1}{H_{\ell;\omega'}
- E'} \delta \Lambda_\ell}^s \right ) \ \le \ C \, \max
(A_1,A_2,A_3)
\end{equation}
with
\begin{equation}
\begin{aligned}
A_1 \ = & \ \delta^{-s} \ell^{2(d-1)} \, \e^{-s\mu \delta^{1/2} \ell} \; , \\
A_2 \ = & \ \ell^{(2+s)(d-1) + (1-s/t) d}
\kappa(\mathcal{J_\delta})^{1-s/t} \; ,\\
A_3 \ = & \ \ell^{(2+s)(d-1) + (1-s/t)(2d - n)} \delta^{-n(1-s/t)}
\; .
\end{aligned}
\end{equation}

By the analogue of Theorem~\ref{thm:1} discussed above, in which
the Dirichlet operator $H^{(\Lambda_\ell)}$ is replaced by the
random quasi-periodic operator $H_\ell$, we see that there is a
fixed quantity $B=B(s,\lambda,E)$ such that if $\max_j(A_j) < B$
then the conclusion of the present theorem holds, i.e., the
interval $\J_{\delta/4}$ is contained in the localization regime.
Note that the infinite volume operator does not depend on the
quasi-momentum $\mathbf{k}$.  Nonetheless, the finite volume
quasi-periodic operators may be used because locally, i.e., for
functions supported in the interior of the cube $\Lambda_\ell$,
they agree with the infinite volume operator.

To obtain a concrete result, we let $\ell \approx \delta^{-r}$ for
some $r > 1$.  Then $A_1 < B$ and $A_3 < B$ for all sufficiently
small $\delta$ provided we choose $n$ sufficiently large. Thus,
there is $\delta_0
>  0$ such that for $\delta < \delta_0$ the condition $\max_j(A_j) < B$
is equivalent to requiring that
\begin{equation}
  A_2 = \delta^{-\xi(1-s/t)} \kappa(\mathcal{J_\delta})^{1-s/t} \ < \
  B \; ,
\end{equation}
where $\xi = rd + r(d-1)(2+s)/(1-s/t)$.   For
$s=0$, $t=1$, and $r=1$, the expression for $\xi$ reduces to $3d-2$,
and hence any value $\xi >  3d -2$  can be attained with some
permissible selection of  $0<s<t<1$ and $r>1$.
This completes the proof of Theorem~\ref{thm:lowdensity}.
\end{proof}

\newpage

\appendix

\section{Technical comments}
\label{sect:technicomm}

Following are some comments of technical nature, which are
intended to supplement the discussion of the assumptions and
results stated in the introduction.

\noindent (1)  {\em (The operator nature of $H_{\omega}$)}
  Under the stated assumptions,
  $H_{\omega}$ is  essentially self adjoint on $C_0^\infty$
  \cite{Leinfelder/Simader}.
  The random potential $V_{\omega}(q)$ is
  non-negative and uniformly bounded by $\lambda b_+$.
The operator      $H_{\omega}$    is  bounded below, with
$\sigma(H_{\omega})
  \subset [E_0, \infty)$.

  \noindent  (2)   For $1\le p < \infty$ let $\I_p$ denote the Schatten
      class of order $p$, i.e., the ideal of bounded operators $A$ on
      $L^2(\R^d)$ with $\|A\|_p := (\tr |A|^p)^{1/p} < \infty$. Then for
      all $z$ in the resolvent set of $H$ and all
      $f\in L^p(\R^d)$ with $p > d/2$,
      \begin{equation} \label{eq-Ipprop1}
    f(q) (H_{\omega}-z)^{-1} \in \I_p
      \end{equation}
Also, for any $a > - E_0$,
      \begin{equation} \label{eq-Ipprop2}
    f(q) (H_{\omega}+a)^{-r} \in \I_p
      \end{equation}
      if $f\in L^p(\R^d)$ and $r>d/2p$.  Proofs can be found, for example, in
      refs.~\cite{Simon:Trace,Simon:semigroups}.
Under the assumptions  ${\mathcal A}2$ and ${\mathcal A}3$, which
imply bounds on the random potential $V_{\omega}(q)$,  the
Schatten class bounds hold uniformly in $\omega$.

\noindent (3) Weak solutions of $H \varphi = z\varphi$, $z\in \C$,
    have the {\it unique continuation property}, i.e.\ if $\varphi$
    vanishes on a non-empty open set, then $\varphi$ vanishes
    identically.  This follows from ref.~\cite{Kalf} which requires
    only local form-boundedness of $V_0^2$, $\A^2$, and $(\grad
    \wedge \A)^2$ with respect to the Laplacian. There is quite a bit
of literature on unique
    continuation allowing more general $\A$ and $V$; c.f.,
    \cite{Berthier,Simon:semigroups, Jerison/Kenig, Koch/Tataru,
    Wolff:1990, Wolff:1992}.

\noindent  (4)  The above three properties hold also for the
  restrictions of $H$ to any open set $\Omega$ with Dirichlet
  boundary conditions and for the restriction to any cube
  $\Lambda_{x,L} :=x + [-L/2,L/2]^d$ with Neumann or quasi-periodic
  boundary conditions (see Section~\ref{sec:smalldensity}).

  \noindent (5) {\em  (The regularization by $\im 0 $)}
For {\it unbounded} regions $\Omega$ one knows that the operator
norm limit
\begin{equation} \label{eq:epslim}
\chi_x (H^{(\Omega)}_{\omega} - E-\im 0)^{-1} \chi_y := \lim_{\eps
\downarrow 0} \chi_x (H^{(\Omega)}_{\omega} -E-\im \eps)^{-1}
\chi_y
\end{equation}
exists almost surely for almost every $E \in \R$. This follows
from Fubini's theorem and the fact that, for fixed $\omega$,
$\chi_x (H_{\omega}^{(\Omega)} -E-\im 0)^{-1} \chi_y$ exists for
almost every $E$. To prove the latter one writes
\begin{multline}
\chi_x (H_{\omega}^{(\Omega)}-E-\im \eps)^{-1} \chi_y \ = \ \chi_x
(H_{\omega}^{(\Omega)} -E-\im \eps)^{-1}
P_I(H_{\omega}^{(\Omega)}) \chi_y \\ + \chi_x
(H_{\omega}^{(\Omega)} -E-\im \eps)^{-1} P_{\R\setminus I}
(H_{\omega}^{(\Omega)}) \chi_y,
\end{multline}
where $I$ is a bounded interval and we denote the spectral
projection onto a measurable set $M$ for a self-adjoint operator
$H$ by $P_M(H)$. If $E$ is in the interior of $I$, then the second
term in the sum trivially has a limit. For the first term, a
polarization argument shows that it suffices to consider
$G_A(E+\im \eps) := \1_A (H_{\omega}^{(\Omega)}-E-\im\eps)^{-1}
P_I(H_{\omega}^{(\Omega)}) \1_A$ for bounded regions $A$.  The
operator function $z \mapsto G_A(z)$ defined for $z$ in the upper
half plane $\C_+$ is trace class valued (see \eq{eq:density}) and
analytic with non-negative imaginary part.  It follows from a
result of de Branges \cite{deBranges}, also used in
Appendix~\ref{sec:Weakbound}, that the limit exists for almost
every $E$. To conclude one exhausts $\R$ with bounded intervals
$I$.

By \eq{eq:epslim} we can extend the bound \dn{eq:weak1-1_1} of Lemma
\ref{lem:finite_smoment} to $\Pr (\| \chi_x (H^{(\Omega)}-E-\im 0)^{-1}
\chi_y\| >t)$ for almost every $E$. However, it is convenient for us to have
\eq{eq:weak1-1_1} for {\it all} $E$, which is why we prefer to work with
$\eps>0$.

\noindent  (6) {\em  (The removal of  $\im 0 $ for bounded
domains)} If we fix a {\em bounded} region $\Omega$, then any
fixed $E\in \R$ is almost surely not in $\sigma(H^{(\Omega)})$, as
follows from the assumptions on the distribution of the random
couplings stated in Section~1.2 (via analytic perturbation theory
and unique continuation of eigenfunctions). Thus, $\norm{\chi_x
(H^{(\Omega)}-E)^{-1}\chi_y}$ is an almost surely finite random
variable, and Fatou's lemma shows that the bound \eq{eq:weak1-1_1}
on $\Pr (\| \chi_x (H^{(\Omega)}-E-\im 0)^{-1} \chi_y\| >t)$
implies also such a bound for $\eps=0$ .

\noindent (7) {\em (Results for large disorder regime)\/ } Readers
familiar with fractional-moment methods for discrete random
operators will likely note the bound we obtain in
Lemma~\ref{lem:finite_smoment} is a bit weaker than the {\em
a-priori} bound derived in that context, which falls off like
$\lambda^{-1}$ for large $\lambda$. We showed\textemdash in the
proof of Theorem~\ref{thm:largedisorder}\textemdash that the
$s$-moments of the resolvent of $H_\omega^{(\Omega)}$ for a
bounded region $\Omega$ are $\mathcal{O}(|\Omega|^s/\lambda^{s})$.
Coupled with Theorem~\ref{thm:1} this allowed us to conclude
localization at ``large disorder.'' However, we do not show here
that the fractional-moments of the infinite volume resolvent tend
to zero for large $\lambda$ (although this may still be true).

\noindent (8) {\em (Possible coexistence of bulk localization with
extended boundary states)\/ }  An operator may exhibit
localization in the bulk (in terms of transition amplitudes) along
with extended {\em boundary states} occurring in certain
geometries.  Such situations have been studied and are of
particular interest for the Quantum Hall Effect, with $H_{o}$ the
Landau operator \cite{DBP,FGW}. Our use of the  domain-adapted
metric, $\dist_\Omega$\textemdash in which exponential decay is
compatible with the above picture\textemdash allows the analysis
of localization to proceed even in such cases. However, it is also
possible to formulate other finite volume criteria which rule out
extended surface states. The input conditions need to be more
restrictive and involve propagators between boundary regions in
arbitrary geometries. For this purpose one may present a modified
version of Theorem~\ref{cor}, changed in  a manner similar to what
was done for discrete models in ref.~\cite[Theorem 1.1]{ASFH}. A
key point is that the domain-adapted metric can be replaced in
\eq{eq:corassump}  by the usual distance, in which case the
conclusions\textemdash \eq{eq:corresult} and pure point
spectrum\textemdash hold in any sufficiently regular region and in
particular, under the stronger assumptions, rule out also extended
boundary states.

\noindent (9) {\em (Energy dependence of the bounds)\/} We note
that Theorem~\ref{cor} only requires exponential
   decay of the energy-averaged Green function. However, typically this will
   be established, for example by Theorem~\ref{thm:1}, through a bound
   which is uniform in energy.

  \noindent (10) {\em (Other norms in \eqref{eq:corresult})\/}
We have used operator norms in stating \eqref{eq:corresult} and
its various consequences, but they extend to arbitrary Schatten
norms. This follows from the fact that the operators involved are
``super-trace class'' in the sense that for every $p>0$ we have
$\tr \abs{\chi_x g(H^{(\Omega)}) P_{\J}(H^{(\Omega)}) \chi_y}^p
\le C_p < \infty$ independent of $x,y, \Omega, g$ and the disorder
(this follows from remark (2) above). Considering, for example,
the trace norm $\|\cdot\|_1$ we find, picking $p<1$,
\begin{multline}
   \Ev{\sup_g \norm{\chi_x g(H^{(\Omega)}) P_{\J}(H^{(\Omega)}) \chi_y}_1}
   \le C_p \Ev{\sup_g \norm{\chi_x
   g(H^{(\Omega)}) P_{\J}(H^{(\Omega)}) \chi_y}^{1-p}} \\
   \le \ C_p \left [ \Ev{ \sup_g \norm{\chi_x
   g(H^{(\Omega)}) P_{\J}(H^{(\Omega)}) \chi_y}}\right ]^{1-p} \ \le \
   \widetilde A \e^{-\widetilde \mu \dist_{\Omega}(x,y)} \; ,
\end{multline}
for appropriate constants $\widetilde A < \infty$ and $\widetilde
\mu > 0$.

\noindent (11) {\em (The bound on Fermi projections)\/} In fact,
the hypothesis of Thm.\ \ref{cor}\textemdash namely
\eq{eq:corassump}\textemdash implies a result somewhat stronger
than \eq{eq:corresult}, namely
\begin{equation} \label{eq:corresultextended}
\Ev {\sup_{g} \|\chi_x g(H^{(\Omega)})
   \chi_y \|} \ \le \widetilde A {e}^{-\widetilde \mu \dist_{\Omega}(x,y)}
   \; ,
\end{equation}
where the supremum is over all Borel measurable functions $g$, with $|g| \le 1$
pointwise and constant on $\J_< = \set{E : E \le \inf \J}$ and $\J_> =\set{E :
E \ge \sup \J}$. The collection of sets consisting of $\J_{<,>}$ and the Borel
subsets of $\J$ is a sigma algebra $\Sigma$, and the supremum is taken over the
unit ball $B_1(\Sigma)$ of the bounded $\Sigma$ measurable functions. A special
case, corresponding to $g(H) = P_{(-\infty, E_F)}(H)$,  is the bound
(\ref{eq:fermi}) on Fermi projections, which may also be derived using a
contour integral representation of the projection operator as in
\cite{Aizenman/Graf}.

To verify \eq{eq:corresultextended}, we fix a $C^\infty$ function
$h$,  $0 \le h \le 1$, supported in $[E_0-2, E_+]$, with $E_+
=\sup \J$, and identically equal to $1$ on $[E_0-1, \inf \J]$.
Note that, for $g \in B_1(\Sigma)$, $g(H^{(\Omega)})$ may be
decomposed as $\alpha \1 + \beta h(H^{(\Omega)}) + \widetilde g
(H_\omega)$ with $\widetilde g$ supported in $\J$. The
contribution from $\widetilde g$ may be estimated by
\eq{eq:corresult}, and the contribution from $\alpha \1$ is
bounded and zero for $|x-y| \ge 2r$. To estimate the contribution
from $h(H^{(\Omega)})$ we write it, using the Helffer-Sj\"ostrand
fromula \cite{Helffer/Sjostrand}, as
\begin{equation}
   h(H^{(\Omega)}) \ = \ \frac{1}{2 \pi} \int_{\C}
   F(z)  \frac{1}{H^{(\Omega)} - z} \di x \di y \; ,
\end{equation}
where $z=x+\im y$ and the bounded function $F(x+\im y)$ satisfies
$F(x + \im y) = \mathcal{O}(y^n)$ for some $n \ge 1$ ($n=1$ will
do) and is supported in the union of the sets $\J +\im [-1,1]$,
$[E_0-2,E_+] +\im [-1,-\frac{1}{2}]$, $[E_0-2,E_+] +\im
[\frac{1}{2},1]$ and $[E_0 -2, E_0-1] +\im [-1,1]$. ($F(x+\im y)=
\partial_{\bar z} \widetilde
   h(z)$ with $\widetilde h(z)$ an almost
analytic extension of $h$; see eq.\ \ref{eq:HS}.) Using that
$\norm{(H-z)^{-1}} \le 1/\Im z$, we get
\begin{equation}
     \Ev{\norm{\chi_x h(H^{(\Omega)}) \chi_y}} \ \le \
     \frac{1}{2 \pi} \int_{\C} \frac{\abs{F(z)}}{|\Im z|^{1-s}}
     \Ev{\norm{\chi_x \frac{1}{H^{(\Omega)}-z} \chi_y}^s} \; .
\end{equation}
Since the support of $F$ approaches the spectrum of $H$ only in
the interval $\J$, the Combes-Thomas estimate and the fractional
moment bound \eq{eq:corassump} (which holds with $\Omega$ in place
of $\Lambda_n$ by strong resolvent convergence) together show that
the integrand here is exponentially small in $\dist_\Omega(x,y)$.
This in turn gives \eqref{eq:corresultextended}.

The above argument combined with the previous remark (10) shows that the
estimate \eqref{eq:corresultextended} also holds in trace norm under the
restriction that $|x-y| \ge 2r$ or if $g$ is required to vanish on $\mathcal
J_>$.

\noindent (12) {\em (The assumption ${\mathcal A}3'$)\/}
   The condition of blow-up regularity of a random variable $X$ requires
  its  probability density to be absolutely continuous, with a bounded density
  $\rho(\cdot)$,  since \eq{eq:decompose} implies (for arbitrary $n$)
   \begin{equation}
     \rho(x) \ = \ n \, \int
     \rho_n(nx-y|y) \mu_n(\di y)  \; ,
   \end{equation}
   where $\mu_n$ is the probability distribution of $Y^{(n)}$.
   The converse is not true: there exist bounded densities such that
   the associated probability measure has infinite blow up norm.
However, if $\ln \rho$ is Lipschitz-continuous
   then $D_\rho < \infty$.  In this case,
   a particularly simple decomposition of the random variable $X$
   with distribution $\rho(x) \di x$ is obtained with $X^{(n)}$
   the fractional part of $n X$, so that $Y^{(n)}$ is integer valued and
   \begin{equation}
     \rho_n(x| j) \ := \ \frac{ \rho( j/n + x/n) }{ n \int_{j/n}^{(j+1)/n}
     \rho(x) \di x} \; .
   \end{equation}
   The Lipschitz condition guarantees the uniform boundedness of
   $\rho_n$ defined in this way.    A particularly simple example,
   for which the results are already of interest, is provided by the
   uniform distribution in $[0,1]$.

  \newpage

  \section{The Birman-Schwinger relation}\label{sec:tools}

Central to our analysis is the consideration of one-parameter
operator families of the form
\begin{equation} \label{eq:onepar}
A_{\xi} = A_0 - \xi V
\end{equation}
with a non-negative operator $V$ and $\xi \in \R$. Such families
arise here when all but one of the random couplings in the random
operator (\ref{eq:RSO}) are considered fixed. Their study replaces
the rank one perturbation arguments which have played a key role
in the analysis of discrete operators, e.g. in
\cite{Simon/Wolff,Aizenman/Molchanov}.

The family (\ref{eq:onepar}) formally satisfies the {\em
Birman-Schwinger relation}:
\begin{equation} \label{eq:BirSchrel}
V^{1/2} A_{\xi}^{-1} V^{1/2} = \left( (V^{1/2} A_0^{-1}
V^{1/2})^{-1} - \xi \1 \right)^{-1}
\end{equation}
to be interpreted as equality of  operators in
$(\mbox{ker}\,V)^{\perp}$, subject to issues of invertibility.  In
the ``classical'' version of this relation \cite{Birman,
Schwinger}, $A_0 = -\Delta+\gamma \1 $ for some $\gamma
>0$,  and $V$ is a non-negative relatively compact potential
perturbation. It is shown, e.g. in \cite[Sect.8]{Simon:FunctInt},
that the number of eigenvalues of $-\Delta -V$ less than $-\gamma
<0$ equals the number of eigenvalues of $V^{1/2}(-\Delta +\gamma
\1 )^{-1} V^{1/2}$ greater than one.  This can be understood
through the observation that the two sides of (\ref{eq:BirSchrel})
become singular for the same values of the parameters $\gamma$ and
$\xi$.

We use here two non-classical, though certainly not new, versions
of the Birman-Schwinger relation which are described below. In our
applications they arise through the two standard procedures for
regularizing the Green function at energies in the infinite volume
spectrum: (1) adding a small imaginary part to the energy or (2)
using finite volume approximations.

The case of complex energy is covered by the following.
\begin{lem}  \label{lem2.1}
Let $A_0 = B+iC$ be an operator on a separable Hilbert space
$\Hi$, where $B$ is self-adjoint and $C$ is bounded with $C\ge
\delta \1 $ for some $\delta>0$. Also, let $V$ be a bounded
non-negative operator in $\Hi$. Then the Birman-Schwinger operator
\begin{equation} \label{eq:BirSchop}
A_{BS} := (V^{1/2} A_0^{-1} V^{1/2})^{-1}
\end{equation}

\noindent is maximally dissipative in $(\mbox{ker}\,V)^{\perp}$,
with ${\mathcal D}(A_{BS}) = {\mathcal
R}(V^{1/2}A_0^{-1}V^{1/2})$, the range of $V^{1/2}
A_0^{-1}V^{1/2}$. Moreover, its resolvent set $\rho(A_{BS})$
includes the closed lower half plane $\overline{\C}_- $, and the
Birman-Schwinger relation
\begin{equation} \label{eq:BirSchrel2}
(A_{BS}-\xi \1)^{-1} = V^{1/2} (A_0-\xi V)^{-1} V^{1/2}
\end{equation}

\noindent holds in $(\mbox{ker}\,V)^{\perp}$ for every $\xi \in
\overline{\C}_-$.
\end{lem}

\noindent {\bf Remark:} In our applications $V$ appears as a
non-negative potential and $A_0=z-H$ with $\Im \,z >0$ and $H$ a
self-adjoint Schr\"odinger operator on $L^2(\Omega)$. In this case
$(\mbox{ker}\,V)^{\perp} = L^2(\{x:V(x)>0\})$.

\begin{proof}
Note that $A_0$ is boundedly invertible with $\|A_0^{-1}\|\le
1/\delta$. To show that the restriction of $V^{1/2} A_0^{-1}
V^{1/2}$ to $(\mbox{ker}\,V)^{\perp}$ is invertible, suppose we
are given $f\in (\mbox{ker}\,V)^{\perp}$ such that $V^{1/2}
A_0^{-1} V^{1/2}f=0$. Then $g:= A_0^{-1} V^{1/2}f \in {\mathcal
D}(A_0)$ with $V^{1/2}g=0$. Thus $0= \langle V^{1/2}g,f\rangle =
\langle g, V^{1/2}f \rangle = \langle g, A_0 g\rangle$. Taking
imaginary parts and using $C\ge \delta \1 $ we find that $g=0$.
Thus $V^{1/2}f=A_0 g=0$ and therefore $f\in (\mbox{ker}\,V) \cap
(\mbox{ker}\,V)^{\perp} = \{0\}$.

We conclude that $A_{BS}$ exists as an operator in
$(\mbox{ker}\,V)^{\perp}$. It is densely defined, since if
$\langle f, V^{1/2} A_0^{-1} V^{1/2} g \rangle =0$ for all $g\in
(\mbox{ker}\,V)^{\perp}$, then $V^{1/2}(A_0^{-1})^* V^{1/2}f=0$
and, by the same argument as above, $f=0$.

For $\xi \in \overline{\C}_-$ one verifies explicitly that $$
(A_{BS}-\xi \1) V^{1/2} A_0^{-1} V^{1/2} = \1 - \xi V^{1/2}
A_0^{-1} V^{1/2}$$ is the inverse of $\1+ \xi V^{1/2} (A_0-\xi
V)^{-1} V^{1/2}$ in $(\mbox{ker}\,V)^{\perp}$. In particular, for
$\xi = -\im$, this shows that $(A_{BS}+\im \1) {\mathcal
D}(A_{BS}) = (\mbox{ker}\,V)^{\perp}$, proving that $A_{BS}$ is
maximally dissipative (see \cite{Sz-Nagy/Foias}). Using the
resolvent identity we also get (\ref{eq:BirSchrel2}).
\end{proof}

The argument in the above proof which showed that
$V^{1/2}A_0^{-1}V^{1/2}$ is invertible does not generally carry
over to the case where $A_0$ is an invertible self-adjoint
operator. If $A_0$ has fixed sign (as in the classical
BS-principle) it does, but not if $0$ is in a spectral gap for
$A$. However, in the case which is of interest to us, namely that
$A_0 = H-E$ with $H$ a finite volume Schr\"odinger operator, we
can make use of the fact that $H$ is {\em local} in the sense that
$H\varphi$ vanishes on an open set $O$ if $\varphi \in {\mathcal
D}(H)$ vanishes on $O$.

Thus, let $\Lambda \subset \R^d$ be open and bounded and let $H_0$
be the Dirichlet restriction of a Schr\"odinger operator $(i\nabla
-A)^2 + V_0$ onto $L^2(\Lambda)$, where $A$ and $V_0$ satisfy the
general assumptions of Section~1. Let $V\ge 0$ be a bounded
non-zero potential of compact support with $|\partial
(\mbox{supp}\,V)|=0$, and for $\xi \in \R$ let
\begin{equation} \label{eq:oneparSch}
H_{\xi} = H_0 - \xi V.
\end{equation}
Then $H_{\xi}$ is self-adjoint with compact resolvent. Thus, by
Theorem VII.3.9 of \cite{Kato}, its repeated eigenvalues
$E_n(\xi)$, $n\in \N$, and corresponding complete set of
orthonormal eigenfunctions $\psi_n(\xi)$ can be labeled such that
$E_n(\cdot)$ and $\psi_n(\cdot)$ are holomorphic (note that
crossings are possible, that is, the $E_n$ may be degenerate and
are not necessarily in increasing order). By the Feynman-Hellmann
Theorem and the unique continuation property of eigenfunctions
(e.g., remark (3) in Appendix A),
\begin{equation} \label{eq:FeynHell}
E_n'(\xi) = - \langle V\psi_n(\xi), \psi_n(\xi) \rangle < 0.
\end{equation}
Thus $\Gamma_n := E_n^{-1}$ exists on the range of $E_n$ and
\begin{equation} \label{eq:FHconseq}
\Gamma_n'(E) = - \frac{1}{\langle V \psi_n(\Gamma_n(E)),
\psi_n(\Gamma_n(E)) \rangle} .
\end{equation}

For real $E\not\in \sigma(H_{\xi})$, we define
\begin{equation} \label{eq:BirSchop2}
K_{\xi,E} = V^{1/2} (H_{\xi}-E)^{-1} V^{1/2}
\end{equation}
as an operator in $L^2(\mbox{supp}\,V)$, where it is compact and
self-adjoint. We claim that $\mathrm{ker} \, K_{\xi,E} = \{0\}$.
Indeed, if $V^{1/2} (H_{\xi}-E)^{-1} V^{1/2} f=0$ for $f\in
L^2(\mbox{supp}\,V)$, then $(H_{\xi}-E)^{-1} V^{1/2}f=0$ on
supp$\,V$ and, since $H_{\xi}$ is local, $V^{1/2}f=0$ in the
interior of supp$\,V$. This implies that $f=0$ since $|\partial
(\mbox{supp}\,V)|=0$.

We conclude that $K_{\xi,E}^{-1}$ exists in $L^2(\mbox{supp}\,V)$
as an unbounded self-adjoint operator with discrete spectrum. By
the arguments used in the proof of Lemma~\ref{lem2.1} it may be
shown that for $E\not\in \sigma(H_{\xi}) \cup \sigma(H_0)$,
\begin{equation} \label{eq:BirSchrel3}
K_{\xi,E} = (K_{0,E}^{-1} -\xi \1)^{-1}.
\end{equation}

\begin{lem} \label{lem2.2}
Let $E\not\in \sigma(H_0)$ and $\xi \not=0$. Then
\begin{enumerate}
\item $\phi$ is a normalized eigenfunction of $K_{0,E}^{-1}$ with
eigenvalue $\xi$ if and only if $\psi := \xi (H_0-E)^{-1} V^{1/2}
\phi$ is an eigenfunction of $H_{\xi}$ with eigenvalue $E$ and
$\langle \psi, V \psi \rangle =1$. \item the repeated eigenvalues
of $K_{0,E}^{-1}$ are given by $\Gamma_n(E)$, $n\in \N$, with
(non-normalized) complete eigenvectors $V^{1/2} \psi_n(\Gamma_n(E))$.
\end{enumerate}
\end{lem}

\begin{proof}
The second claim follows from the first and the definition of
$\Gamma_n(E)$.

To prove the first claim, first suppose that that $K_{0,E}^{-1}
\phi = \xi \phi$ and let $\psi := \xi (H_0-E)^{-1} V^{1/2} \phi$.
Then $V^{1/2} \psi = \xi K_{0,E}\phi = \phi$ and $(H_{\xi}-E)\psi
= \xi (H_{\xi}-E)(H_0-E)^{-1} V^{1/2} \phi = \xi V^{1/2}\phi - \xi
V\psi =0$.

Conversely, if $H_{\xi}\psi = E\psi$ with $\langle V\psi, \psi
\rangle =1$, then $\psi = \xi (H_0-E)^{-1} V\psi$ and thus $\phi:=
V^{1/2}\psi \in {\mathcal R}(V^{1/2} (H_0-E)^{-1} V^{1/2}) =
{\mathcal D}(K_{0,E}^{-1})$ and is normalized. Moreover,
$K_{0,E}^{-1} \phi = K_{0,E}^{-1} V^{1/2}\psi = \xi K_{0,E}^{-1}
K_{0,E} V^{1/2} \psi = \xi \phi$.
\end{proof}

\newpage

  \section{A ``weak $L^{1}$'' bound for resolvents of dissipative operators}
  \label{sec:Weakbound}

  The goal of this appendix is to provide a proof of
  Lemma~\ref{lem:HSweakbound}. All the main arguments are taken from
  \cite{Naboko}.

  We will use here that a maximally dissipative operator
  $A$ in $\Hi$ has a selfadjoint dilation $L$ in a Hilbert space
  $\tilde{\Hi}$ which contains $\Hi$ as a subspace, i.e.\
  \begin{equation} \label{eq:B0}
  (A-\xi)^{-1} = P (L-\xi)^{-1} P^*
  \end{equation}
  for every $\xi\in \C$ with Im$\,\xi < 0$. Here $P$ is the
  orthogonal projection onto $\Hi$ in $\tilde{\Hi}$. For this and
  much more on the general theory of dissipative operators see the
  survey \cite{Pavlov} or the book \cite{Sz-Nagy/Foias} (where the
  equivalent theory of contractions and their unitary dilations is presented).

  We start the proof of Lemma~\ref{lem:HSweakbound} with two reduction
steps. First, we show that
  it is sufficient to deal with the case $M_1 = M = M_2^*$.

  Thus, let us assume that it is proven that
  \begin{equation} \label{eq:B2}
  |\{ v\in \R: \, \|M^*(A-v+i0)^{-1} M\|_{HS} > t\}| \le \frac{C}{t}
\|M\|_{HS}^2
  \end{equation}
  for all self-adjoint $A$ in $\Hi$ and Hilbert-Schmidt operators $M:
\Hi_1 \to \Hi$.

  The estimate (\ref{eq:weakbound}) follows from (\ref{eq:B2}) by a
  polarization and a scaling argument. For this let us temporarily
  write $T=(A-v+i0)^{-1}$. One checks that
  \begin{eqnarray} \label{eq:B3}
  M_2 T M_1 & = & \frac{1}{2} (M_2+M_1^*) T (M_2^*+M_1) - \frac{i}{2}
(M_2-iM_1^*) T
  (M_2^* + iM_1) \nonumber \\
  & & \mbox{} - \frac{1-i}{2} M_2 T M_2^* - \frac{1-i}{2} M_1^* T M_1.
  \end{eqnarray}
  All four terms on the r.h.s.\ of (\ref{eq:B3}) are of the type which
is covered by
  (\ref{eq:B2}). The set $\{v: \|M_2 T M_1\|_{HS} > t\}$ is contained
in the union of
  $\{ v: \|\frac{1}{2}(M_2+M_1^*) T (M_2^* + M_1) \|_{HS} > t/4\}$ and
three similar
  sets. Applying (\ref{eq:B2}) to all of them gives
  \begin{equation} \label{eq:B4}
  |\{ v: \|M_2 T M_1\|_{HS} >t\}| \le \frac{C_1}{t} (\|M_2\|_{HS}^2 +
\|M_1\|_{HS}^2)
  \end{equation}
  with a suitable constant $C_1$. Scaling of (\ref{eq:B4}) yields
  \begin{eqnarray} \label{eq:B5}
  |\{v: \, \|M_2 T M_1\|_{HS} > t\}| & = & \left|\left\{v: \, \left\|
  \frac{M_2}{\|M_2\|_{HS}} T \frac{M_1}{\|M_1\|_{HS}} \right\|_{HS} >
  \frac{t}{\|M_1\|_{HS} \|M_2\|_{HS}} \right\} \right| \nonumber \\
  & \le & \frac{2C_1}{t} \|M_1\|_{HS} \|M_2\|_{HS},
  \end{eqnarray}
  and thus (\ref{eq:weakbound}).

  It remains to show (\ref{eq:B2}), which follows if we can show that
  \begin{equation} \label{eq:B6}
  |\{ v\in \R: \, \|M^*(A-v+i\delta)^{-1} M\|_{HS} > t\}| \le
\frac{C}{t}\|M\|_{HS}^2
  \end{equation}
  for all $\delta > 0$, with $C<\infty$ independent of $\delta$.

  To show that (\ref{eq:B6}) implies (\ref{eq:B2}), consider the
  function $\Phi$ defined on $\C^- = \{\mbox{Im}\,\xi<0\}$ by
  $\Phi(\xi) = M^* (A-\xi)^{-1} M$. The function $\Phi$ is analytic
  and takes values in the trace class operators on $\Hi$ with
  non-negative imaginary part. By a result of de Branges
  \cite{deBranges}, later proven independently in \cite{Asano} and
  \cite{Birman/Entina}, $\Phi(v-i0) := \lim_{\delta \downarrow 0}
  \Phi(v-i\delta)$ exists in Hilbert-Schmidt norm for almost every
  $v\in \R$. Together with (\ref{eq:B3}) this implies the existence
  statement in part (1) of Lemma~\ref{lem:HSweakbound}. For $\delta
  \ge 0$ let $g_{\delta}$ denote the characteristic function of the
  set $\{v\in\R: \, \|M^*(A-v+i\delta)^{-1} M \|_{HS} > t\}$. One
  checks that $g_0(v) \le \lim\inf_{\delta\downarrow 0}
  g_{\delta}(v)$ for almost every $v$. Therefore (\ref{eq:B2}) is a
  consequence of (\ref{eq:B6}) and Fatou's lemma.

  \vspace{.1in}

  Before we proceed with the remaining proof of (\ref{eq:B6}), we
state two classical facts
  on Hilbert transforms which will be used.

  The Hilbert transform of a function $f:\R \to \C$ is defined by the
principle-value integral

  \begin{equation} \label{eq:B7}
  Hf(x) = \frac{1}{\pi} \lim_{\varepsilon \downarrow 0} \int_{\R
\setminus [x-\varepsilon,
  x+\varepsilon]} \frac{f(y)}{x-y} \,dy,
  \end{equation}
  whenever this limit exists. The same definition applies when $f$
takes values in a
  Hilbert space, in which case the r.h.s.\ of (\ref{eq:B7}) is
interpreted as a Bochner
  integral.

  \begin{prop} \label{prop:B2}
  Suppose that $\Phi \in H^2(\C^-)$, i.e.\ $\Phi: \C^- \to \C$ is analytic and
  \begin{equation} \label{eq:B8}
  \sup_{y>0} \int_{\R} |\Phi(x-iy)|^2\,dx < \infty.
  \end{equation}
  Then the boundary value $\Phi(x) = \lim_{y \downarrow 0}
  \Phi(x-iy)$ exists for almost every $x\in \R$, $\Phi \in L^2(\R)$,
  and its real and imaginary parts are conjugate, i.e.\
  \begin{equation} \label{eq:B9}
  \mathrm{Re}\,\Phi(x) = H(\mathrm{Im}\,\Phi)(x) \quad \mbox{for
  a.e.\ $x\in \R$}.
  \end{equation}
  \end{prop}

  \begin{prop} \label{prop:B3}
  Let $\Hi$ be a separable Hilbert space and $f\in L^1(\R,\Hi)$ in the
sense of Bochner
  integration. Then the Hilbert transform $Hf(y) \in \Hi$ exists for
a.e.\ $y\in \R$,
  and there exists a constant $C<\infty$, independent of $f$, such
that for all $t>0$
  \begin{equation} \label{eq:B10}
  | \{ y\in \R:\: \|Hf(y)\|_{\Hi} > t \} | \le \frac{C}{t} \int_{\R}
\|f(x)\|_{\Hi} \,dx.
  \end{equation}
  \end{prop}

  A modern proof of Proposition~\ref{prop:B2} can be found in
  \cite{Garnett}. Proposition~\ref{prop:B3}, i.e.\ the
  weak-$L^1$-property of the Hilbert transform, is well known for
  the case $\Hi = \C$. A proof in the context of more general
  Calderon-Zygmund inequalities can be found in \cite[Chapter
  2]{Stein}, where it is remarked that the result extends to the
  $\Hi$-valued case. Detailed proofs of such results for
  vector-valued functions, which contain Proposition~\ref{prop:B3}
  as a special case, can be found in \cite{Schwartz}.

  We now apply these facts to $T_{\delta}(v) :=
  M^*(A-v+i\delta)^{-1}M$. The trace class norm will be denoted
  $\|\cdot\|_1$. The real and imaginary parts of bounded operators
  are defined as usual by Re$\,C = \frac{1}{2}(C+C^*)$ and Im$\,C =
  \frac{1}{2i}(C-C^*)$.

  \begin{lem} \label{lem:B4}
  For every $\delta>0$ it holds that
  \begin{equation} \label{eq:B11}
  \int_{\R} \|\mathrm{Im}\,T_{\delta}(v)\|_{HS}\,dv \le \int_{\R}
  \|\mathrm{Im} T_{\delta}(v)\|_1\,dv = \pi \|M\|_{HS}^2.
  \end{equation}
  \end{lem}

  \begin{proof}
  The first part of (\ref{eq:B11}) follows from $\|\cdot\|_{HS} \le
  \|\cdot\|_1$. Let $E(t)$ be the spectral resolution of the
  selfadjoint dilation $L$ of $A$ and $\phi \in \Hi_1$. Then by
  (\ref{eq:B0}), the spectral theorem, and Fubini
  \begin{eqnarray} \label{eq:B12}
  \int \langle \mbox{Im}\,T_{\delta}(v) \phi,\phi \rangle_{\Hi_1}
  \,dv & = & \int \int \frac{\delta}{(x-v)^2+\delta^2}
  \,d\|E(x)P^*M\phi\|_{\tilde{\Hi}}^2 \,dv \nonumber \\ & = & \pi
  \|M\phi\|_{\Hi_1}^2.
  \end{eqnarray}
  Let $(\phi_n)$ be an orthonormal basis in $\Hi_1$. We have
Im$\,T_{\delta}(v) \ge 0$
  and thus by (\ref{eq:B12})
  \begin{eqnarray} \label{eq:B13}
  \int \| \mbox{Im}\,T_{\delta}(v)\|_1\,dv & = & \int \tr (\mbox{Im}\,
T_{\delta}(v))\,
  dv \nonumber \\
  & = & \int \sum_n \langle \mbox{Im}\,T_{\delta}(v) \phi_n, \phi_n
\rangle_{\Hi_1} \,dv
  \nonumber \\
  & = & \pi \|M\|_{HS}^2.
  \end{eqnarray}
  \end{proof}

  \begin{lem} \label{lem:B5}
  Let $\Hi_{HS}$ denote the separable Hilbert space of all
  Hilbert-Schmidt operators on $\Hi_1$.

  (a) If $\phi \in \Hi_1$, then $\langle T_{\delta}(\cdot) \phi, \phi
\rangle \in H^2(\C^-)$ for each fixed $\delta >0$.

  (b) For almost every $v\in \R$ one has
  \begin{equation} \label{eq:B14}
  \mathrm{Re}\,T_{\delta}(v) = H(\mathrm{Im}\,T_{\delta})(v)
  \end{equation}
  in the sense of Hilbert transforms of $\Hi_{HS}$-valued functions.
  \end{lem}

  \begin{proof}
  Consider arbitrary $\phi$ and $\psi$ in $\Hi_1$ and use (\ref{eq:B0}) and the
  spectral theorem to estimate
  \begin{eqnarray} \label{eq:B15}
  \int |\langle T_{\delta}(v-iy)\phi, \psi\rangle|^2\, dv & \le &
  \int \|(L-v-i(y+\delta))^{-1} P^* M\phi\|^2 \,dv \, \|M\psi\|^2
  \nonumber \\ & = & \int \int \frac{1}{(x-v)^2+(y+\delta)^2} \,dv
  \, d\|E(x)P^*M\phi\|^2 \|M\psi\|^2 \nonumber \\ & = &
  \frac{\pi}{y+\delta} \|M\phi\|^2 \|M\psi\|^2 \le
  \frac{\pi}{\delta} \|M\phi\|^2 \|M\psi\|^2.
  \end{eqnarray}

  Summing this over an orthonormal basis of vectors $\psi$ and then
  over an orthonormal basis of vectors $\phi$ leads to
  \begin{equation} \label{eq:B15a}
  \int \|T_{\delta}(v-iy)\|_{HS}^2 \,dy \le \frac{\pi}{\delta}
  \|M\|_{HS}^4.
  \end{equation}

  This implies (a). In fact, we have proven the stronger result that
  $T_{\delta}(\cdot)$ is a Hilbert-Schmidt-valued $H^2$-function in
  the lower half plane.

  By Lemma~\ref{lem:B4} and Proposition~\ref{prop:B3},
  $H(\mbox{Im}\,T_{\delta})$ exists almost everywhere as a
  Hilbert-Schmidt operator. Since the strong topology is weaker than
  the Hilbert-Schmidt topology, this implies the existence of
  $H\langle \mbox{Im}\, T_{\delta}(\cdot) \phi, \phi \rangle = \langle
  H(\mbox{Im}\, T_{\delta})(\cdot) \phi, \phi \rangle$ for every
  $\phi \in \Hi_1$. By (a) and Proposition~\ref{prop:B2} the latter
  is equal to $\langle \mbox{Re}\, T_{\delta}(\cdot) \phi, \phi
  \rangle$. We conclude (\ref{eq:B14}) since bounded operators are
  determined by their quadratic form.
  \end{proof}

  We are now prepared to prove (\ref{eq:B6}) and thereby complete
  the proof of Lemma~\ref{lem:HSweakbound}. We have, using
  Lemma~\ref{lem:B5}(b),
  \begin{eqnarray} \label{eq:B21}
  |\{ v:\, \|T_{\delta}(v)\|_{HS} >t \}| & \le & |\{ v:\, \|\mbox{Re}\,
  T_{\delta}(v)\|_{HS} > \frac{t}{2}\}| + |\{v:
\|\mbox{Im}\,T_{\delta}(v)\|_{HS}
  > \frac{t}{2}\}| \nonumber \\
  & = & |\{ v:\, \|H(\mbox{Im}\,T_{\delta})(v)\|_{HS} > \frac{t}{2}\}|
  + |\{v: \|\mbox{Im}\,T_{\delta}(v)\|_{HS} > \frac{t}{2}\}| \nonumber \\
  & \le & \frac{2(C+1)}{t} \int \|\mbox{Im}\,T_{\delta}(v)\|_{HS}\,dv,
  \end{eqnarray}
  where in the end Proposition~\ref{prop:B3} and Chebychev's
inequality were used.
  Thus (\ref{eq:B6}) is a consequence of Lemma~\ref{lem:B4}.

  \newpage

  \section{A disorder-averaged spectral shift bound}
  \label{sect:app-spect-shift}
  Here we present a new result, which amounts to boundedness at fixed energy
  of a fractional moment\textemdash under averaging over local
disorder\textemdash
  of the spectral shift associated with the addition to
  a Schr\"odinger operator of a local potential.

  As was explained in the introduction, some of the difficulties which
  have in the past impeded the extention of fractional moment methods to
  the continuum can be traced to the lack of uniform bounds on
  such spectral shifts.    The following result is enabled by the methods of
  Section~\ref{sec:s-moments}.  While the analysis presented above does
  not proceed through this bound, the issues involved are closely related,
  and the result may provide a useful  tool.

  \begin{thm}\label{thm:spectral-shift}
  Let $H_t = \widehat H + t V$ where $\widehat H$ satisfies ${\mathcal A}1$
  and $V$ is a non-negative bounded function with compact support.
  Let $U$ be a non-negative bounded function such that $V$ is
  strictly positive throughout the set $Q =\set{q : \dist(q,
  \supp(U)) < \delta}$ with some $\delta > 0$ and set $v_- = \inf_{x
  \in Q} V(x)$.  Then, for any $0 < s < \min(2/d, 1/2)$ there is
  $C_{s,\delta} < \infty$ such that the spectral shift function,
  defined as
  \begin{equation}\label{eq:ssdefn}
    \xi(t,E) \ = \ \tr \left [ P \left ( H_t < E \right )
      - P \left ( H_t + U < E \right ) \right] \; ,
  \end{equation}
  satisfies, for any $E \ge \inf \sigma(\widehat H)$:
  \begin{equation}\label{integ}
    \int_0^1 \abs{\xi(t,E)}^s \di t \ \le \ C_{s,\delta} \norm{U}_\infty
    (1+|E-E_0| +\norm{V}_\infty )^{s(2d + 2)}\; ,
  \end{equation}
  with $E_0= \inf \sigma(\widehat H)$.
  \end{thm}
  \begin{proof}
  We claim that it suffices to prove \eq{integ} for operators $H_t$
  restricted to bounded regions with a constant $C_{s,\delta}$ which is
  independent of the region.  To verify this, note that strong resolvent
  convergence and lower semi-continuity of the trace norm imply that
  \begin{equation}
    \xi(t,E) \ \le \ \liminf_{L \rightarrow \infty} \xi_L(t,E)
  \end{equation}
  where $\xi_L(t,E)$ is computed with $H_t^{(\Lambda_L)}$ in place
  of $H_t$ with $\Lambda_L = [-L,L]^d$.  It is useful to note that,
  because $U$ is non-negative, the difference of projections
  appearing in \eq{eq:ssdefn} is a positive semi-definite operator
  so its trace is equal to its trace norm.  An application of
  Fatou's lemma yields \eq{integ} for $H_t$ provided it holds for
  $H_t^{(\Lambda_L)}$.

  Throughout the rest of the proof we fix $L > 0$ and write $H_t$
  and $\widehat H$ for the restrictions of these operators to
  $[-L,L]^d$ with Dirichlet boundary conditions.  We begin with the
  observation that
  \begin{equation}
   \xi(t,E) \ = \ \Tr P_t \; ,
  \end{equation}
  where $P_t$ is the spectral projection to the interval $(-\infty,-1]$
  of the Birman Schwinger operator
  \begin{equation}
    K_t \ = \ U^{1/2} \frac{1}{H_t  - E} U^{1/2} \; .
  \end{equation}
  This fact follows from the Birman-Schwinger representation since
  $E$ becomes an eigenvalue of $H_t + \eta U$ precisely when $\eta$
  is equal to an eigenvalue of $ - K_t^{-1}$.

  Note that for any $n \ge 1$,
  \begin{equation}
     \norm{K_t^{-n} P_t (1 + K_t^{-1})^n} \ \le 1 \; .
  \end{equation}
  Thus by the H\"older inequality for trace norms
  \begin{equation}
    \xi(t,E) \ \le \  \norm{K_t^n}_{HS} \,  \norm{(1 + K_t^{-1})^{-n}}_{HS} \;
    .
  \end{equation}
  Noting that
  \begin{equation}
    (1 + K_t^{-1} )^{-1} \ = \ U^{1/2} \frac{1}{H_t + U - E} U^{1/2} \;
    ,
  \end{equation}
  we find
  \begin{multline}\label{eq:cauchy}
  \int_0^1 \di t \abs{\xi(t,E)}^s \ \le \ \left ( \int_0^1
    \norm{ U^{1/2} \frac{1}{H_t - E} U^{1/2} }_{2n}^{2ns} \di t \right )^{1/2}
  \\ \times \ \left (
    \int_0^1 \norm{ U^{1/2} \frac{1}{H_t + U - E} U^{1/2}}_{2n}^{2ns} \di t
  \right )^{1/2} \; ,
  \end{multline}
  where $\norm{A}_m = (\tr{\abs{A}^m})^{1/m}$.

  Lemma~\ref{lem:HSweakbound} and the representation \eq{repr2} used
in the proof
  of Lemma~\ref{lem:finite_smoment} may be used to show that the integrals
  on the right hand side of \eq{eq:cauchy} are bounded if (1) $2ns < 1$
  and (2) $4n > d$ ($d$ is the dimension).  We now outline the proof
  of this assertion.

  The arguments used in Lemma~\ref{lem:finite_smoment} can be used to produce a
  representation \begin{equation}
     U^{1/2} \frac{1}{H_t - E} U^{1/2} \ = \ U^{1/2} T \Theta \frac{1}{H_t - E}
     \Theta T^\dag U^{1/2} \ + \ B
  \end{equation}
  with $T,T^\dag$ Hilbert-Schmidt.  In the proof of
  Lemma~\ref{lem:finite_smoment} it was noted that $B$ is bounded. In
  addition, using ${\mathcal A}1$, $B$ can be seen to be in $\I_p$ for any
  $p > d/2$, with uniform bounds on its $\I_p$ norm. Since
  $\norm{\cdot}_{2n} \le \norm{\cdot}_{HS}$ we find that
  \begin{equation}
    \int_0^1
    \norm{ U^{1/2} \frac{1}{H_t - E} U^{1/2} }_{2n}^{2ns} \di t \ \le \ \int_0^1
    \norm{ U^{1/2} T \Theta \frac{1}{H_t - E} \Theta T^\dag U^{1/2} }_{HS}^{2ns}
    \ + \ {\mathrm{O}(1)} \; ,
  \end{equation}
  with a similar expression for the other factor in \eq{eq:cauchy}.
  The weak $L^1$ bound can be used to bound this final integral
  since
  \begin{equation}
    \Theta \frac{1}{H_t - E} \Theta \ = \
    \Theta V^{-1/2} \frac{1}{t + \widehat K^{-1}} V^{-1/2} \Theta \; ,
  \end{equation}
  with appropriate $\widehat K$.  Note that $\Theta V^{-1/2}$ is
  bounded since $\Theta$ is supported in $Q$. As in
  Sec.~\ref{sec:s-moments} there is slight complication due to the
  fact that $T$ (and $T^\dag$) depend on $t$.  However, as before,
  the dependence is polynomial and may be handled in the same way.
  \end{proof}

  \bigskip

  \section*{Acknowledgments}
  In the course of this project, the authors' work was supported in part by
  NSF grants PHY-9971149 (MA, and AE),  DMS 0070343 and
  0245210 (GS), INT-0204308 (JS),
  NSF postdoctoral research fellowship (JS), and a NATO
  Collaborative Linkage Grant PST.CLG.976441 (GS and SN).
  The authors also thank Caltech, Universit\'{e} Paris 7, and
  the Institute Mittag-Leffler for hospitality which has facilitated
  this collaboration.

  \newpage


\end{document}